\documentclass[paper=a4,fontsize=11pt,DIV=10,numbers=noendperiod,bibliography=totoc,longbibliography]{scrartcl}

\usepackage[utf8]{inputenc}
\usepackage[english]{babel}

\usepackage{microtype}

\usepackage{amsmath}
\usepackage{amsthm}
\usepackage{amssymb}

\usepackage[T1]{fontenc}
\usepackage{lmodern}
\usepackage{textcomp}
\usepackage{bbm}

\usepackage{graphicx}

\usepackage[sort]{natbib}

\usepackage[breaklinks=true,colorlinks]{hyperref}

\KOMAoptions{DIV=last}

\newcommand{\cf}{cf.\ }

\newcommand{\coloneq}{\mathrel{\mathop:}=}
\newcommand{\eqcolon}{=\mathrel{\mathop:}}
\newcommand{\dd}{\mathrm{d}}
\newcommand{\Tr}{\operatorname{Tr}}
\newcommand{\realt}{\operatorname{Re}}

\newcommand{\bkew}[3]{\left\langle{#1}\middle|{#2}\middle|{#3}\right\rangle}
\newcommand{\ket}[1]{\left|{#1}\right\rangle}

\newcommand{\ketbra}[2]{\left|{#1}\middle\rangle\middle\langle{#2}\right|}
\newcommand{\proj}[1]{\ketbra{#1}{#1}}
\newcommand{\ew}[1]{\left\langle{#1}\right\rangle}

\newcommand{\kB}{k_\mathrm{B}}
\newcommand{\sigmax}{\sigma_x}

\newcommand{\sigmaz}{\sigma_z}
\newcommand{\sminus}{\sigma_-}
\newcommand{\splus}{\sigma_+}

\newcommand{\indexc}{\mathrm{c}}
\newcommand{\indexh}{\mathrm{h}}
\newcommand{\Gc}{G_\indexc}
\newcommand{\Gh}{G_\indexh}
\newcommand{\betaeff}{\beta_\mathrm{eff}}
\newcommand{\betac}{\beta_\indexc}
\newcommand{\betah}{\beta_\indexh}

\newcommand{\Tc}{T_\indexc}
\newcommand{\Th}{T_\indexh}
\newcommand{\Jc}{J_\indexc}
\newcommand{\Jh}{J_\indexh}

\newcommand{\psib}{\psi_\mathrm{b}}
\newcommand{\psid}{\psi_\mathrm{d}}
\newcommand{\Neff}{N_\mathrm{eff}}

\newcommand{\Pidew}{\ew{\Pi_\mathrm{d}}_{\rho(0)}}
\newcommand{\Omegacrit}{\Omega_\mathrm{crit}}
\newcommand{\Omegamax}{\Omega_\mathrm{max}}

\newcommand{\rhop}{\rho_\mathrm{P}}
\newcommand{\rhoptilde}{\tilde{\rho}_\mathrm{P}}

\newcommand{\rhoeq}{\rho_\mathrm{eq}}

\newcommand{\rhosb}{\rho_\mathrm{SB}}

\newcommand{\Hs}{H_\mathrm{S}}
\newcommand{\Hp}{H_\mathrm{P}}
\newcommand{\Hb}{H_\mathrm{B}}
\newcommand{\Hsp}{H_\mathrm{SP}}
\newcommand{\Hsb}{H_\mathrm{SB}}
\newcommand{\Htot}{H_\mathrm{tot}}
\newcommand{\Tp}{T_\mathrm{P}}

\newcommand{\Sp}{\mathcal{S}_\mathrm{P}}
\newcommand{\Es}{E_\mathrm{S}}
\newcommand{\sinc}{\operatorname{sinc}}

\newcommand{\Wmax}{W_\mathrm{max}}
\newcommand{\tc}{t_\mathrm{c}}

\newcommand{\cv}{C_\mathrm{V}}

\title{Thermodynamics of quantum systems under dynamical control}

\author{David Gelbwaser-Klimovsky\thanks{Department of Chemical Physics, Weizmann Institute of Science, Rehovot 7610001, Israel}~\thanks{These authors contributed equally to this work.} \and Wolfgang Niedenzu\footnotemark[1]~\footnotemark[2] \and Gershon Kurizki\footnotemark[1]}

\date{\today}

\hypersetup{pdfauthor={David Gelbwaser-Klimovsky, Wolfgang Niedenzu and Gershon Kurizki}, pdftitle={Thermodynamics of quantum systems under dynamical control}}

\begin{document}

\maketitle

\tableofcontents

\begin{abstract}
In this review the debated rapport between thermodynamics and quantum mechanics is addressed in the framework of the theory of periodically-driven/controlled quantum-thermodynamic machines. The basic model studied here is that of a two-level system (TLS), whose energy is periodically modulated while the system is coupled to thermal baths. When the modulation interval is short compared to the bath memory time, the system-bath correlations are affected, thereby causing cooling or heating of the TLS, depending on the interval. In steady state, a periodically-modulated TLS coupled to two distinct baths constitutes the simplest quantum heat machine (QHM) that may operate as either an engine or a refrigerator, depending on the modulation rate. We find their efficiency and power-output bounds and the conditions for attaining these bounds. An extension of this model to multilevel systems shows that the QHM power output can be boosted by the multilevel degeneracy. 
\par
These results are used to scrutinize basic thermodynamic principles: (i) Externally-driven/modulated QHMs may attain the Carnot efficiency bound, but when the driving is done by a quantum device (``piston''), the efficiency strongly depends on its initial quantum state. Such dependence has been unknown thus far. (ii) The refrigeration rate effected by QHMs does not vanish as the temperature approaches absolute zero for certain quantized baths, e.g., magnons, thous challenging Nernst's unattainability principle. (iii) System-bath correlations allow more work extraction under periodic control than that expected from the Szilard-Landauer principle, provided the period is in the non-Markovian domain. Thus, dynamically-controlled QHMs may benefit from hitherto unexploited thermodynamic resources.
\end{abstract}

\section{Introduction}\label{sec_introduction}

To what extent can dynamical control of open quantum systems enhance the ability to extract work or cooling from available thermodynamic resources? If it can, then we may be able to pursue new or improved quantum-thermodynamic functionalities by harnessing such control to our advantage. Here we review our past and present research that has been dedicated to the elucidation of this issue. In what follows, we briefly summarize the principles of thermodynamics, discuss their applicability in the quantum domain, then revisit some of its tenets that may be challenged for quantum systems under dynamical control.

\subsection{The principles of thermodynamics for quantum systems}

The theoretical framework of thermodynamics rests on its four basic laws~\citep{callenbook,gemmerbook}. The \emph{zeroth law} allows the definition of temperature and the \emph{first law} is that of energy conservation. Some processes may be allowed by the first law, but are forbidden by the \emph{second law} of thermodynamics. A prime example is the flow of heat between cold and hot bodies. While the first law does not set any preferred direction to the flow, the second law requires heat to flow from the hot to the cold body. The \emph{third law} forbids any cooling process to attain the absolute zero.

\par

Despite efforts to reconcile quantum mechanics and thermodynamics over the years, their compatibility is still an open fundamental problem, with crucial bearing on the validity of these thermodynamic laws and the performance bounds of heat machines in the quantum domain~\citep{gemmerbook,kosloff2013quantum}. A priori, it is unclear why these two disciplines should be related: While thermodynamics was developed as a theory that limits possible macroscopic processes, quantum mechanics describes mainly microscopic systems. Nevertheless, thermodynamic principles have played an essential r\^ole in the development of quantum mechanics. Planck proposed the existence of quanta~\citep{planck1900theorie} by employing a thermodynamic approach to explain the blackbody radiation spectrum. Based on thermodynamic arguments, Einstein predicted stimulated emission~\citep{einstein1916strahlung}. In the same way, a thermodynamic approach may yet provide new insights into the ability to exploit quantum processes in machines designed for the performance of work or cooling. 

\subsection{Why study quantum heat machines?}\label{sec_why_study_qhm}

Notwithstanding the long-standing rapport of thermodynamics and quantum mechanics, the basic question remains open: What limits does thermodynamics set to the performance of quantum-mechanical devices, particularly, quantum heat machines? 

\par

This fundamental issue has been feeding for years the lively interest in quantum thermodynamics that has culminated in a multitude of quantum heat machine models~\citep{geusic1967quantum,alicki1979quantum,ford1985quantum,geva2002irreversible,lin2003performance,kieu2004second,allahverdyan2005minimal,vandenbroeck2005thermodynamic,ford2006quantum,jarzynski2007comparison,quan2007quantum,talkner2007fluctuation,campisi2009fluctuation,esposito2009universality,blickle2011realization,campisi2011colloquium,scully2011quantum,vandenbroeck2012efficiency,linden2010how,gemmerbook,esposito2010entropy,kosloff2013quantum,martinez2013dynamics,abah2012single,chen2012quantum,cleuren2012cooling,esposito2012stochastically,venturelli2013minimal}. Many of these models have reproduced the classic thermodynamic bounds, such as the Carnot efficiency limit. On the other hand, there has been a growing number of claims that those bounds can be violated at the quantum level~\citep{scully2001extracting,scully2003extracting,dillenschneider2009energetics,galve2009nonequilibrium,huang2012effects,boukobza2013breaking,correa2014quantum,rossnagel2014nanoscale,abah2014efficiency}. These claims call for a clarification of the general principles of quantum heat machines, beyond specific realizations, and for the development of tools that would safeguard us against inconsistencies with the laws of thermodynamics. These motivations underlie the present study of simple dynamically-controlled quantum heat machines, their efficiency and power bounds.
 
\par

The theoretical study of the quantum-thermodynamic models mentioned above has not yet been followed by experimental demonstrations. The models for quantum thermodynamic machines that have emerged from our research presented here~\citep{gelbwaser2013minimal,gelbwaser2013work,gelbwaser2013workquantized,gelbwaser2014heat,gelbwaser2014power,kolar2012quantum,niedenzu2015performance,gelbwaser2015work} also await experimental verification. Such verification would not only advance the state of the art of thermodynamics but would also benefit the ongoing effort aimed at device miniaturization towards the nanoscale~\citep{cui2001functional,gudiksen2002growth,blais2004cavity,hu2007heterostructure,burek2012free,maletinsky2012robust,aspelmeyer2014cavity}. As the device size approaches the nanoscale, we are necessarily confronted with quantum thermodynamic issues~\citep{segal2008stochastic,blickle2011realization,scully2011quantum}. This prompts the need for studying quantum heat machines and formulating their thermodynamic performance limits and design principles that are acutely needed for any application where cooling rate, power and size constraints are very severe. Particularly, the miniaturization of refrigerators becomes important, in view of the growing density of transistors on microchips that increases heat production. These developments require smaller and more efficient coolers and power sources~\citep{pop2006heat,epstein1995observation,clark1996laser,mungan1997laser,genes2009micromechanical,linden2010how}.

\par

The considerations outlined above underscore the need to resolve the following issues: (i) Is the Carnot bound on efficiency upheld in quantum heat machines? (ii) Are the design principles of quantum heat machines as regards work or cooling power different from those of their classical counterparts, especially when their operation is non-adiabatic (and hence far from the Carnot limit cycle)? Here we show that for the simple but generic models we analyze, the answer to the first question is affirmative (Secs.~\ref{sec_periodically-modulated} and~\ref{sec_nlevels}), but quantum-state preparation of the ``piston'' in the heat machine may temporarily cause it to outperform the standard Carnot bound, since the quantum state is an extra thermodynamic resource~\citep{gelbwaser2014heat} (Secs.~\ref{sec_quantum_piston} and~\ref{sec_refrigerator}). The answer to the second question is also affirmative (Secs.~\ref{sec_periodically-modulated} and~\ref{sec_nlevels}): The highly non-adiabatic operation of such quantum heat machines~\citep{gelbwaser2013minimal} may yield much higher power than their non-adiabatic classical-like counterparts~\citep{geva1992quantum,geva1992classical}.

\subsection{Quantum heat machines: Basics and outstanding issues}\label{sec_introduction_basics}

At the macroscopic level, thermodynamic laws govern the exchange of work and heat, bounding the efficiency of heat engines and refrigerators~\citep{callenbook,kondepudibook}. In standard models of such machines, work extraction or cooling are based on a ``working fluid'' (system $S$) that is weakly coupled to two infinite, Markovian, heat baths, satisfying system-bath separability. The two heat baths are at equilibrium at different temperatures. The system is driven in a cyclic way by a piston ($P$) that modulates the system energy while giving or extracting work. These workhorse models of heat machines are used to test their compatibility with the laws of thermodynamics.

\par

A microscopic quantum description of a heat engine or refrigerator~\citep{spohn1978entropy,geva1992quantum,levy2012quantum,alicki2004thermodynamics} requires the formulation of thermodynamic laws within a dynamical context, wherein the working fluid is a quantum system, driven by an external force (field) or by quantum non-demolition (QND) energy measurements or phase shifts of the system, while the system is coupled to reservoirs (heat baths): At this level of description one may hope to derive, rather than postulate, the constraints and bounds imposed by thermodynamics. This amounts to the analysis of the following thermodynamic quantities:

\par

(i) The average energy of the system in state $\rho$ subject to the time-dependent Hamiltonian $\Hs(t)$
\begin{subequations}\label{eq_energy_heat_work}
  \begin{equation}
    \ew{\Hs(t)} = \Es(t)=\Tr\left[\rho(t)\Hs(t)\right],
  \end{equation}
  (ii) the heat 
  \begin{equation}
    Q(t)=\int_{0}^{t} \Tr \left[\frac{\dd\rho(t^\prime)}{\dd t^\prime} \Hs(t^\prime) \right] \dd t^\prime,
  \end{equation}
  and (iii) the work
  \begin{equation}
    W(t)=\int_{0}^{t} \Tr \left[\rho(t^\prime) \frac{\dd \Hs(t^\prime)}{\dd t^\prime}\right] \dd t^\prime,
  \end{equation}
\end{subequations}
are related by the first law of thermodynamics
\begin{equation}
 \frac{\dd \Es}{\dd t}=\frac{\dd W}{\dd t}+\frac{\dd Q}{\dd t}.
\end{equation}
In a closed cycle, the standard division of energy exchange between heat and work under \emph{classical (parametric)} driving of the reduced state of the system, $\rho(t)$, via a \emph{cyclic} system Hamiltonian $\Hs(t)$, is~\citep{alicki1979quantum}
\begin{subequations}\label{eq_heat_work_general_definition}
 \begin{align}
  Q(t)&=\oint_\mathrm{cycle} \Tr\left[\dot{\rho}(t) \Hs(t)\right] \dd t\\
  W(t)&=\oint_\mathrm{cycle} \Tr\left[\rho(t)\dot{H}_\mathrm{S}(t)\right]\dd t.\label{eq_work_general_definition}
 \end{align}
\end{subequations}
(iv) The system (von~Neumann) entropy ($\kB$ denotes the Boltzmann constant)
\begin{equation}\label{eq_entropy_general_definition}
 \mathcal{S}(\rho)=-\kB\Tr (\rho\ln \rho)
\end{equation}
varies partly due to heat exchange with a bath at temperature $T$, and partly due to the so-called entropy production $\sigma(t)$, satisfying~\citep{alicki1979quantum}
\begin{equation}\label{eq_introduction_entropy_production}
 \frac{\dd\mathcal{S}}{\dd t} = \frac{1}{\kB T} \frac{\dd Q}{\dd t} +\sigma(t).
\end{equation}
The first term on the r.h.s. of Eq.~\eqref{eq_introduction_entropy_production} can be either negative or positive, corresponding to cooling or heating of the system, respectively. The second term is \emph{always non-negative} according to Spohn's theorem~\citep{spohn1978entropy}, considered fundamental in the theory of open quantum systems~\citep{breuerbook}, i.e., 
\begin{equation}\label{eq_entropy_production_positive}
 \sigma(t)\geq 0,
\end{equation}
expressing the monotonic evolution of the entropy relative to its equilibrium value~\citep{spohn1978entropy,lindblad1974expectations}. Equation~\eqref{eq_entropy_production_positive} is commonly held to be a statement of the second law or the time arrow (directionality) towards entropy increase~\citep{alicki2004thermodynamics}. Yet its validity crucially depends on the Markovian approximation (Sec.~\ref{sec_nonmarkovian}).

\par

\begin{figure}
  \centering
  \includegraphics[width=9cm]{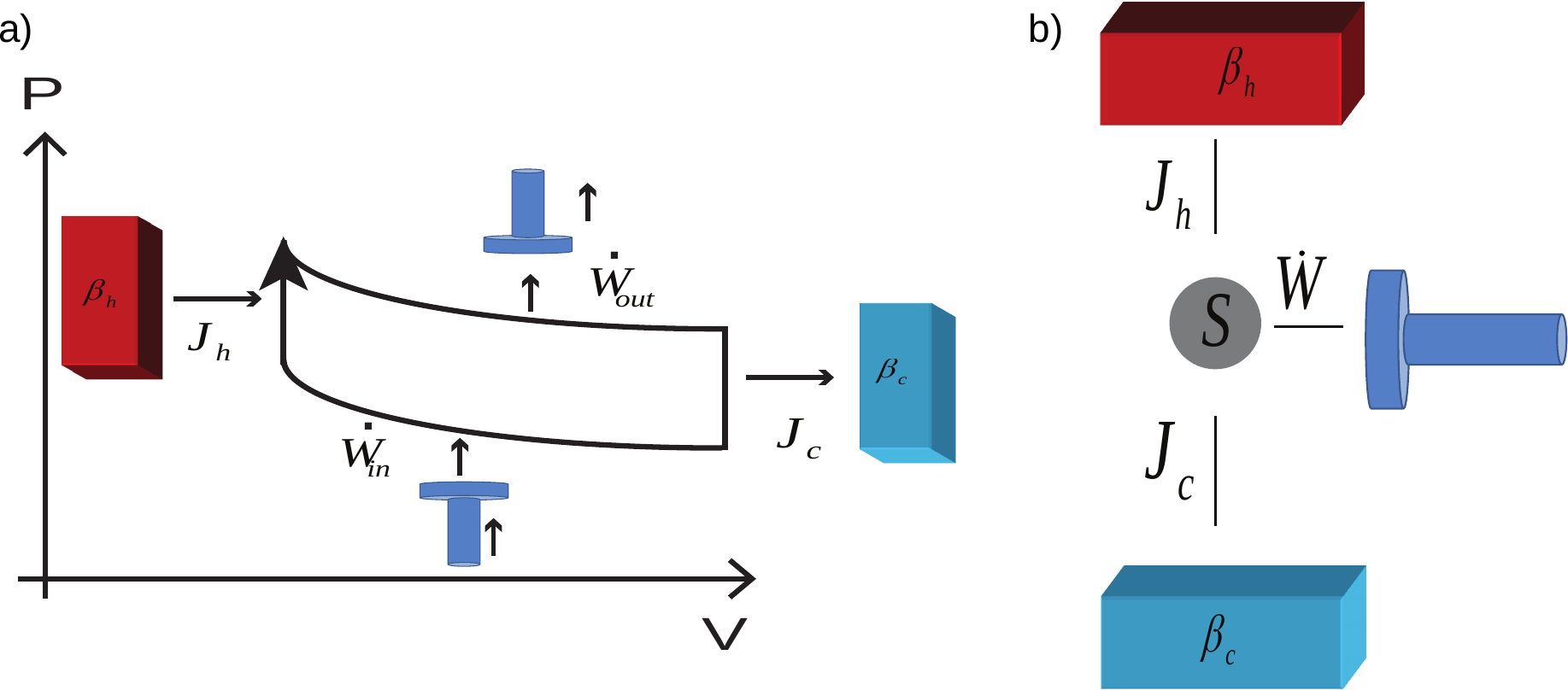}
  \caption{(a) A reciprocating (Otto) cycle heat machine operates in 4 strokes: Alternating coupling to the hot and the cold baths, as well as alternating compression and decompression by the piston. (b) A continuous-cycle quantum heat machine: The working fluid (system $S$) is permanently coupled to a cold heat bath at inverse temperature $\betac$ and a hot heat bath at $\betah$. The piston modulates the system and allows for power ($\dot{W}$) extraction or supply.}\label{fig_heat_machines}
\end{figure}

Most of the quantum heat machine models that have been recently proposed and studied~\citep{ford2006quantum,ford1985quantum,vandenbroeck2005thermodynamic,esposito2009universality,kieu2004second,scully2011quantum,blickle2011realization,allahverdyan2005minimal,talkner2007fluctuation,campisi2009fluctuation,campisi2011colloquium,jarzynski2007comparison,geusic1967quantum,geva2002irreversible,quan2007quantum,lin2003performance,bender2000quantum,alicki2004thermodynamics,alicki1979quantum,segal2006molecular,segal2008stochastic,geva1992quantum,bender2002entropy,he2002quantum,feldmann2000performance,wang2009thermal,henrich2007quantum,jahnke2008nature,allahverdyan2008work,feldmann2004characteristics,rezek2006irreversible,abah2012single,thomas2011coupled,he2012thermal,wang2013efficiency,feldmann1996heat,feldmann2010minimal,rempp2007cyclic,kaufman2012cooling} are the quantum counterparts of the classical \emph{reciprocating-cycle} engine. A reciprocating cycle consists of ``strokes'' in which $S$ alternates between coupling to the ``hot'' and ``cold'' heat baths and may also be modulated by $P$. Typically, it has four strokes, as in the case of the Carnot and Otto cycles (see Fig.~\ref{fig_heat_machines}a). Both are composed of two adiabatic strokes in which the working medium is isolated from the environment while being driven by the piston, and two heat transfer strokes in which the working medium is alternately coupled to one of the heat baths. The latter strokes are isotherms for the Carnot cycle and isochores for the Otto cycle. At the macroscopic level, reciprocating models describe most of the existing machines. For example, a car engine performs an Otto cycle, while standard refrigerators execute an absorption-compression refrigerator cycle~\citep{callenbook}.

\par
\par

\par

Yet, in microscopic or nanoscopic devices, especially when they operate quantum mechanically, reciprocating cycles pose a serious problem: \emph{On-off switching of system-bath interactions and their non-adiabaticity (finite duration) may strongly affect energy and entropy exchange}, which casts doubts on the validity of commonly discussed models that ignore such effects. Shortcuts to adiabaticity have been motivated by these concerns~\citep{torrontegui2013shortcuts,deng2013boosting}. In any case, for nanoscale systems totally embedded in thermal baths, such system-bath decoupling may not be possible~\citep{levy2014local}.

\par

For these reasons, it is expedient to consider \emph{continuous-cycle} heat machines~\citep{levy2012quantumrefrigerators,kosloff2014quantum,gelbwaser2013minimal,gelbwaser2013workquantized,gelbwaser2014heat,gelbwaser2014power,niedenzu2015performance,gelbwaser2015work,alicki2014quantum}. In a continuous cycle there are no strokes, neither coupling nor decoupling to and from the baths. Instead, the working fluid $S$ is \emph{permanently} coupled to the two heat baths while being modulated by $P$ (see Fig.~\ref{fig_heat_machines}b). These machines describe scenarios that are common at quantum scales. The analysis of simple continuous-cycle quantum heat machines is the subject of Secs.~\ref{sec_periodically-modulated},~\ref{sec_nlevels}, and~\ref{sec_quantum_cooler}. To appreciate the merit of continuous-cycle models, it is instructive to consider (Sec.~\ref{sec_quantum_cooler}) a gas mixture of two atomic species, one of which is driven by a laser which plays the r\^ole of the piston $P$. The driven atoms represent the system $S$. They interact through collisions with another atomic species (a ``buffer gas''), which represents one of the baths. Concurrently, the $S$ atoms are coupled to the electromagnetic vacuum (the second bath) through spontaneous emission. The impossibility of avoiding spontaneous emission and collisions with the buffer gas makes the coupling to or decoupling from the baths impossible, so that the reciprocating-cycle models are irrelevant in such scenarios, as opposed to continuous-cycle models.

\subsection{The Carnot and maximum-power bounds for quantum heat machines}\label{sec_introduction_carnot}

 The importance of the reciprocating cycle stems from its ability to set a universal efficiency limit, known as the Carnot bound~\citep{carnotbook}, which applies to any work extraction process. Just like Kelvin's formulation of the second law of thermodynamics, the Carnot bound restricts work extraction in cyclic processes involving a system (working fluid) and baths at different temperatures. The bound is reached only by the ideal, reversible, infinitely-slow Carnot cycle whose efficiency is described by the simple formula
\begin{equation}\label{eq_carnot_eta}
 \eta_\mathrm{Carnot}=1-\frac{\Tc}{\Th},
\end{equation}
where $T_{\indexc(\indexh)}$ is the temperature of the cold (hot) heat bath. 

\par

In the case of refrigerators and heat pumps (HP) the corresponding bounds are set by the coefficient of performance (COP) and are 
\begin{equation}\label{eq_carnot_cop}
 \mathrm{COP}^\mathrm{refrigerator}_\mathrm{Carnot}=\frac{\Tc}{\Th-\Tc}
\end{equation}
and
\begin{equation}
 \mathrm{COP}^\mathrm{HP}_\mathrm{Carnot}=\frac{\Th}{\Th-\Tc},
\end{equation}
respectively.

\par
	
The efficiency bound for a work cycle at \emph{maximum power} is usually of greater practical interest than the Carnot efficiency, which can only be achieved in the limit of vanishing power. Under typical conditions (such as linear heat transfer) the efficiency at maximum power obeys the so-called Curzon-Ahlborn bound~\citep{curzon1975efficiency,esposito2010efficiency,novikov1958efficiency,chambadalbook,yvonbook}
\begin{equation}\label{eq_curzohn-ahlborn}
 \eta_\mathrm{CA}=1-\sqrt{\frac{\Tc}{\Th}}.
\end{equation}
However, contrary to the Carnot bound, it is not universal and may be surpassed by suitable designs that do not contradict any thermodynamic law~\citep{gelbwaser2013minimal,esposito2010efficiency,birjukov2008quantum}. 

\par

Although the Carnot bound is considered universal in certain models of quantum heat machines~\citep{alicki1979quantum,spohn1978entropy,davies1974markovian,lindblad1975completely,alicki2012periodically}, it has been repeatedly contested, as mentioned in Sec.~\ref{sec_why_study_qhm}~\citep{scully2001extracting,scully2003extracting,boukobza2013breaking,correa2014quantum,rossnagel2014nanoscale}, based on the assertion that there may be non-classical thermodynamic resources, e.g., quantum coherence~\citep{scully2003extracting}, bath squeezing~\citep{correa2014quantum,galve2009nonequilibrium,abah2014efficiency}, or negative temperature of the bath~\citep{dunkel2013consistent} that may enhance work extraction or cooling power. But is it indeed possible to use quantum effects to transgress the Carnot bound? 
	
\par

At least in some models of quantum heat machines, a careful thermodynamic analysis, which considers the cycle-initialization costs as well as their non-equilibrium dynamics, has shown that the resulting performance bounds are not higher than Carnot~\citep{gelbwaser2013work,gelbwaser2013workquantized,gelbwaser2014heat,zubairy2002photo,opatrny2005maser}. Yet, the thermodynamic implications of quantum effects, such as coherence/interference, as well as the work or cooling capacity of quantum states, are still not fully understood, so that such potential quantum resources must be further scrutinized.

\par

Here we put forward (Secs.~\ref{sec_cycles},~\ref{sec_periodically-modulated}, and~\ref{sec_nlevels}) a rigorous approach to continuous cycles in quantum heat machines: We study the steady-state dynamics of \emph{periodically-driven} open quantum systems that are \emph{permanently} coupled to heat baths. We ask: Are the restrictions set by thermodynamics on such continuous cycles of quantum machines the same as for their classical counterparts or for reciprocating-cycle quantum machines? As shown in Sec.~\ref{sec_periodically-modulated}, continuous-cycle quantum heat machines require spectrally separated baths and reach their maximal efficiency/COP, the Carnot bound, when modulated at a critical frequency (rate) yielding zero power.

\par

A major impediment towards finding the ultimate performance limits of quantum heat machines is the unresolved question: If two quantum systems exchange energy, how is it in general divided into work and heat? The second law of thermodynamics draws an essential distinction between heat and work: Work can be completely transformed into heat in a cyclic process, yet the opposite is not true. While at the classical level there is a standard and well-established definition of work, this is not the case at the quantum level. In recent years there have been many different proposals for defining the work performed by quantum heat machines~\citep{gelbwaser2013workquantized,boukobza2006thermodynamics,schroeder2010work,horodecki2013fundamental,goswani2013thermodynamics,gelbwaser2014heat,skrzypczyk2013extracting}. Nevertheless, a consensus has not yet been reached.

\par

In order to answer the questions above, we put forward a fully quantum heat machine model, where the periodic driving is replaced by a quantum piston. This model differs from the vast majority of quantum heat machine models that have been proposed to date, which do not address this issue~\citep{gemmerbook,alicki1979quantum,kosloff2013quantum,geusic1967quantum,ford2006quantum,geva2002irreversible,quan2007quantum,vandenbroeck2005thermodynamic,lin2003performance,esposito2009universality,kieu2004second,agarwal2013quantum,scully2011quantum,blickle2011realization,allahverdyan2005minimal,campisi2009fluctuation,campisi2011colloquium,brunner2014entanglement} as they employ classical fields or forces to drive the working fluid and thus may be deemed \emph{semiclassical}. As a result, their size and energy consumption may well impede overall miniaturization of the device, limiting potential applications and experimental realizations at the nanoscale. In analogy to light-matter interaction~\citep{scullybook}, where the quantization of light results in new effects, the same may be true for quantum thermal machines. Would the performance bound of a quantum heat machine be the same if the piston is considered to be a quantum device instead of an external classical modulation?

\par

As discussed in Secs.~\ref{sec_quantum_piston} and~\ref{sec_refrigerator}, the study of fully-quantized heat machines must rely on a physically sound definition of work and take into account the quantum nature of the piston. Consequently, the efficiency of fully quantized heat engines and the coefficient of performance (COP) of their refrigerator counterparts are shown to strongly depend on the initial quantum piston state, its subsequent thermalization and entropy production. These properties of the state, nicknamed \emph{non-passivity} by us, are shown to determine the ability of the quantized piston to serve as a thermodynamic resource.

\subsection{The third law as absolute-zero unattainability}

Quantum refrigerators may be used to study the third law of thermodynamics in the quantum domain. The \emph{dynamical} formulation of the third law of thermodynamics~\citep{nernst1906ueber,landsberg1956foundations,belgiorno2003notes}, Nernst's \emph{unattainability principle}, forbids cooling to attain the absolute zero ($T=0$) in a finite number of steps, or, more generally, \emph{in finite time}. However, the universality of this principle has been postulated rather than proven. It is also debatable whether this formulation is always equivalent to Nernst's \emph{heat theorem}, whereby the \emph{entropy vanishes} at $T=0$~\citep{landsberg1956foundations,levy2012quantum}. Does the unattainability principle also apply to quantum processes? As shown in Sec.~\ref{sec_cooling_speed} for a minimal model of a quantum refrigerator consisting of a qubit driven by $\pi$-flips, the cooling rate of certain realistic quantized baths does not vanish as $T \rightarrow 0$, thus challenging the unattainability principle~\citep{kolar2012quantum}.

\subsection{Work-information tradeoff and its Szilard-Landauer bound}

Apart from issues related to quantum heat machines, quantum thermodynamics is concerned with the fundamental problems of thermal equilibration and the work-information tradeoff. The following tenets represent the prevailing views of these problems:

\begin{itemize}

\item The thermal equilibrium of a system and a bath cannot be changed by mere observation of the system, unless some action is taken (by a ``demon'') pursuant to the observation~\citep{moorebook,szilard1929ueber,landauer1961irreversibility,caves1990quantitative,scully2001extracting,maruyama2009colloquium,toyabe2010experimental,delrio2011thermodynamic,jacobs2012quantum,deffner2013information}. If observations are of the QND kind, so that they affect neither the state of the system nor that of the bath, then any such change may be suspected to be a violation of either the first or the second law, as in the case of~\citep{allahverdyan2005minimal}.

\item Equivalently, work cannot be performed in a closed cycle by a system coupled to a bath solely as a result of observations: The first law implies that observations affect neither the energy nor the heat exchange between the system and the bath~\citep{alicki1979quantum}. Furthermore, Kelvin's formulation of the second law precludes the extraction of work in a cycle by a system that is coupled to a single bath. So does Lindblad's formulation of the second law~\citep{lindbladbook}. 

\item In a Szilard engine, by contrast, work extraction is obtained through specific actions depending on the result of the observation~\citep{szilard1929ueber}. If information gathered by observations is commuted into work by a ``demon'', this work must obey the Szilard-Landauer (SL) bound~\citep{moorebook,landauer1961irreversibility,caves1990quantitative,scully2001extracting,maruyama2009colloquium,toyabe2010experimental,delrio2011thermodynamic,jacobs2012quantum}: It must not exceed the cost incurred by the erasure (resetting) of the demon's memory. If one has a $d$-level quantum system in a pure state, one can draw $\kB T \ln d$ work out of the heat bath. If the system is in a mixed state $\rho$ with entropy $\mathcal{S}(\rho)$, then the amount of work that can be extracted is~\citep{alicki2004thermodynamics}
\begin{equation}\label{eq_work_information_sl}
  W_\mathrm{ext}=\kB T \ln d - T \mathcal{S}(\rho).
\end{equation}
Here we follow the convention that \emph{extractable} work has a positive sign for an engine. The function $f(\rho)=\ln d -\mathcal{S}(\rho)/\kB$ can be viewed as the information content of the state. This bound is commonly construed to follow from the second law.

\end{itemize}

Yet, our recent research~\citep{erez2008thermodynamic,gordon2009cooling,alvarez2010zeno,gordon2010equilibration,gelbwaser2013minimal,chapin2008quantum} [see also~\citep{jahnke2010quantum}] indicates that the foregoing tenets do not hold if the quantum system, as simple as a qubit, is subject to dynamical control consisting of frequent observations or phase shifts, even if they do not directly affect the system. The basic requirements for such anomalies are that these acts of control should be shorter than all the relevant time scales and their intervals should be within the memory time of the bath. Namely, the violation of these tenets is intimately connected with the breakdown of both the Markovian and the adiabatic approximations. As shown in Secs.~\ref{sec_nonmarkovian} and~\ref{sec_work_information}, frequent unread measurements of a quantum system coupled to a thermal bath may yield work in a closed cycle from the system-bath interaction (correlation) energy, unaccounted for by the SL bound. Whereas work and information are related by Eq.~\eqref{eq_work_information_sl} that draws upon Markovian assumptions, their relation is changed by effects of system-bath correlations within non-Markovian time intervals. Nevertheless, the basic laws of thermodynamics, in this case the first and second laws, remain intact despite these violations and may be colloquially summarized by a single tenet: \emph{``There are no free lunches''} (Sec.~\ref{sec_work_information}).

\subsection{Main results}

The results presented in this review may be summarized as follows, according to its sections.

\begin{enumerate}
\item A master equation is presented for an open quantum system weakly coupled to two heat baths and periodically modulated by an external drive, based on the Floquet expansion of the Lindblad generator. This master equation is the prerequisite for studying steady-state thermal machines that are periodically driven by external fields. It serves to infer the heat currents and power output of these machines that (by construction) satisfy the second law of thermodynamics (Sec.~\ref{sec_cycles}). This general theoretical framework is used to analyze several models, as detailed below.

\item A two level system (TLS) that is simultaneously coupled to hot and cold baths via off-diagonal ($\sigmax$ Pauli-operator) coupling is studied. The TLS transition frequency is periodically modulated by an external field. It is shown that this setup is a minimal model of a quantum heat machine with two operation modes (engine and refrigerator) that can be interchanged by varying the modulation rate. Under spectral separation of the two baths, this machine reaches the Carnot bound. The maximum efficiency at maximum power is also calculated and shown to be larger than the Curzon-Ahlborn efficiency (Sec.~\ref{sec_periodically-modulated}).

\item These results are extended to multilevel systems with upper-state degeneracy (Sec.~\ref{sec_nlevels}). This degeneracy is shown to be a power-boosting resource. Although coherences between degenerate levels may affect this power boost, it is primarily dependent on the ability of the initial state to thermalize.

\item Work extraction and cooling are studied in a self-contained model of a TLS-based heat machine driven by a quantum device (quantum ``piston''/``battery''). The standard definition of work is shown to fail in such fully-quantum setups. The correct extractable work (based on the notion of non-passivity), or cooling, and their efficiency bound are shown to crucially depend on the initial quantum state of the piston/battery. The operation mode of the machine (engine or refrigerator) and performance depend on the quantum-piston frequency and non-passivity. In Sec.~\ref{sec_quantum_piston} the efficiency bound of the self-contained quantum engine is shown to be allowed to temporarily exceed the standard Carnot bound, whereas in Sec.~\ref{sec_refrigerator} its quantum-refrigerator counterpart is shown to be able to surpass the Carnot bound on its coefficient of performance (COP). In both cases the compliance with the second law is ensured, and the extra efficiency reveals the presence of hitherto unknown thermodynamic resources embodied by the quantum states.

\item Another model consists of a laser field that modulates a TLS coupled to a dephasing bath by the $\sigmaz$ Pauli operator and to the electromagnetic-field (vacuum) bath by $\sigmax$-coupling. The dephasing bath may be heated up or cooled down depending on the laser-field detuning from the TLS resonance. This model explains the cooling of a gas by collisional redistribution of radiation that was experimentally observed in~\citep{vogl2009laser} (Sec.~\ref{sec_quantum_cooler}).

\item A minimal model of a quantum refrigerator (QR), i.e., a periodically phase-flipped TLS permanently coupled to a finite-capacity cold bath and to an infinite-capacity heat dump (hot bath), is used to investigate the cooling of the cold bath towards absolute zero. The cold-bath cooling rate is shown not to vanish as $T\rightarrow0$ for quantized baths comprised of magnons in the Heisenberg spin-chain model. This result challenges Nernst's third-law formulation known as the unattainability principle (Sec.~\ref{sec_cooling_speed}).

\item Frequent measurements can control the temperature and entropy of a quantum system immersed in a heat bath if performed at intervals shorter than the bath memory (correlation) time, i.e., within the non-Markovian time domain (Sec.~\ref{sec_nonmarkovian}). Such measurements may enable the system to do more work in a closed cycle than permitted by the Szilard-Landauer principle. They allow the extraction of extra work from the system-bath correlations, a hitherto unexploited work resource. This is possible even if no information is gathered or the bath is at zero temperature, provided the cycle is non Markovian. This resource may be the basis of quantum engines embedded in a bath with long memory-time (Sec.~\ref{sec_work_information}).

\item Sec.~\ref{sec_discussion} presents an outlook that charts anticipated developments in quantum thermodynamics, with a focus on possible realizations of the models discussed here.
\end{enumerate}

\section{Steady-state cycles under periodic modulation}\label{sec_cycles}

\subsection{Model}
 
The basic setup of a continuous quantum heat machine consists of a periodically modulated system (working fluid) \emph{permanently} coupled to cold and hot baths in equilibrium (see Fig.~\ref{fig_heat_machines}b). The interaction with the baths and the modulation triggers the dynamics of the working fluid, which eventually reaches a periodic steady state or limit cycle~\citep{alicki2012periodically}. The periodicity endows the machine with the ability to sustain its operation indefinitely. Our thermodynamic analysis of such machines rests on a dynamical approach---we formulate the Hamiltonian and the dynamic equations, solve them to obtain the evolution and calculate the key thermodynamic variables, i.e., the heat currents and power at steady state, upon imposing the constraints of the first and second laws on these variables.

\par

The total Hamiltonian of such a continuous quantum heat machine reads
\begin{equation}\label{eq_cycle_H}
  \Htot=\Hs(t)+\sum_{j\in\{\indexh,\indexc\}} \left(\Hsb^j+\Hb^j\right).
\end{equation}
Here
\begin{equation}
  \Hs(t)=\Hs\left(t+\frac{2\pi}{\Omega}\right)
\end{equation}
is the time-periodic modulated system (working-fluid) Hamiltonian and
\begin{equation}\label{eq_cycle_Hsb}
  \Hsb^j=S_j\otimes B_j
\end{equation} 
is the coupling of a working-fluid operator $S_j$ to $B_j$, an operator of the $j$th bath. Finally, $\Hb^j$ is the free Hamiltonian of the $j$th bath. The form of the modulation or the couplings may vary from one model to another. 

\par

The time dependence of the system Hamiltonian $\Hs(t)$ renders the derivation of the master equation non-trivial. In this section we derive the general equations for the evolution of the working-fluid density matrix $\rho$, which will be used in subsequent sections to infer the relevant thermodynamic properties of the machine in different scenarios. 

\subsection{Floquet expansion of the Markovian (Lindblad) master equation}

In this subsection we extend the standard derivation of the Lindblad generators in the Markovian master equation~\citep{breuerbook,kryszewski2011master} for the reduced density matrix of the system by allowing for its Floquet expansion under periodic driving~\citep{alicki2012periodically,gelbwaser2013minimal,kosloff2013quantum}. Here we assume weak coupling between the system and the baths. The main steps of this derivation are summarized as follows.
\begin{enumerate}
\item The system Hamiltonian is periodic in $\tau$, $\Hs(t)= \Hs(t + \tau)$. Its associated time-evolution operator reads
  \begin{equation}\label{eq_time_evol_op}
    U(t,0) \equiv \mathcal{T}\exp\left(-\frac{i}{\hbar}\int_0^t H_S(s)\dd s\right),
  \end{equation}
  where $\mathcal{T}$ denotes the time ordering operator. According to the Floquet theorem~\citep{floquet1883sur}, this time-evolution operator can be decomposed as $U(t,0)=P(t)e^{Rt}$, where $P(t)$ is $\tau$-periodic and $R$ is a constant operator. From $U(0,0)=\mathbbm{1}$ it follows that $P(0)=\mathbbm{1}$ and hence $U(\tau,0)=P(\tau)e^{R\tau}=P(0)e^{R\tau}=e^{R\tau}$ due to the periodicity of $P(t)$. We now identify the constant operator $R$ with an effective Hamiltonian $H_\mathrm{eff}$ via $U(\tau,0)=e^{R\tau}\eqcolon e^{-iH_\mathrm{eff}\tau/\hbar}$. This effective Hamiltonian---the average over a period of $\Hs(t)$---defines quasi-energies $\hbar\omega_k$ via its spectrum,
  \begin{equation}
    H_\mathrm{eff}=\sum_{k}\hbar\omega_k\proj{k}.
  \end{equation}
  Hence, the part of the time-evolution operator~\eqref{eq_time_evol_op} associated with $H_\mathrm{eff}$ can be decomposed as
  \begin{equation}
    e^{-iH_\mathrm{eff}t/\hbar}=\sum_k e^{-i\omega_k t}\proj{k}.
  \end{equation}
  Likewise, the expansion of the periodic $P(t)$ reads
  \begin{equation}
    P(t)=\sum_{q\in\mathbb{Z}}\tilde{P}(q)e^{-iq\Omega t},
  \end{equation}
  with $\Omega\coloneq 2\pi/\tau$ and
  \begin{equation}
    \tilde{P}(q)=\frac{1}{\tau}\int_0^\tau P(t) e^{iq\Omega t}\dd t.
  \end{equation}
  The Fourier components of the system operator in Eq.~\eqref{eq_cycle_Hsb} are then given in the interaction picture by
  \begin{equation}\label{eq_floquet_systemoperator}
    S(t)=U^\dagger(t,0)S U(t,0)=\sum_{q\in\mathbb{Z}}\sum_{\{\omega\}}S_{q\omega}e^{-i(\omega+q\Omega)t}.
  \end{equation}
  Here $\{\omega\}$ is defined as the set of all transition (Bohr) frequencies $\omega_k-\omega_l$ between the levels of $H_\mathrm{eff}$ and the operators $S_{q\omega}$ are the $q$th-harmonic transition operators between those levels~\citep{szczygielski2013markovian}.
\item The Floquet-expanded Lindblad generator acting on the system-state $\rho$ then assumes the following form in the interaction picture,
  \begin{equation}\label{eq_ss_L} 
    \mathcal{L} = \sum_{q\in\mathbb{Z}}\sum_{\{\omega>0\}}\mathcal{L}_{\omega q}\equiv\sum_{q\in\mathbb{Z}}\sum_{\{\omega>0\}}\sum_{j\in\{\indexc,\indexh\}}\mathcal{L}^j_{\omega q},
  \end{equation}
  in terms of the $q$th-harmonic Liouvillians (Lindblad generators) associated with the $j$th bath,
  \begin{multline}\label{eq_ss_L_sub-bath} 
    \mathcal{L}^j_{\omega q}\rho = \frac{1}{2}G_j(\omega+q\Omega)\left(2S_{\omega q}\rho S_{\omega q}^\dagger - S_{\omega q}^\dagger S_{\omega q}\rho - \rho S_{\omega q}^\dagger S_{\omega q} \right)\\
    + \frac{1}{2}G_j(-\omega-q\Omega)\left(2S_{\omega q}^\dagger\rho S_{\omega q} - S_{\omega q} S_{\omega q}^\dagger\rho - \rho S_{\omega q} S_{\omega q}^\dagger\right).
  \end{multline}
  Here we have defined the temperature-dependent bath response spectra 
  \begin{equation}\label{eq_cycle_Gj}
    G_j(\omega)\coloneq\int_{-\infty}^{\infty}e^{i\omega t}\langle B_j(t)B_j(0) \rangle \dd t,
  \end{equation}
  which fulfill the detailed-balance KMS relation $G_j(-\omega)=e^{-\beta_j\hbar\omega}G_j(\omega)$~\citep{breuerbook} with the inverse temperature $\beta_j=1/\kB T_j$. The expansion~\eqref{eq_ss_L_sub-bath} decomposes the effects of physical baths (labeled by $j$) into those of multiple ``sub-baths'' (labeled by $q$) which interact with the system. By virtue of the KMS theorem, each sub-bath Lindblad generator~\eqref{eq_ss_L_sub-bath} possesses a Gibbs-like stationary state 
  \begin{equation}\label{eq_cycle_sub_gibbs}
    \tilde{\rho}^{jq\omega} = Z^{-1} \exp\left(-\frac{\omega + q\Omega}{\omega}\beta_j H_\mathrm{eff}\right).
  \end{equation}
\end{enumerate} 
As an example, for a modulated qubit energy $\hbar\omega_0+\hbar\omega(t)$, the operators $S_{q\omega}$ correspond to the phase-modulated Pauli raising and lowering operators according to~\citep{alicki2012periodically,alicki2014quantum}
\begin{equation}
  \sminus(t)=\sum_{q\in\mathbb{Z}}\xi(q)\sminus e^{-i(\omega_0+q\Omega)t},
\end{equation}
with
\begin{equation}
  \xi(q)\coloneq\frac{1}{\tau}\int_0^\tau\dd t^\prime\exp\left(-i\int_0^{t^\prime}\omega(s)\dd s\right)e^{iq\Omega t^\prime}.
\end{equation}
The sub-bath Lindblad generators then assume the form
\begin{multline}\label{eq_periodically_lindblad}
  \mathcal{L}_q^j\rho=\frac{1}{2}P(q)G_j(\omega_0+q\Omega)\left(2\sminus\rho\splus-\splus\sminus\rho-\rho\splus\sminus\right)\\
  +\frac{1}{2}P(q)G_j(-\omega_0-q\Omega)\left(2\splus\rho\sminus-\sminus\splus\rho-\rho\sminus\splus\right),
\end{multline}
with $P(q)\coloneq|\xi(q)|^2$ being the weight of the $q$th (Floquet harmonic) ``sub-bath''.

\subsection{Heat currents and the first and second law}

Upon taking the time derivative of the von~Neumann entropy $\mathcal{S}(\rho(t))=-\kB\Tr\left[\rho(t)\ln\rho(t)\right]$, one obtains
\begin{equation}\label{eq_vn_dot}
  \frac{\dd}{\dd t} \mathcal{S}\bigl(\rho(t)\bigr)=-\kB\Tr\left[\dot\rho(t)\ln\rho(t)\right]\equiv-\kB\sum_{q\in\mathbb{Z}}\sum_{\{\omega>0\}}\sum_{j\in\{\indexc,\indexh\}}\Tr\left[\mathcal{L}^j_{\omega q}\rho(t)\ln\rho(t)\right],
\end{equation}
where we have used the Liouville (master) equation $\dot\rho(t)=\mathcal{L}\rho(t)$, expansion~\eqref{eq_ss_L}, and $\Tr[\dot\rho(t)]=0$. We now make use of the Spohn inequality~\citep{spohn1978entropy}
\begin{equation}
  \Tr\left[\mathcal{L}\rho(\ln\rho-\ln\rho^\mathrm{ss})\right]\leq0,
\end{equation}
which is the expression of the second law for any Lindblad superoperator $\mathcal{L}$ and its stationary state $\rho^\mathrm{ss}$ ($\mathcal{L}\rho^\mathrm{ss}=0$). This inequality is applied to every term of the sum in Eq.~\eqref{eq_vn_dot}, resorting to the stationary state of $\mathcal{L}^j_{\omega q}$, $\tilde{\rho}^{jqw}$, yielding
\begin{equation}\label{eq_vn_dot_after_spohn}
  \frac{\dd}{\dd t} \mathcal{S}\bigl(\rho(t)\bigr)+\kB\sum_{q\in\mathbb{Z}}\sum_{\{\omega>0\}}\sum_{j\in\{\indexc,\indexh\}}\Tr\left[\mathcal{L}^j_{\omega q}\rho(t)\ln\tilde{\rho}^{jqw}\right]\geq 0.
\end{equation}
This inequality is then compared to the dynamical version of the second law of thermodynamics~\citep{kosloff2013quantum,gelbwaser2013minimal}
\begin{equation}\label{eq_cycle_second_law}
  \frac{\dd}{\dd t} \mathcal{S}\bigl(\rho(t)\bigr) - \sum_j \frac{1}{T_j}J_j(t)\geq 0.
\end{equation}
This comparison of Eqs.~\eqref{eq_vn_dot_after_spohn} and~\eqref{eq_cycle_second_law} allows us to identify the \emph{heat current} $J_j(t)$ (energy-flow rate) between the working fluid and the $j$th bath as~\citep{alicki2012periodically,kolar2012quantum,gelbwaser2013minimal,kosloff2013quantum}
\begin{equation}\label{eq_cycle_currents}
  J_j(t) = -\frac{1}{\beta_j}\sum_{q\in \mathbb{Z}}\sum_{\{\omega\geq 0\}}\Tr\left[\mathcal{L}^j_{q\omega}\rho(t)\ln \tilde{\rho}^{jq\omega}\right].
\end{equation}
For steady-state operation (the limit cycle in the Schr\"odinger picture), wherein the system is in state $\rho^\mathrm{ss}$, these heat currents simplify, upon inserting Eq.~\eqref{eq_cycle_sub_gibbs} into Eq.~\eqref{eq_cycle_currents}, to the following form,
\begin{equation}\label{eq_cycle_currents_ss}
  J_j = \sum_{q\in \mathbb{Z}}\sum_{\{{\omega}\geq 0\}}\frac{\omega+q\Omega}{\omega}\Tr\left[\left(\mathcal{L}^j_{q{\omega}}\rho^\mathrm{ss}\right)H_\mathrm{eff}\right].
\end{equation}
Here $\mathcal{L}^j_{q\omega}\rho^\mathrm{ss}$ expresses the energy-flow rate exchanged with the sub-bath $(j,q)$ and $\frac{\omega+q\Omega}{\omega}H_\mathrm{eff}$ corresponds to the $q$th harmonic energy. The product of these two quantities is the energy exchanged with the respective sub-bath.

\par

In this steady-state regime, where transient effects have died out, the second law~\eqref{eq_cycle_second_law} adopts the form
\begin{equation}
  \sum_j \frac{J_j}{T_j}\leq 0.
\end{equation}
According to the first law (of energy conservation), the stationary power (the time derivative of the work) of an external periodic force acting on the system is
\begin{equation}\label{eq_cycle_power_first_law}
  \dot{W}= - \sum_j J_j.
\end{equation}
A negative sign of the power means that work is extracted from the heat machine, i.e., that it is operated as a heat engine.

\subsection{Coarse-grained evolution of a periodically-modulated qubit: From non-Markovian to Markovian dynamics}

The dynamics obtained from the Floquet-expanded Lindblad superoperator~\eqref{eq_periodically_lindblad} can also be derived upon coarse-graining over a cycle (modulation period) the time-dependent evolution generated by a periodically-driven non-Markovian master equation~\citep{kofman2004unified}. In what follows we outline this procedure.

\par

We assume that the system-bath interaction is weak, which allows for the Born approximation wherein the system and the bath are approximately in a product state. Zwanzig's projection operator technique then allows to derive the following non-Markovian master equation in the interaction picture for the reduced system density matrix $\rho$~\citep{kofman2004unified,gordon2007universal,gordon2008optimal},
\begin{equation}\label{eq_nm_master}
 \dot\rho(t)=\int_0^t\dd t^\prime\left(\Phi_T(t-t^\prime)\left[S(t-t^\prime)\rho(t),S\right]+\text{H.c.}\right),
\end{equation}
where
\begin{equation}
 \Phi_T(t)=\ew{e^{i\Hb t/\hbar}Be^{-i\Hb t/\hbar}B}_\mathrm{B}
\end{equation}
is the ``memory'' or correlation function of the bath and the time dependence of the system operator $S(t)$ is evaluated in the interaction picture. 

\par

For a periodically modulated qubit, the master equation~\eqref{eq_nm_master} yields the rate equations~\citep{kofman2004unified}
\label{eq_nm_rate_equation}
\begin{equation}
  \dot{\rho}_{ee}=-\dot{\rho}_{gg}=-R_e(t)\rho_{ee}+R_g(t)\rho_{gg},
\end{equation}
where the $\ket{e}\rightarrow\ket{g}$ and $\ket{g}\rightarrow\ket{e}$ transition rates are, respectively,
\begin{subequations}\label{eq_nm_rates}
  \begin{align}
    R_e(t)&=2\realt\int_0^t\dd t^\prime e^{i\omega_0(t-t^\prime)}\xi(t)\xi^*(t^\prime)\Phi_T(t-t^\prime)\\
    R_g(t)&=2\realt\int_0^t\dd t^\prime e^{-i\omega_0(t-t^\prime)}\xi^*(t)\xi(t^\prime)\Phi_T(t-t^\prime),
  \end{align}
\end{subequations}
and the time-dependent phase factors are defined as
\begin{equation}\label{eq_nm_epsilon}
  \xi(t)\coloneq\exp\left(-i\int_0^t\omega(s)\dd s\right).
\end{equation}
The periodicity of the transition rates~\eqref{eq_nm_rates} at times much longer than the bath correlation (memory) time $\tc$ is determined by the external modulation rate $\Omega=2\pi/\tau$. Owing to the weak coupling of the qubit to the bath, the qubit evolves much more slowly than $\tau$, so that the rates can be averaged (coarse-grained) as follows~\citep{kolar2012quantum,kofman2004unified}
\begin{subequations}\label{eq_cycles_transition_rates_averaged}
  \begin{align}
    \overline{R}_e&\approx\frac{1}{\tau}\lim_{n\rightarrow\infty}\int_{n\tau}^{(n+1)\tau}R_e(t)\dd t=\sum_{q\in\mathbb{Z}} |\xi(q)|^2 G_T(\omega_0+q\Omega)\\
    \overline{R}_g&\approx\frac{1}{\tau}\lim_{n\rightarrow\infty}\int_{n\tau}^{(n+1)\tau}R_g(t)\dd t=\sum_{q\in\mathbb{Z}} |\xi(q)|^2 G_T(-\omega_0-q\Omega).
  \end{align}
\end{subequations}
Here the right-hand equality involves the Fourier expansion
\begin{subequations}
  \begin{align}
    \xi(t)&=\sum_{q\in\mathbb{Z}} \xi(q)e^{-iq\Omega t},\\
    \xi(q) &\coloneq\frac{1}{\tau}\int_0^{\tau}\dd t^\prime \exp\left(-i\int_0^t\omega(s)\dd s\right)e^{iq\Omega t^\prime}, 
  \end{align}
\end{subequations}
and the Fourier-transformed autocorrelation function of the bath [\cf Eq.~\eqref{eq_cycle_Gj}]
\begin{equation}
  G_T(\omega)= \int_{-\infty}^{+\infty} e^{i\omega t}\Phi_T(t)\dd t.
\end{equation}
These time-averaged rates~\eqref{eq_cycles_transition_rates_averaged}, obtained by coarse-graining the non-Markovian master equation~\eqref{eq_nm_master}, are identical to those derived from the Floquet-expanded Markovian master equation~\eqref{eq_periodically_lindblad}.

\subsection{Summary}

In this section we have derived the basic dynamic equations and thermodynamic variables (heat currents and power) of a QHM that we will apply to different cases in the remainder of this review. 

\par

The Lindblad master equation for a periodically-driven (controlled) system interacting with two heat baths in the interaction picture can be decomposed into ``sub-bath'' contributions according to Floquet's theorem. These ``sub-bath'' Lindblad operators describe the interaction of the periodically-modulated system with a single heat bath at a shifted Bohr frequency (corresponding to a harmonic Floquet sideband). The heat currents can likewise be decomposed into Floquet harmonics and are (by construction) compatible with the second law of thermodynamics.

\section{Periodically-modulated qubit-based heat machine}\label{sec_periodically-modulated}

\subsection{Steady-state operation of a continuous-cycle single-qubit heat machine}

\par
\begin{figure}
  \centering
  \includegraphics[width=9cm]{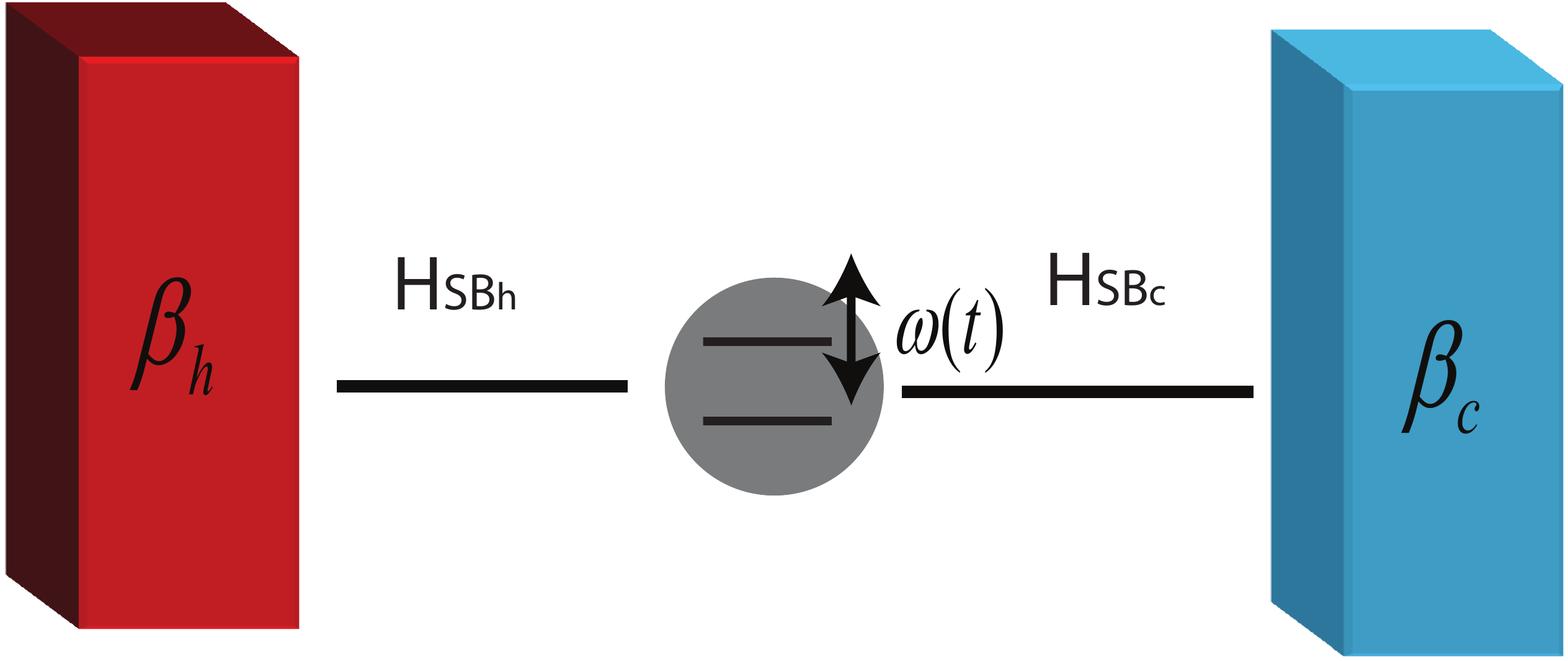}
  \caption{Periodically-modulated qubit coupled to hot and cold heat baths.}\label{fig_external_modulation}
\end{figure}
\par

The Floquet decomposition of the Markovian master equation presented in Sec.~\ref{sec_cycles} provides a general framework for the analysis of continuously-operated periodically-driven heat machines whose working fluid is treated quantum-mechanically. We here review a simple machine of this kind based on a single qubit (two-level system---TLS), in order to determine its efficiency or COP and compare it with reciprocating-cycle (stroke-based) heat machines.

\par

The transition energy of the working-fluid TLS is periodically modulated by an external field according to (see also Fig.~\ref{fig_external_modulation})~\citep{gelbwaser2013minimal}
\begin{equation}\label{eq_periodically_Hs}
  \Hs(t)=\frac{1}{2}\hbar\omega(t)\sigmaz
\end{equation}
with $\omega\left(t+\frac{2\pi}{\Omega}\right)=\omega(t)$. In this semiclassical model the driving field plays the r\^ole of a piston, i.e., it allows for extracting/supplying work from/to the heat machine. At the same time, this TLS is coupled to cold and hot heat baths via
\begin{equation}
  \Hsb=\sigmax\otimes(B_\indexh+B_\indexc),
\end{equation}
where $B_j$ are the respective bath operators. 

\par

The reduced density operator for the TLS evolves according to the Floquet-expanded master equation [Eqs.~\eqref{eq_ss_L} and~\eqref{eq_ss_L_sub-bath}]
\begin{equation}\label{eq_periodically_master}
  \dot\rho=\mathcal{L}\rho=\sum_{q\in\mathbb{Z}}\sum_{j\in\{\indexc,\indexh\}}\mathcal{L}_q^j\rho,
\end{equation}
with the Lindblad operators~\eqref{eq_periodically_lindblad} derived in Sec.~\ref{sec_cycles}. The steady-state solution of the master equation~\eqref{eq_periodically_master} is diagonal, with the ratio of excited- and ground-state populations satisfying~\citep{gelbwaser2013minimal}
\begin{equation}
  w=\frac{\rho^\mathrm{ss}_{ee}}{\rho^\mathrm{ss}_{gg}}=\frac{\sum_{q,j}P(q)G_{j}(\omega_{0}+q\Omega)e^{-\beta_j\hbar(\omega_0+q\Omega)}}{\sum_{q,j}P(q)G_{j}(\omega_{0}+q\Omega)},
\end{equation}
where $P(q)$ and $G(\omega_0+q\Omega)$ are as in Eq.~\eqref{eq_periodically_lindblad}. According to Eqs.~\eqref{eq_cycle_currents} and~\eqref{eq_cycle_currents_ss}, the cold $(\indexc)$ and hot ($\indexh$) heat currents between the TLS and the respective baths are then given by
\begin{equation}\label{eq_periodically_currents}
  J_{\indexc(\indexh)}=\sum_{q\in\mathbb{Z}}\hbar(\omega_{0}+q\Omega)P(q)G_{\indexc(\indexh)}(\omega_{0}+q\Omega)\frac{e^{-\beta_{\indexc(\indexh)}\hbar(\omega_0+q\Omega)}-w}{w+1}.
\end{equation}
The power is given, according to Eq.~\eqref{eq_cycle_power_first_law} (the first law), by
\begin{equation}\label{eq_periodically_power}
  \dot{W}= \sum_{q\in\mathbb{Z}}\sum_{j\in\{\indexc,\indexh\}}\hbar(\omega_0+q\Omega)P(q)G_j(\omega_0+q\Omega)\frac{w-e^{-\beta_j\hbar(\omega_0+q\Omega)}}{w+1}.
\end{equation}
In what follows we investigate the conditions for a negative sign of $\dot{W}$ that corresponds to power extraction (at expense of the hot bath), i.e, to the heat machine being operated as an engine, and, conversely, a positive sign of $\dot{W}$ that may yield refrigeration of the cold bath (at the expense of power invested by the piston).

\subsection{Periodic modulation and bath spectra}

The preceding expressions for the heat currents [Eq.~\eqref{eq_periodically_currents}] and the power [Eq.~\eqref{eq_periodically_power}] are not easy to interpret. We therefore review special choices of the modulation form $\omega(t)$ [Eq.~\eqref{eq_periodically_Hs}] as well as of the bath response spectra $\Gc(\omega)$ and $\Gh(\omega)$ discussed in~\citep{gelbwaser2013minimal}. We will see that the modulation rate $\Omega$ allows us to operate the thermodynamic machine on demand as either an engine or a refrigerator.

\par

\par
\begin{figure}
  \centering
  \includegraphics[width=9cm]{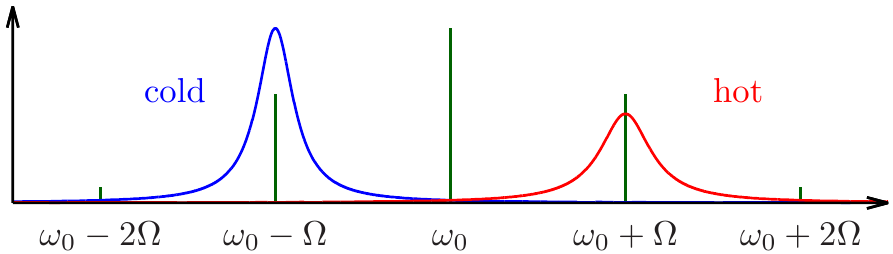}
  \caption{An example of spectrally separated baths required for high-efficiency continuous-cycle heat machine operation in Fig.~\ref{fig_external_modulation}: The two-level system with resonance frequency $\omega_0$ is coupled to the cold (hot) bath at the left (right) Floquet sideband only. The green vertical lines indicate the coupling weights $P(q)$ to these sidebands.}\label{fig_spectral_separation}
\end{figure}
\par

The Carnot efficiency bound can be reached if only two harmonics contribute to the sums in Eqs.~\eqref{eq_periodically_currents} and~\eqref{eq_periodically_power}~\citep{gelbwaser2013minimal}. This requires a \emph{spectral separation} of the two baths, i.e., the system must only be coupled to the cold bath at a specific sideband $\omega_0+q_1\Omega$ and to the hot bath only at \emph{another} sideband $\omega_0+q_2\Omega$ (\cf Fig.~\ref{fig_spectral_separation}). As an example, for a sinusoidally-modulated TLS transition frequency
\begin{equation}
  \omega(t) = \omega_0 + \kappa\sin(\Omega t) 
\end{equation}
three harmonics ($q=0,\pm1$) contribute to the coupling in the weak-modulation limit $0\leq \kappa\ll \Omega$, with respective weights
\begin{equation}
P(0) \simeq 1- \frac{1}{2}\left(\frac{\kappa}{\Omega}\right)^2, \quad P(\pm1) \simeq \frac{1}{4}\left(\frac{\kappa}{\Omega}\right)^2.
\end{equation}
The restriction of the coupling to only two sidebands can be achieved for specially designed bath response spectra, e.g., 
\begin{equation}\label{eq_periodically_sin_case2}
  \Gc(\omega)\simeq 0 \text{ for } \omega \approx \omega_0 \pm \Omega, \quad \Gh(\omega)\simeq 0\text{ for } \omega \leq \omega_0.
\end{equation}
For this specific choice the general equations~\eqref{eq_periodically_currents} and~\eqref{eq_periodically_power} for the heat currents and the power are reduced to~\citep{gelbwaser2013minimal}
\begin{subequations}\label{eq_periodically_sin_case_2_currents}
  \begin{align}
    \Jh&=\hbar(\omega_0+\Omega)\left[e^{-\betah\hbar(\omega_0+\Omega)}-e^{-\betac\hbar\omega_0}\right]N\\
    \Jc&=-\hbar\omega_0\left[e^{-\betah\hbar(\omega_0+\Omega)}-e^{-\betac\hbar\omega_0}\right]N\\
    \dot{W}&=-\hbar\Omega\left[e^{-\betah\hbar(\omega_0+\Omega)}-e^{-\betac\hbar\omega_0}\right]N,
\end{align}
\end{subequations}
where the normalization factor $N$ is a \emph{positive} function of $\kappa$, $\Omega$, the bath response spectra $G_j(\omega)$ and the inverse bath temperatures $\beta_j=1/\kB T_j$, where $\kB$ is the Boltzmann constant. 

\par
\begin{figure}
  \centering
  \includegraphics[width=9cm]{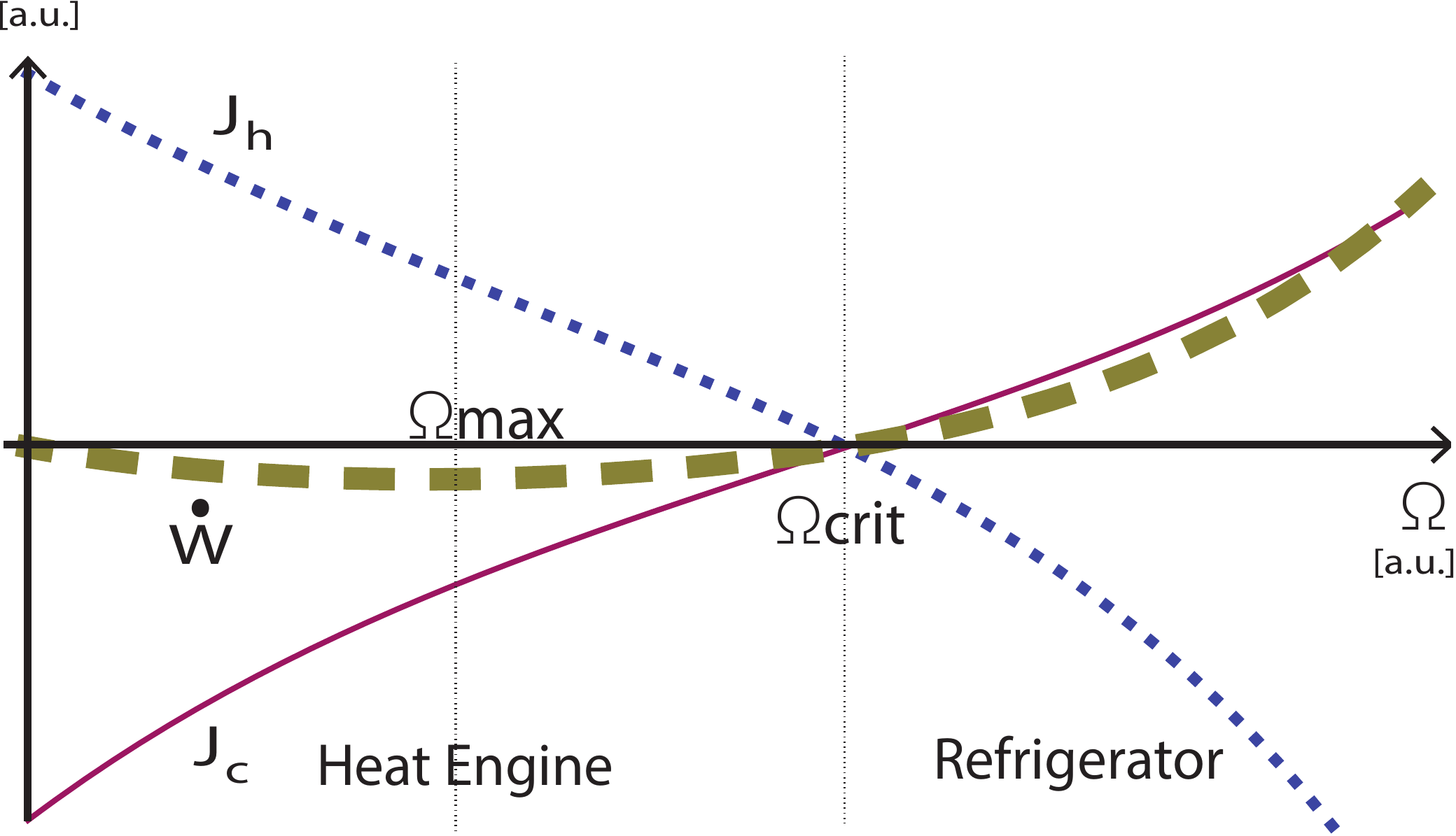}
  \caption{Heat currents for the QHM described by Figs.~\ref{fig_external_modulation} and~\ref{fig_spectral_separation} as a function of the modulation frequency. For low modulation frequencies (below $\Omegacrit$) the QHM operates as an engine (negative power $\dot{W}<0$) and for high frequencies (above $\Omegacrit$) as a refrigerator ($\dot{W}>0$). The heat currents change their signs accordingly. The maximum power is obtained at modulation frequency $\Omegamax$.} \label{fig_universal}
\end{figure}
\par
\begin{figure}
  \centering
  \includegraphics[width=9cm]{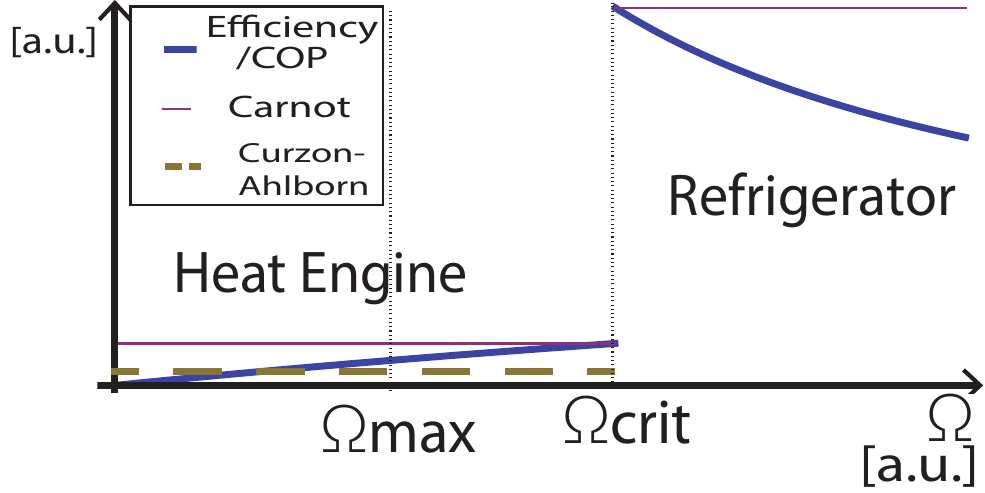}
  \caption{Efficiency/COP in the two operational regimes of Fig.~\ref{fig_universal} as a function of the modulation rate $\Omega$. At $\Omegamax$, where the power is maximized, the efficiency exceeds the Curzon-Ahlborn bound. At $\Omegacrit$ the zero-power Carnot bound is attained.} \label{fig_universal_efficiency}
\end{figure}
\par

Equations~\eqref{eq_periodically_sin_case_2_currents} reveal the existence of a critical modulation frequency~\citep{gelbwaser2013minimal}
\begin{equation}
  \Omegacrit =\omega_{0}\frac{\Th-\Tc}{\Tc}.
\end{equation}
According to Eqs.~\eqref{eq_periodically_sin_case_2_currents} the machine acts as a heat engine (i.e., $\dot{W}<0$, $\Jc<0$, $\Jh>0$) when driven below this critical frequency ($\Omega < \Omegacrit$), and its efficiency is found to be
\begin{equation}
  \eta=\frac{\Omega}{\omega_{0}+\Omega}.
\end{equation}
At $\Omega=\Omegacrit$ the efficiency attains the Carnot bound, where the power and the heat currents vanish, $\dot{W}=\Jc=\Jh=0$. For higher modulation frequencies ($\Omega > \Omegacrit$) the signs of the heat currents and the power in Eqs.~\eqref{eq_periodically_sin_case_2_currents} are reversed, implying that the machine operates as a refrigerator of the cold bath, $\Jc>0$ (at the expense of invested power, $\dot{W}>0$), with the coefficient of performance
\begin{equation}
  \mathrm{COP}=\frac{\omega_{0}-\Omega}{\Omega}.
\end{equation}
Thus, the Carnot efficiency bound reached at $\Omega=\Omegacrit$ cannot be surpassed (see Figs.~\ref{fig_universal} and~\ref{fig_universal_efficiency}). Numerical simulations of a somewhat similar heat machine also revealed such a universal behavior~\citep{birjukov2008quantum}.

For many applications the maximally achievable efficiency is less important than the heat engine's efficiency at \emph{maximum} power output. For macroscopic Carnot-type engines the latter is given by the Curzon-Ahlborn efficiency bound~\citep{curzon1975efficiency} [Eq.~\eqref{eq_curzohn-ahlborn}], which, however, can be surpassed in the presented continuous-cycle heat machine, provided that the bath response spectra fulfill Eq.~\eqref{eq_periodically_sin_case2}~\citep{gelbwaser2013minimal}.

\subsection{Realization considerations}

To efficiently operate the machine described above, a necessary requirement on the bath coupling spectra is their separation. For hot and cold phonon baths associated with two different materials that have a Debye spectrum~\citep{gelbwaser2013minimal}
\begin{equation}
  G_j(\omega) = f_j\left(\frac{\omega}{\omega_\mathrm{D}^j}\right)^3\frac{1}{1-e^{-\beta_j\hbar\omega}} \Theta\left(\omega_\mathrm{D}^j-|\omega|\right),
\end{equation}
where $\Theta$ denotes the Heaviside step function, $\omega_\mathrm{D}^j$ are the Debye frequencies of the two baths, and $f_j$ are bath-specific functions, the desired separation can be achieved if the respective Debye frequencies lie in the vicinity of the material temperatures,
\begin{equation}\label{eq_periodically_debye_cutoff}
\hbar\omega_\mathrm{D}^\indexc \simeq \frac{1}{\betac}, \quad \hbar\omega_\mathrm{D}^\indexh \simeq \frac{1}{\betah}.
\end{equation}

\begin{figure}
  \centering
  \includegraphics[width=9cm]{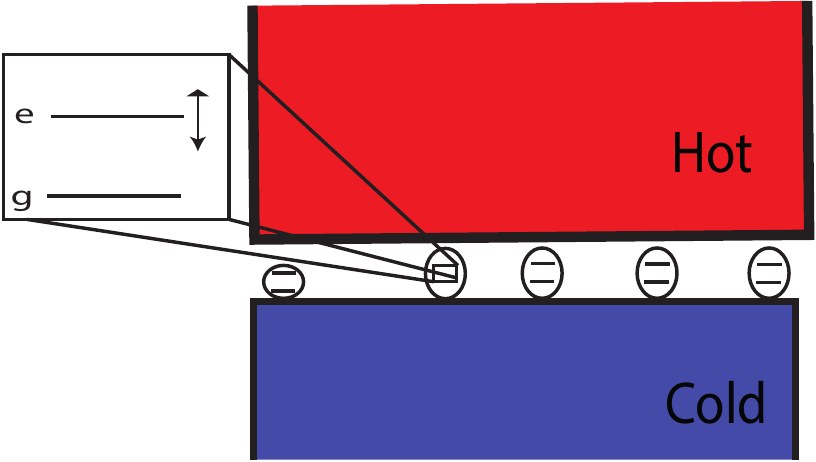}
  \caption{Schematic view of a quantum heat machine comprised of many TLS (e.g., quantum dots) sandwiched between layers of different materials characterized by different phonon spectra and temperatures. The inset shows the periodic modulation (ac Stark shift) of the TLS level distance by an off-resonant electromagnetic field.}\label{fig_qdot_heat_machine}
\end{figure}

In this scenario, the TLS working fluid may be realized by a quantum dot sandwiched between two different materials (as depicted in Fig.~\ref{fig_qdot_heat_machine}). Further implementations, e.g., by means of a double-well potential are discussed in~\citep{gelbwaser2013minimal,alicki2014quantum}. The classical drive modulating the qubit's resonance frequency may be provided by an off-resonant microwave or optical field via the ac Stark effect.

\par

A general way of obtaining separated bath response spectra, discussed in~\citep{gelbwaser2013minimal}, consists of constructing ``filter'' modes through which the system interacts with the baths. These filter modes can be easily tuned to create the desired form (or \emph{effective} bath spectra) without the need to directly engineer the materials' autocorrelation functions.

\subsection{Summary}

The presented ``minimal'' realization of a continuous, periodically-driven heat machine based on a single qubit displays a wide range of interesting features. It can be universally operated, i.e., be switched on demand from an engine to a refrigerator regime by tuning the modulation rate. The engine's efficiency at \emph{maximal} power output can surpass the Curzon-Ahlborn bound at an appropriate modulation rate. The Carnot bound, however, remains intact as this model (by construction) adheres to the first and second laws of thermodynamics. When approaching the critical driving frequency from below, the engine reaches Carnot efficiency but at the same time the heat currents and the power output vanish. Any further increase of the driving field turns the engine into a refrigerator. Spectral separation of the two baths is essential for attaining high efficiency.

\par

This inherently non-adiabatic model circumvents the difficulty of breaking a finite-time (reciprocating) cycle into strokes: Abrupt on-off switching of system-bath interactions in alternating strokes of such cycles may strongly affect their energy and entropy exchange and thereby their quantum state, which casts doubts on the validity of existing models of finite-time engines that ignore such effects.

\section{Quantum heat machines based on periodically-modulated multilevel systems}\label{sec_nlevels}

\subsection{Introduction and model}

\begin{figure}
  \centering
  \includegraphics[width=9cm]{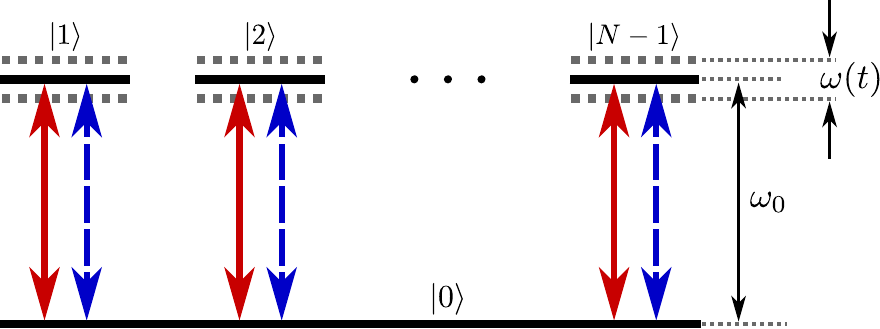}
  \caption{A degenerate multilevel QHM: An $N$-level system permanently coupled to a cold (dashed blue arrows) and a hot (solid red arrows) heat bath. The degenerate upper states are periodically modulated by an external field.}\label{fig_system}
\end{figure}

Having presented in Secs.~\ref{sec_cycles} and~\ref{sec_periodically-modulated} the \emph{universal} features of periodically-driven, continuous-cycle, steady-state quantum heat machines, we now extend the theory from the two-level to the $N$-level case~\citep{gelbwaser2014power,niedenzu2015performance}. The main motivation for this study stems from the pioneering work that identified quantum coherences as a favorable asset in quantum heat engines~\citep{scully2003extracting,scully2011quantum}. To this end we consider an $N$-level system with a common ground state and $N-1$ degenerate upper states (Fig.~\ref{fig_system}). We show that in such a system both the heat currents and the work can be strongly enhanced with respect to a single two-level system (TLS) by means of bath-mediated dipolar couplings between different transitions. However, the efficiency (in the case of an engine) or the COP (in the case of a refrigerator) remain the same as for their two-level counterparts.

\par
\par

As in the TLS case, the upper states are assumed to be periodically modulated (by $\sigmaz$-coupling modulation as in the two-level case) according to
\begin{equation}\label{eq_H_S}
  H_\mathrm{S}(t)=\frac{1}{2}\hbar[\omega_0+\omega(t)]\sum_{j=1}^{N-1}\sigmaz^j,
\end{equation}
where $\omega(t+\frac{2\pi}{\Omega})=\omega(t)$, the modulation rate $\Omega$ being as in Eq.~\eqref{eq_periodically_Hs} and $\sigmaz^j\coloneq\proj{j}-\proj{0}$. Assuming dipole coupling of the system to the baths, the Floquet-expanded master equation becomes the following generalization of Eq.~\eqref{eq_periodically_lindblad}~\citep{gelbwaser2014power,niedenzu2015performance},
\begin{subequations}\label{eq_master_L}
  \begin{equation}\label{eq_master}
    \dot\rho=\sum_{q\in\mathbb{Z}}\sum_{i=\{\indexc,\indexh\}}\mathcal{L}_i^q\rho,
  \end{equation}
  in terms of the ``sub-bath'' Liouvillians
  \begin{multline}\label{eq_L}
    \mathcal{L}_i^q\rho=\frac{1}{2}P(q)G_i(\omega_0+q\Omega)\sum_{j=1}^{N-1}\left[\mathcal{D}\left(\sminus^j,\splus^j\right)+\sum_{\substack{j^\prime\neq j}}\mathfrak{p}_{jj^\prime}\mathcal{D}\left(\sminus^j,\splus^{j^\prime}\right)\right]+\\
    +\frac{1}{2}P(q)G_i(-\omega_0-q\Omega)\sum_{j=1}^{N-1}\left[\mathcal{D}\left(\splus^j,\sminus^j\right)+\sum_{\substack{j^\prime\neq j}}\mathfrak{p}_{jj^\prime}\mathcal{D}\left(\splus^j,\sminus^{j^\prime}\right)\right],
  \end{multline}
\end{subequations}
where we have used the definition $\mathcal{D}(a,b)\coloneq 2a\rho b-ba\rho-\rho ba$ of the dissipator. The rates of the absorption and emission processes in~\eqref{eq_L} are determined by the Floquet coefficients $P(q)$ for the $q$th harmonic sideband, the bath temperatures $\beta_i=1/\kB T_i$ and the bath response spectra. For simplicity we have here assumed real and equally strong dipoles [the general case is analyzed in~\citep{niedenzu2015performance}]. The geometrical configuration of the $N-1$ dipole-transition vectors is described by the (Hermitian) dipole-alignment matrix
\begin{equation}
  \mathfrak{p}_{ij}\coloneq\frac{\mathbf{d}_{i}\cdot\mathbf{d}_{j}}{|\mathbf{d}_{i}||\mathbf{d}_{j}|}\equiv \cos\measuredangle(\mathbf{d}_{i},\mathbf{d}_{j}).
\end{equation}
The ``sub-bath'' Liouvillians in Eq.~\eqref{eq_L} have a clear physical interpretation: Emission and absorption processes involving a single transition are represented by the terms $\mathcal{D}\left(\sminus^j,\splus^j\right)$ and $\mathcal{D}\left(\splus^j,\sminus^j\right)$. These terms also describe effective population transfer between the different excited states via the common ground state. The remaining dissipators $\mathcal{D}\left(\sminus^j,\splus^{j^\prime}\right)$ and $\mathcal{D}\left(\splus^j,\sminus^{j^\prime}\right)$ are cross-correlations between the transitions that create time-dependent coherences and population transfer among the excited states. These bath-mediated correlated interactions between two excited states are largest if their corresponding dipole-transition vectors $\mathbf{d}_j$ and $\mathbf{d}_{j^\prime}$ are aligned ($\mathfrak{p}_{jj^\prime}=1$) and vanish for orthogonal transitions ($\mathfrak{p}_{jj^\prime}=0$). However, even for orthogonal transitions, the existence of a common ground state is expected to give rise to correlations between populations of different excited states, but not to coherences between them. Namely, even in the latter case the presence of multiple excited states will affect the steady-state populations.

\subsection{Steady-state solution}

The steady-state solution of the master equation~\eqref{eq_master} strongly depends on the relative orientations of the transition dipoles. For the sake of clarity we first present the result for a three-level system. The system dynamics is then strongly influenced by the relative orientation of the two dipole-transition vectors. For non-aligned dipoles ($\mathfrak{p}\neq 1$), the three-level system thermalizes with the baths with an effective temperature~\citep{gelbwaser2014power,niedenzu2015performance},
\begin{equation}\label{eq_ode_solution_pneq1}
  \rho_{11}^\mathrm{ss}=\rho_{22}^\mathrm{ss}=e^{-\betaeff\hbar\omega_0}\rho_{00}^\mathrm{ss},
\end{equation}
where the effective temperature is defined by the Boltzmann factor
\begin{equation}\label{eq_betaeff}
  e^{-\betaeff\hbar\omega_0}\coloneq\frac{\sum_{q\in\mathbb{Z}}\sum_{i\in\{\indexc,\indexh\}} P(q)G_i(-\omega_0-q\Omega)}{\sum_{q\in\mathbb{Z}}\sum_{i\in\{\indexc,\indexh\}} P(q)G_i(\omega_0+q\Omega)}.
\end{equation}
The steady-state solution~\eqref{eq_ode_solution_pneq1} is thermalized (diagonal) and does not contain any coherences (off-diagonal matrix elements).

\par

For aligned dipoles, however, the steady-state solution drastically changes. One can show that the system-bath coupling Hamiltonian [and thus the master equation~\eqref{eq_master}] possesses a \emph{dark state} $\ket\psid$ (a linear combination of the excited states)~\citep{gelbwaser2014power,niedenzu2015performance}, which does not interact with the environment. Consequently, this state is removed from the dynamics and the three-level system behaves as a two-level system with an enhanced transition-dipole strength. This effective TLS is formed by the ground state and a ``bright'' state $\ket\psib$ (orthogonal to the dark state). As the overlap of the initial state with this dark state is a constant of motion, it is clear that the steady-state solution depends on their overlap. Accordingly, we find~\citep{gelbwaser2014power,niedenzu2015performance}
\begin{subequations}\label{eq_eigenvalues}
  \begin{align}
    \rho_{\mathrm{b}\mathrm{b}}^\mathrm{ss}&=\frac{1}{1+e^{\betaeff\hbar\omega_0}}\left[1-\rho_\mathrm{dd}(0)\right]\equiv e^{-\betaeff\hbar\omega_0}\rho_{00}^\mathrm{ss}\\
    \rho_{\mathrm{d}\mathrm{d}}^\mathrm{ss}&=\bkew{\psid}{\rho(0)}{\psid}.
  \end{align}
\end{subequations}
This is a (diagonal) \emph{partially thermalized state} characterized by its \emph{thermalization capability} $\left[1-\rho_\mathrm{dd}(0)\right]$. This steady-state solution~\eqref{eq_eigenvalues} is only diagonal in the basis formed by the dark, the bright, and the ground states. When it is transformed back to the ``bare'' basis of states $\ket1$ and $\ket2$, coherences between the excited states are revealed.

\par

The solution can be generalized to $N-1$ excited states. The number of different dipole directions $n$ then determines the dimensionality $\Neff<N$ of an effective multilevel system ($\Neff=n+1$). The steady-state solution of the master equation~\eqref{eq_master} is a partially thermalized state [like Eq.~\eqref{eq_eigenvalues}] for this $\Neff$-level system, where the $\Neff-1$ upper states and the ground state $\ket{0}$ thermalize with the two baths to the effective inverse temperature $\betaeff$~\citep{niedenzu2015performance},
\begin{subequations}\label{eq_ss_nlevels_general}
  \begin{align}
    \rho_{00}^\mathrm{ss}&=\frac{1}{1+(\Neff-1)e^{-\betaeff\hbar\omega_0}}\left[1-\Pidew\right]\label{eq_ss_nlevels_general_ground_state}\\
    \rho_{ii}^\mathrm{ss}&=e^{-\betaeff\hbar\omega_0}\rho_{00}^\mathrm{ss}\quad\text{for }i=1,\dots,\Neff-1.
  \end{align}
\end{subequations}
Here the projector $\Pidew$ measures the system's initial overlap with the multidimensional dark subspace, which is excluded from the thermalization capability. The remaining diagonal elements correspond to these dark states.

\subsection{Heat currents and power}

The heat currents and the power associated to the steady-state solution~\eqref{eq_ss_nlevels_general} can be derived from the general formulae~\eqref{eq_cycle_currents} and~\eqref{eq_cycle_power_first_law}. The resulting expressions are quite cumbersome, but can be simplified when comparing them to their TLS counterparts from Sec.~\ref{sec_periodically-modulated}, which yields

\begin{equation}\label{eq_heat_currents_general_tls}
  \frac{J_i}{J_i^\mathrm{TLS}}\equiv\frac{\dot{W}}{\dot{W}^{\mathrm{TLS}}}=\left(N-1\right)\Big[1-\Pidew\Big]\frac{1+e^{-\betaeff\hbar\omega_0}}{1+(\Neff-1)e^{-\betaeff\hbar\omega_0}}.
\end{equation}
The first factor on the r.h.s.\ describes the enhancement of the heat currents and the power that stems from the multiple \emph{thermalization pathways}, i.e., the number of transitions compared to a single TLS. The second factor is the thermalization capability that measures to what extent the initial state undergoes thermalization. This factor differs from unity if two or more transition-dipole vectors are parallel. The last factor depends on the modulation type, the baths' temperatures and their response spectra through the effective temperature $\betaeff$. A special case is $\Neff=2$, i.e., a TLS or an $N$-level system whose transitions are all parallel: The effective-temperature-dependent factor then becomes unity. Further insight into Eq.~\eqref{eq_heat_currents_general_tls} can be obtained from the equivalent form~\citep{gelbwaser2014power,niedenzu2015performance}
\begin{equation}\label{eq_heat_currents_general_tls_rho00}
   \frac{J_i}{J_i^\mathrm{TLS}}\equiv\frac{\dot{W}}{\dot{W}^{\mathrm{TLS}}}=\left(N-1\right)\frac{\rho_{00}^\mathrm{ss}}{\rho_{00}^\mathrm{TLS}}.
\end{equation}
Hence, the steady-state \emph{ground-state} population determines the power enhancement. 

\par

Another important consequence of Eq.~\eqref{eq_heat_currents_general_tls_rho00} concerns the efficiency $\eta=-\dot{W}/\Jh$ (when operating as an engine) or the $\mathrm{COP}=\Jc/\dot{W}$ (for a refrigerator). The currents and the power are equally modified with respect to a TLS, so that these figures of merit are not altered: A multilevel quantum heat engine performs as efficiently as its two-level counterpart. Notably, the Carnot bound is adhered to~\citep{gelbwaser2013minimal}, as the heat currents by definition obey the second law.

\par

\begin{figure}
  \centering
  \includegraphics[width=9cm]{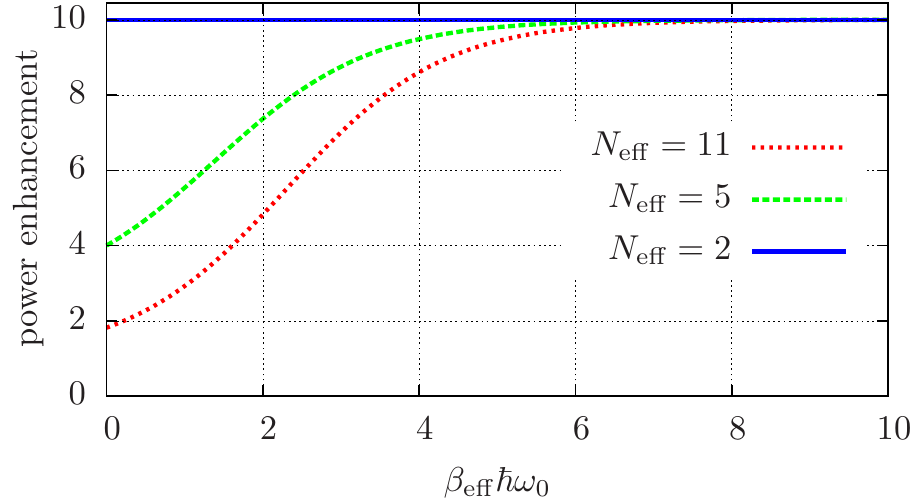}
  \caption{Power enhancement for a working fluid realized by an 11-level system relative to its two-level counterpart in the QHM of Fig.~\ref{fig_system} assuming optimal power enhancement conditions (no initial overlap with dark states). The three curves represent (i) no parallel dipoles ($\Neff=N=11$), (ii) some parallel dipoles ($\Neff=5$), and (iii) all dipoles being parallel ($\Neff=2$).}\label{fig_currents_nlevels}
\end{figure}

To better understand how the dipole alignment affects the power enhancement factor, let us assume an initial state amenable to full thermalization (i.e., no dark-state population), e.g., $\rho(0)=\proj{0}$. As shown in Fig.~\ref{fig_currents_nlevels}, in the low-temperature regime ($\betaeff\hbar\omega_0\gg1$), the enhancement is the same for any alignment. This behavior can be understood from Eqs.~\eqref{eq_heat_currents_general_tls_rho00} and~\eqref{eq_ss_nlevels_general}: At such temperatures the ground-state population in steady state, $\rho_{00}^\mathrm{ss}$, is very close to unity and is independent of $\Neff$, i.e., the number of available levels. Therefore, the heat currents and the power are both enhanced by a factor of $N-1$. In the high-temperature regime ($\betaeff\hbar\omega_0\ll1$), however, a difference is expected between aligned and non-aligned configurations due to the last factor in Eq.~\eqref{eq_heat_currents_general_tls}. At such temperatures, $e^{-\betaeff\hbar\omega_0}\approx 1$ and Eq.~\eqref{eq_ss_nlevels_general} corresponds to an equipartition amongst all available states. The fewer levels are available, the more will the ground state be populated. Therefore, systems with more parallel dipole transitions (and hence fewer available states) are favorable compared to the entirely non-aligned system. If all dipoles are parallel, the $N$-level system corresponds to a single two-level system with an enhanced dipole moment. The power boost is then independent of the modulation type, bath temperatures or bath spectra.
\par
\par
The steady-state solution~\eqref{eq_ss_nlevels_general} is diagonal in the basis spanned by bright and dark states. In the bare-state basis, however, this diagonality may be lost and coherences within the excited-state manifold appear, if at least two dipoles are parallel. However, the criteria for enhanced power extraction (and thermalization capability) are obscure in that basis. In particular, it is impossible to attribute power enhancement neither to the mere presence of coherences, nor to their magnitude $\sum_{i,j=1 (i\neq j)}^{N-1}|{\tilde\rho}_{ij}^\mathrm{ss}|$, ${\tilde\rho}^\mathrm{ss}$ being the steady-state density matrix in the bare-state basis: It is clear from Eq.~\eqref{eq_ss_nlevels_general} that the largest amount of steady-state coherences corresponds to a dark initial state---and hence to zero power. Only such an initial condition corresponds to a population-inverted steady state, as in all other cases the ground state is populated via emission into the bath [Eq.~\eqref{eq_ss_nlevels_general_ground_state}]. On the other hand, if the ground state is populated, the possible amount of coherences between the excited states, $|{\tilde\rho_{ij}}^\mathrm{ss}|$, is inevitably reduced.

\subsection{Summary}

Excited-states degeneracy of the working fluid is a resource in a periodically-driven steady-state quantum heat machine, in the sense that it can boost heat currents and power compared to its two-level counterpart. The maximal power enhancement factor for equally strong transition dipoles is $N-1$. Yet, the efficiency or the coefficient of performance (depending on the operational regime) remains unchanged. The key factor determining the power enhancement is the steady-state \emph{ground-state population}, regardless of the dipole orientation. If there are at least two parallel transition dipoles, the \emph{thermalization capability} of the initial state becomes important: Initial states that do not fully thermalize reduce the heat currents to the two baths and the generated power.

\section{Quantum heat engines driven by a quantum piston}\label{sec_quantum_piston}

\subsection{Introduction and motivation}

As part of our endeavor to reconcile quantum mechanics and thermodynamics, we have introduced in the preceding sections simple models of quantum heat machines and analyzed their compliance with thermodynamic laws. However, these heat machines are not fully quantum-mechanical but rather semiclassical, since their modulation or drive (the piston) is generated by an external classical field.

\par

The quest for a fully quantum heat machine brings out interesting results as well as challenges. As we show below, a new definition of work is required, which allows us to study the thermodynamic nature of different quantum states and classify them according to their contribution to power extraction and cooling.

\subsection{Work in fully quantized setups}\label{sec_work_fully_quantized}

\par
\begin{figure}
  \centering
  \includegraphics[width=9cm]{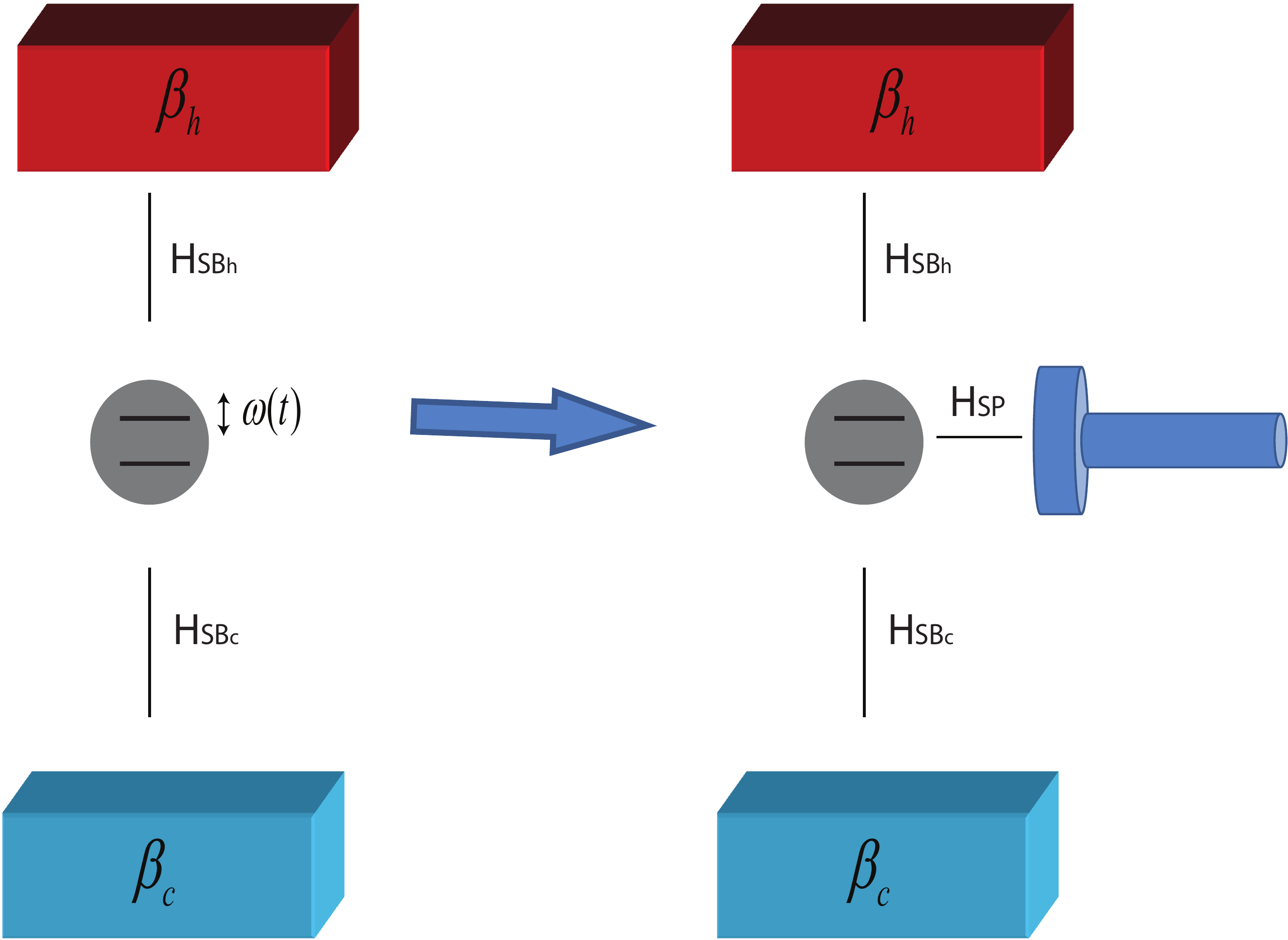}
  \caption{The semiclassical driving field (left) in the model of Fig.~\ref{fig_external_modulation} is replaced in a fully-quantized heat machine by an additional quantum system---the quantized piston (right).}\label{fig_qpiston}
\end{figure}
\par

The introduction of a quantum piston instead of an external modulation (sketched in Fig.~\ref{fig_qpiston}) requires a fundamental change in the total Hamiltonian [Eq.~\eqref{eq_cycle_H}] we have used. For a fully quantized heat machine the total Hamiltonian is
\begin{equation}\label{eq_qp_Htot}
  H_\mathrm{tot}=\Hs+\Hp+\Hsp+\sum_{j=\indexc,\indexh}\left(\Hb^j+\Hsb^j\right).
\end{equation}
Here the system (working fluid) Hamiltonian $\Hs$ no longer depends on time and the Hamiltonian contains two additional terms: The (free) Hamiltonian of the piston, $\Hp$, and the interaction Hamiltonian between the working fluid, $S$, and the piston, $\Hsp$.

\par

In semiclassical models (see Sec.~\ref{sec_cycles}), the first law was used to define the power [Eq.~\eqref{eq_cycle_power_first_law}], implying that all the energy interchanged with the external field is considered to be work. If the piston is quantized, it may seem natural to do the same, i.e., to identify any increase of its energy with work. However, this may lead to contradictions with the second law of thermodynamics. As an example, suppose that a piston is initially in a thermal state, and remains there through its evolution, but with an increasing temperature. The piston gains energy, but this energy cannot be considered work: This can be proven by coupling only one hot bath to the working fluid, which will increase the piston's energy. But if this energy is considered to be work, we may be able to extract work from a single bath, in clear violation of the second law of thermodynamics. This shows the need for a careful analysis of work in a fully quantized setup, which should be free of spurious violations of the second law.

\par

A proper definition is based on the concept of passivity~\citep{lenard1978thermodynamical,pusz1978passive} explained below. Consider a quantum state $\rhop$. If $\rhop$ undergoes an entropy-preserving (thermally adiabatic) process (e.g., a unitary transformation), the resulting energy change is purely work. In particular, it is possible to find the appropriate unitary transformation
\begin{equation}\label{eq_qp_unitary_transformation}
  \rhop\mapsto U\rhop U^\dagger \eqcolon\rhoptilde
\end{equation}
that maximizes the energy change while preserving the entropy, $\mathcal{S}(\rhop)=\mathcal{S}(\rhoptilde)$, thereby maximizing the \emph{extractable} work,
\begin{equation}\label{eq_qp_maxwork}
 \Wmax^\mathrm{ext} (\rhop)= \ew{\Hp}_{\rhop} - \ew{\Hp}_{\rhoptilde}.
\end{equation}
This is the maximum work that can be extracted from the state $\rhop$ (for a given $\Hp$), or its \emph{work capacity}. Equation~\eqref{eq_qp_maxwork} requires $\rhoptilde$ to be passive: States are called passive, if they do not allow for work extraction, so that $\Wmax^\mathrm{ext}(\rhoptilde)=0$.

\par

We may apply this concept to the analysis of a quantized heat machine (QHM). Let us assume that initially the piston is prepared in a state $\rhop(0)$ with work capacity $\Wmax^\mathrm{ext}(\rhop(0))$. The QHM operation changes the state of the piston, so that at time $t_m$ it will be in the state $\rhop(t_m)$ with work capacity $\Wmax^\mathrm{ext}(\rhop(t_m))$. The work extracted by the piston is then the increase of the piston's work capacity,
\begin{equation}\label{eq:wext}
  W_\mathrm{ext}=\Wmax^\mathrm{ext}(\rhop(t_m))-\Wmax^\mathrm{ext}(\rhop(0)).
\end{equation} 
A heat engine is expected to deliver $W_\mathrm{ext}>0$. Note that here we use a different sign convention than previously in order to be consistent with the work capacity definition.

\par

Since a Gibbs state has the minimal energy at a given entropy, Eq.~\eqref{eq_qp_maxwork} is maximized for
\begin{equation}\label{eq_qp_rhoptilde}
  \rhoptilde\coloneq\frac{1}{Z}\exp\left(-\frac{\Hp}{\kB \Tp(t)}\right).
\end{equation}
Namely, we identify $\rhoptilde$ with an \emph{effective} Gibbs state with a \emph{time-dependent} (effective) temperature $\Tp(t)$~\citep{gelbwaser2013workquantized,gelbwaser2014heat}. Using this definition, one can calculate the maximum extractable power~\citep{gelbwaser2013workquantized,gelbwaser2014heat}
\begin{equation}\label{eq_qp_maxpower}
 \left(\frac{\dd W}{\dd t}\right)_\mathrm{max}=\frac{\dd}{\dd t}\ew{\Hp}_{\rhop}-\Tp\frac{\dd}{\dd t}\Sp,
\end{equation}
where $\Sp$ is the piston's entropy. 

\par

Equation~\eqref{eq_qp_maxpower} clarifies the reason that not all of the energy interchanged by the piston is work---part of it involves an entropy change, heating up the piston. If the piston state is always a thermal (Gibbs) state, then $\frac{\dd}{\dd t}\ew{\Hp}_{\rhop}=\Tp\frac{\dd}{\dd t}\Sp$ and hence $\left(\frac{\dd W}{\dd t}\right)_\mathrm{max}=0$. This shows that a thermal state will not be able to extract work, thus resolving the apparent paradox proposed above.

\par

The complete QHM dynamics (including the piston) may be analyzed under the assumption that the system-piston coupling is weak, in addition to the assumption that the working fluid and the baths are weakly coupled as before~\citep{gelbwaser2013workquantized}. Under these assumptions, the working fluid will quickly reach its steady state, while the piston will slowly evolve.

\par

\begin{figure}
  \centering
  \includegraphics[width=9cm]{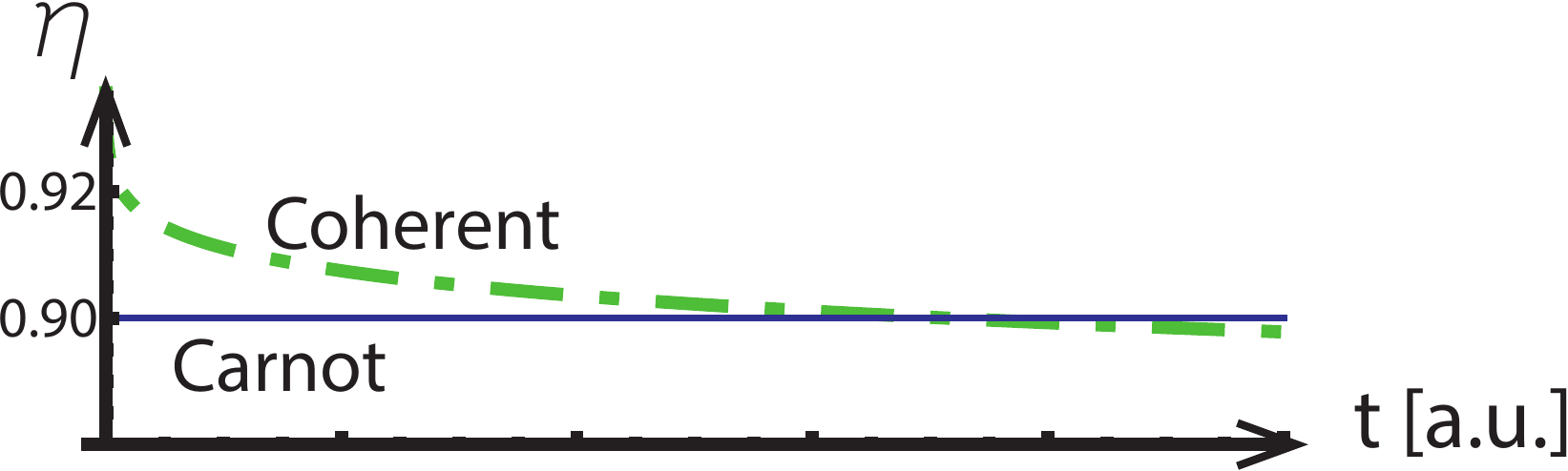}
  \caption{A low-entropy piston is a thermodynamic resource that may temporarily boost the efficiency above the standard Carnot bound in the QHM of Fig.~\ref{fig_qpiston}.}\label{fig_quantum_piston_efficiency}
\end{figure}

The Spohn inequality for fully quantized machines [\cf Eq.~\eqref{eq_cycle_second_law}] then reads 
\begin{equation}\label{eq_qp_spohn_second_law}
 \frac{\dd}{\dd t}\Sp \geq \frac{\Jh}{\Th}+\frac{\Jc}{\Tc}
\end{equation}
and the efficiency, in the engine operation regime, is bounded by 
\begin{equation}\label{eq_qp_eff}
\eta \leq
\begin{cases}
   1- \frac{\Tc}{\Th} & \text{for }\Tp>\Tc \\ 
   1- \frac{\Tp}{\Th} & \text{for }\Tc>\Tp
\end{cases}
.
\end{equation}
Clearly, the second line exceeds the Carnot bound. This striking result shows that the quantum state of the piston is an hitherto unexplored thermodynamic resource---its negentropy may boost the QHM efficiency (see Fig.~\ref{fig_quantum_piston_efficiency}). This resource is outside the standard assumptions that lead to the Carnot bound, therefore the extra efficiency is not forbidden by the second law. This resource may be thought of as an effective third bath, although the piston is not a real bath. A Carnot engine operating with three baths will bound the fully-quantum heat-engine efficiency~\eqref{eq_qp_eff}, thus showing that this effect may be explained by classic thermodynamic arguments, if one considers the thermal resource provided by the piston state~\citep{gelbwaser2014heat}.

\par
\par

\begin{figure}
  \centering
  \includegraphics[width=9cm]{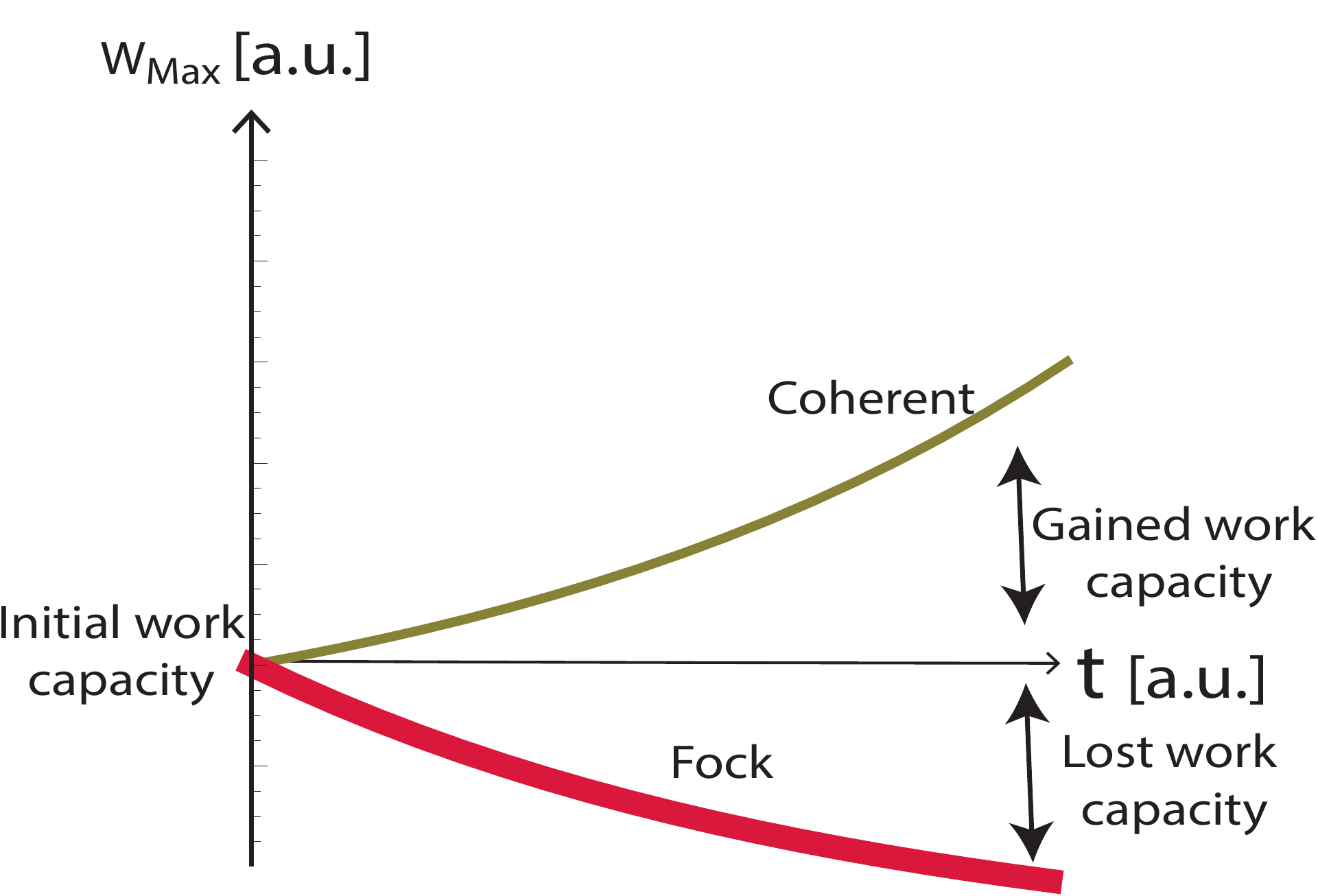}
  \caption{Comparison of work capacity in the QHM of Fig.~\ref{fig_qpiston} for two different initial quantum states of the piston (with the same initial energy): While a coherent state may increase its work capacity, a Fock state will inevitably lose it.}\label{fig_quantum_piston_work_capacity}
\end{figure}

In the semiclassical regime we have studied in Secs.~\ref{sec_cycles},~\ref{sec_periodically-modulated}, and~\ref{sec_nlevels}, as in nearly all QHM models, the state of the external modulation field is ignored, i.e., it is assumed that its change is negligible, even if the QHM acts over infinite time. Clearly, this is a crude approximation if this field is appreciably amplified or depleted in the process. By contrast, in the present model that transgresses the semiclassical approximation, work extraction depends on the initial quantum state of the piston according to the following classification~\citep{gelbwaser2013workquantized,gelbwaser2014heat}:

\begin{itemize}
\item An initially thermal piston state remains thermal at all times and does not allow for work extraction.
\item An initial Fock state of the piston rapidly loses its non-passivity and thus its work capacity by thermalization (see Fig.~\ref{fig_quantum_piston_work_capacity}). This ``fragility'' of Fock states does not allow them to increase their work capacity with time, so they do not extract any work at all.
\item The work capacity (non-passivity) of an initial coherent state of the piston increases exponentially with time, so that work extraction is possible at all times (see Fig.~\ref{fig_quantum_piston_work_capacity}).
\item A quadrature-squeezed state remains non-passive at all times, but it produces higher entropy [\cf Eq.~\eqref{eq_qp_maxpower}] and therefore yields less work than a coherent state with the same mean energy.
\item If the piston is initially in a passive state, it will not extract work. Markovian evolution tends to preserve passivity. Nevertheless, it can be ``ignited'' by a small displacement in the phase plane away from zero energy. This displacement will render the state non-passive and will then allow for work extraction.
\end{itemize}
We may conclude from Eq.~\eqref{eq_qp_maxpower} that the less entropy is produced by a state, the better this state is suited for work extraction. Resilience against thermalization is thus the relevant property for work extraction. The best state among the pure states for work extraction is the most ``classical'' one, the coherent state. However, it is remarkable that even if its phase is averaged out, it still retains a higher capacity for work extraction than other, non-classical, states~\citep{gelbwaser2014heat}.

\subsection{Summary}

The definition of work and the concepts of work capacity and non-passivity of the piston state that we have invoked in a fully quantized heat machine revealed the conclusion that quantum states constitute thermodynamical resources that have been hitherto ignored. They can be used to boost the efficiency of quantum thermal engines without breaking the classical thermodynamic bounds. The key to work extraction is the resilience of the state to thermalization. Higher resilience ensures more non-passivity and therefore higher work extraction.

\par

For a large-amplitude coherent state of the piston, the results presented in Sec.~\ref{sec_periodically-modulated} are recovered~\citep{gelbwaser2014heat}. Hence, the periodically-driven model is a legitimate approximation in the case of semiclassical pistons (external fields). As in the continuous-cycle semiclassical heat engines of Sec.~\ref{sec_periodically-modulated}, the present fully quantum heat engines require spectral separation of the two baths.

\par

It is instructive to compare the present analysis to that of a maser, which can also be described as a heat machine~\citep{scovil1959three}. Its thermodynamic efficiency is given by the ratio of the output (signal) to the pump frequencies, which is the same as the Carnot bound. In our scenario there is no population inversion in the system---the gain is provided by the hot bath, but the analogy is complete, as detailed in~\citep{gelbwaser2014heat}. Yet, the traditional maser analysis does not discern between coherent- and Fock-state gain underscored here.

\section{Self-contained quantum refrigerator with a quantum piston}\label{sec_refrigerator}

\subsection{Motivation}

In this section we study---in analogy to the previous section---a fully quantum heat refrigerator with a quantized piston. The power source is thus also included in the model. Therefore, the QR is fully autonomous and may be miniaturized to the extreme. By contrast, most quantum refrigerator (QR) models~\citep{palao2001quantum,linden2010how,levy2012quantum,levy2012quantumrefrigerators,skrzypczyk2011smallest,correa2013performance,chen2012quantum,venturelli2013minimal,brunner2014entanglement,geusic1959three,gordonbook,gemmerbook,velasco1997new,gieseler2012subkelvin,guo2012performance,pekola2007normal} are semiclassical, i.e., they require an external modulation (or an external heat bath) to power the refrigerator.

\par

In the preceding section we have found that work extraction by a fully quantized heat engine requires the quantum piston to be in a state with high resilience to thermalization, e.g., a coherent state. Here we show that efficient cooling (refrigeration) requires a different property of the piston quantum state.

\subsection{Refrigeration efficiency bound with a quantized piston}

From Spohn's inequality for fully quantum heat machines [Eq.~\eqref{eq_qp_spohn_second_law}], the coefficient of performance (COP) of the refrigerator is found to be~\citep{gelbwaser2014heat}
\begin{equation}\label{eq_qr_cop}
  \mathrm{COP}=\frac{\Jc}{-\frac{\dd}{\dd t}\ew{\Hp}_{\rhop}}\leq\frac{\Tc}{\Th-\Tc}\left(1-\Th\frac{\frac{\dd}{\dd t}\Sp}{\frac{\dd}{\dd t}\ew{\Hp}_{\rhop}}\right).
\end{equation}
To calculate the COP, the refrigerator may be assumed to be powered by heat, $\ew{\Hp}_{\rhop}=\Tp\Sp$, as in the case of an absorption refrigerator, or by work power, $\ew{\Hp}_{\rhop}=\frac{\dd W}{\dd t}$, as in a Carnot refrigerator, or by a combination of both. The balance between heat flow and work power that sets the maximum COP depends on the initial state of the piston. 

\par

The first factor on the right-hand side of Eq.~\eqref{eq_qr_cop} is the standard Carnot bound~\eqref{eq_carnot_cop} for a refrigerator. However, the extra factor may increase the COP above this standard bound provided that the sign of the extra factor is positive---this requires an increase of the piston entropy, i.e., $\frac{\dd}{\dd t}\Sp>0$, while its energy is being reduced (thereby powering the refrigerator, i.e., $\frac{\dd}{\dd t}\ew{\Hp}_{\rhop}<0$).  Only non-passive states may achieve this boost, because a passive state is a minimal-energy state for a fixed entropy (see Sec.~\ref{sec_work_fully_quantized}). Hence, any decrease in the energy of a passive state corresponds to its entropy reduction.

\par

Namely, only a quantum piston in a non-passive state may simultaneously power the QR and absorb heat from the cold bath by increasing its own entropy. The ability of the piston to absorb heat is not considered in the standard Carnot derivation, which therefore does not forbid the boost of the COP.
\par

\begin{figure}
  \centering
  \includegraphics[width=9cm]{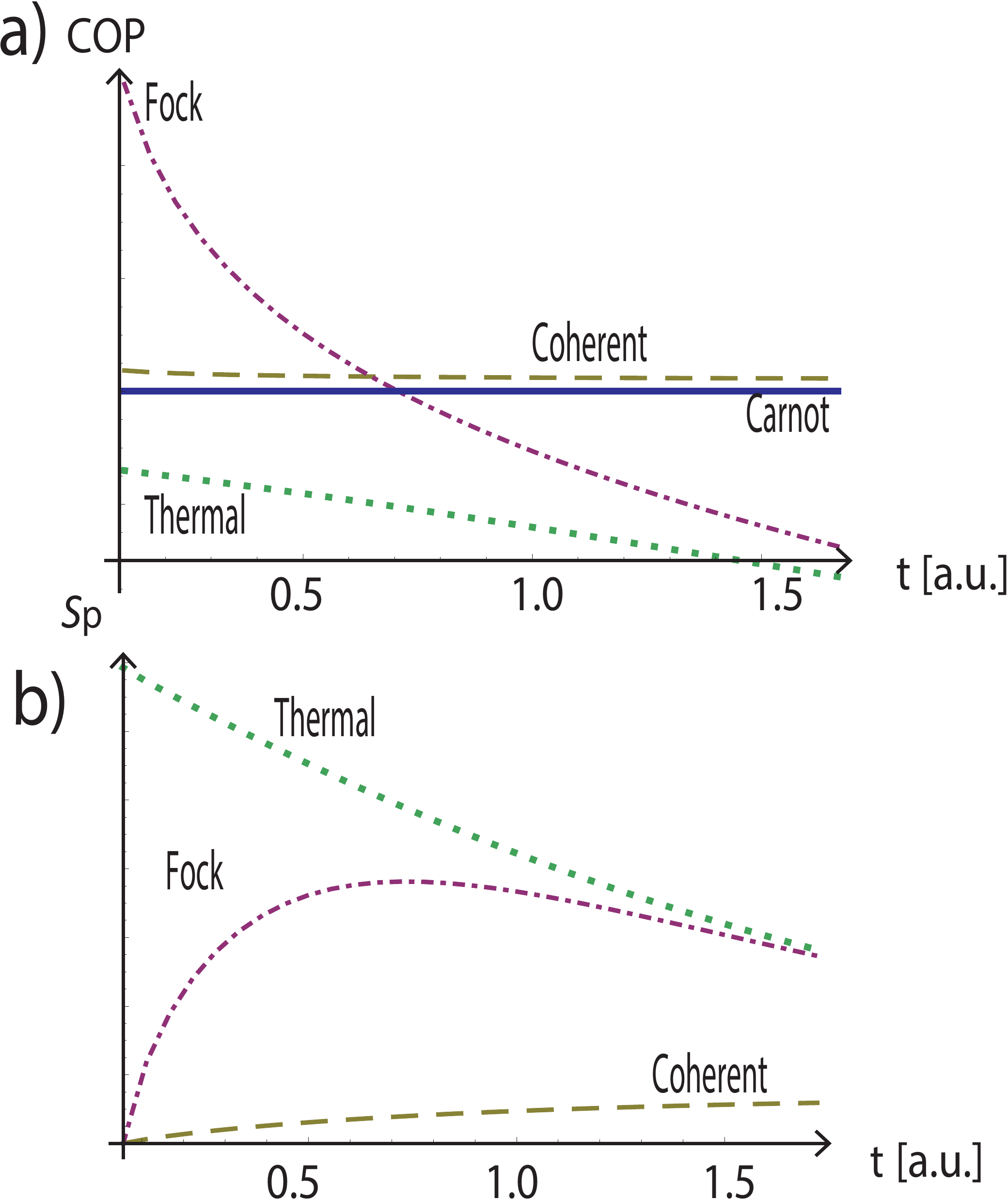}
  \caption{(a) COP and (b) entropy of the piston in a QR (described by Fig.~\ref{fig_qpiston}) for different initial states with the same initial mean energy as a function of time. A Fock state outperforms a coherent state at short times because its entropy change is larger than that of a coherent state.}\label{fig_quantum_refrigerator}
\end{figure}

Amongst the non-passive states, the most suitable for cooling according to Eq.~\eqref{eq_qr_cop}, namely those with the highest COP, are the most unstable states, i.e., the ones that maximize $\frac{\dd}{\dd t}\Sp$, as opposed to states that maximize work extraction by minimizing $\frac{\dd}{\dd t}\Sp$. We may classify the COP of non-passive states as follows~\citep{gelbwaser2014heat}:
\begin{itemize}
\item An initial coherent state, owing to its stability, i.e., the slow increase of its entropy, provides a COP just above the Carnot bound. Whilst their stability prevents coherent states from producing an excessive COP, it still allows them to surpass the Carnot bound at long times.
\item For an initial Fock state, the COP surpasses that of a coherent state with the same initial energy for a short time. The reason is that the Fock state is much more prone to thermalization, and its entropy initially increases very fast (see Fig.~\ref{fig_quantum_refrigerator}). At longer times its entropy saturates and starts decreasing, reducing the COP below the Carnot bound. Amongst all initially pure states with the same energy, the Fock state has the largest COP at short times.
\item For an initial thermal state of the piston at a high temperature ($\Tp>\Th$), the COP will be the same as for an absorption refrigerator~\citep{palao2001quantum}. The piston then plays the r\^ole of an ``ultrahot'' bath that powers the refrigerator. The piston will remain in a thermal state, but its temperature $\Tp$ will decrease until it reaches a critical temperature at which the cooling stops.
\end{itemize}

\subsection{Summary}

In Sec.~\ref{sec_quantum_piston} we showed how to maximize the work capacity of a quantum piston. This work can be used to power a QR, in the same way that a battery is used: The piston that was ``charged'' by the fully quantum heat engine may be subsequently used to power the fully quantum refrigerator. This ``quantum battery'', once charged, does not require any external energy source to drive the QR.

\par

Non-passivity allows a state to power the QR while absorbing heat from the cold bath, thereby cooling it further. As opposed to work extraction, the more unstable a state becomes, the higher its COP at short times. Thus, we find again (\cf Sec.~\ref{sec_quantum_piston}) that non-passive quantum states are thermodynamic resources that may improve the QR performance. As in the case of the engine, the extra efficiency originates from the properties of the state and not from the breakdown of any thermodynamic principle. 

\section{Continuously-driven qubit as quantum cooler}\label{sec_quantum_cooler}

\subsection{Motivation and model}

Up to now we have considered $\sigmaz$-coupling of the working-fluid TLS to the modulating field (piston), resulting in its energy modulation, as opposed to its $\sigmax$-coupling to the thermal baths. In this section we apply our general Floquet theory to a setup based on a piston that drives the working-fluid TLS through $\sigmax$-coupling, $\sigmaz$-coupling between the working fluid and a \emph{dephasing bath}~\citep{rao2011zeno}, and $\sigmax$-coupling of the TLS to another bath. This model allows us to explore the universality of the surveyed principles. In particular, we analyze the case of laser-induced cooling of a bath (buffer gas) at collisional equilibrium with the TLS~\citep{vogl2009laser,vogl2011collisional,szczygielski2013markovian}. This QHM allows \emph{cooling the bath down to an extremely low temperature}~\citep{gelbwaser2015laser}, without restricting the overlap between the bath spectra (unlike Secs.~\ref{sec_periodically-modulated} and~\ref{sec_nlevels}).

\par

The  working-fluid TLS is driven by a laser (the piston), and is permanently coupled to two heat baths: By elastic (dephasing) collisions to the hot ($\indexh$) bath, associated with the buffer gas (BG), and by spontaneous emission to the cold ($\indexc$) bath, the electromagnetic (EM) vacuum.

\par

The laser-driven system Hamiltonian is assumed to be
\begin{equation}
 \Hs(t)=\frac{1}{2}\hbar\omega_{0}\sigmaz+\hbar g\bigl(\splus e^{-i\nu t}+\sminus e^{i\nu t}\bigr),
\end{equation}
where $\omega_{0}$ is the (resonance) frequency of the TLS, $\nu$ is the laser frequency and $g$ is the coupling strength between the laser and the TLS. The laser detuning is $\Delta=\omega_0-\nu$. The TLS coupling to the BG is via elastic collisions that do not change the TLS level populations (thus constituting a dephasing bath), described by
\begin{equation}
 \Hsb^\indexh=\sigmaz \otimes B_\indexh,
\end{equation}
 where $B_\indexh$ is the dephasing (BG) bath operator. The coupling to the cold bath (EM vacuum) via spontaneous emission is given by
\begin{equation}
 \Hsb^\indexc=\sigmax \otimes B_\indexc,
\end{equation}
where $B_\indexc$ is the EM bath operator. 

\par

Using the Floquet method developed in Sec.~\ref{sec_cycles}, a master equation is derived for the laser-dressed TLS coupled to the baths. The transitions induced by the EM ($B_\indexc$) bath in the TLS are at frequencies $\nu_{\pm}=\nu \pm \Omega_\mathrm{G}$, where $\Omega_\mathrm{G}=\sqrt{\Delta^2+4g^2}$ and the BG ($B_\indexh$) bath induces transitions at Rabi frequency $\Omega_\mathrm{G}$. For typical realizations $\beta_\mathrm{EM}\hbar \nu_{\pm} \gg 1$, so that we may consider the EM bath to be effectively at zero temperature. For the BG, as shown below, cooling requires $\beta_\mathrm{BG}\hbar \Omega_\mathrm{G} \sim 1$, where $\beta_\mathrm{BG}$ is the BG inverse equilibrium temperature. This condition can be satisfied by appropriately adjusting the detuning $\Delta$ and the coupling $g$ of the laser-field strengths.

\par

The evolution (master) equation yields for the TLS excited- and ground-state populations $\rho_{ee}$ and $\rho_{gg}$, respectively,
\begin{subequations}\label{eq_cooler_tls}
  \begin{equation}
    \dot{\rho}_{ee}=-\left(r_{0}+r_+\right)\rho_{ee}+\left(e^{-\hbar\beta_\mathrm{BG} \Omega_\mathrm{G}}r_{0}+r_-\right)\rho_{gg},
  \end{equation}
where the transition rates
  \begin{equation}
    r_{\pm}= \left(\frac{\Omega_\mathrm{G} \pm \Delta}{2\Omega_\mathrm{G}} \right)^2 \Gc(\nu_{\pm}), \quad
    r_0=  \left(\frac{2g}{\Omega_\mathrm{G}} \right)^2 \Gh(\Omega_\mathrm{G})
\end{equation}
\end{subequations}
are determined by $\Gc$ and $\Gh$, the EM and buffer gas coupling spectra, respectively.

\par

\begin{figure}
  \centering
  \includegraphics[width=5cm]{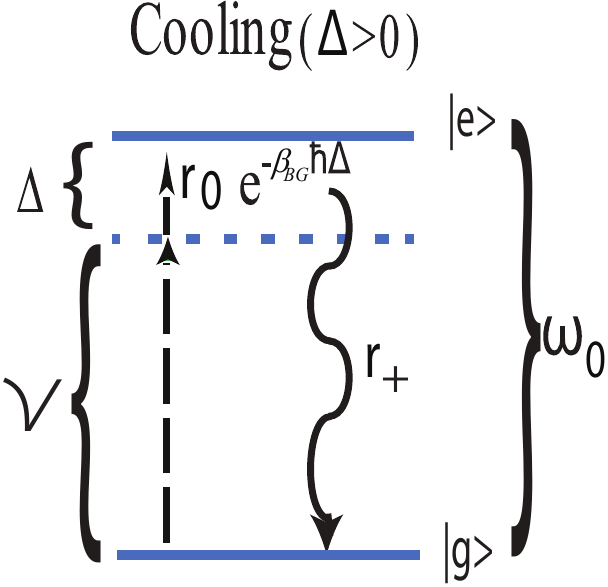}
  \caption{Scheme of cooling a hot dephasing bath by a weak red-detuned laser acting on a TLS: Energy absorbed from the bath compensates the energy mismatch and yields spontaneous emission that cools the hot bath.}\label{fig_quantum_cooler_levels}
\end{figure}

The present cooling process can be understood as follows. The collisions with the BG broaden the TLS levels, allowing it to absorb a photon with less energy than its resonance frequency ($\omega_0>\nu$). This process is accompanied by energy flow from the BG (hot bath) that compensates for the energy mismatch between absorption at $\nu$ and spontaneous emission at $\omega_0$ to the EM vacuum (cold bath). The BG temperature is thereby reduced (\cf Fig.~\ref{fig_quantum_cooler_levels}).

\par
\par

\subsection{Cooling and heating as a function of laser parameters}

At long times the TLS, governed by Eqs.~\eqref{eq_cooler_tls}, reaches a steady state with an effective inverse temperature $\beta_\mathrm{TLS}$, which is a measure of the TLS stationary level-population ratio $\frac{\rho_{ee}}{\rho_{gg}}$ and depends on the laser detuning. This inverse temperature, and therefore the laser detuning, allow us to determine the heat flow between the BG bath and the TLS. 

\par

The heat current flowing from the BG (hot bath) to a single TLS is [see the derivation in~\citep{szczygielski2013markovian}]
\begin{equation}\label{eq:coolpow}
  \Jh =A\left (r_+ e^{- \hbar \beta_\mathrm{BG} \Omega_\mathrm{G}}- r_- \right).
\end{equation}
Here $A$ is a positive process-dependent constant and the current is determined by the interplay of two different rates, cooling ($r_+ e^{-\beta_\mathrm{BG}\hbar \Omega_\mathrm{G}}$) and heating ($r_-$). Their balance determines whether the BG bath is cooled down or heated up. 

\par

In Sec.~\ref{sec_periodically-modulated}, highly efficient quantum heat machines required spectrally separated baths. Here, however, cooling depends on the coupling spectrum of the EM bath at two \emph{different frequencies}, making the spectral separation irrelevant. Namely, we can ensure the cooling \emph{independently} of the shape of the EM coupling spectrum. This advantageous property is clearly seen when the driving is weak ($g \ll |\Delta|$), since one of the rates $r_{\pm}$ is then zero: For $\Delta>0$, $r_-/r_+\rightarrow 0$, $\Jh >0$, thus cooling the BG, while for $\Delta <0$, $r_-/r_+\rightarrow \infty$, $\Jh <0$, heating the BG.

\par

It can be seen from Eq.~\eqref{eq:coolpow} that cooling requires $\beta_\mathrm{BG}\hbar \Omega_\mathrm{G} \sim 1$, otherwise $\Jh$ may be negative. As the BG is cooled, $\beta_\mathrm{BG}$ increases, reducing the cooling power until it vanishes. By changing the detuning (and also the driving power), $\beta_\mathrm{BG}\hbar \Omega_\mathrm{G}$ may be kept constant, thereby maintaining the cooling power. This remarkable property of the present model allows changing controllable parameters ($\Delta$ and $g$) to maintain the cooling down to ultralow temperatures. 

\par

As a consequence of the second law of thermodynamics (through the KMS condition) there is a relation between the cooling and the heating power of the BG gas under weak driving,
\begin{equation}\label{eq:ks}
  \frac{-\Jh (-\Delta)}{\Jh (\Delta)}=e^{\beta_\mathrm{BG}\hbar \Delta}.
\end{equation}

\par

\begin{figure}
  \centering
  \includegraphics[width=9cm]{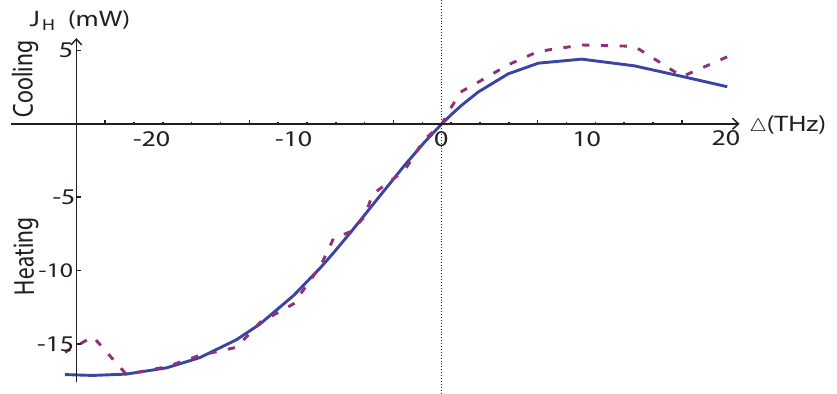}
  \caption{Theoretical (solid) vs. experimental (dashed) cooling power in the scheme of Fig.~\ref{fig_quantum_cooler_levels}, realized by laser-driven TLS in contact with a hot buffer gas (at 600\,\textcelsius).}\label{fig_quantum_cooler}
\end{figure}

As shown in Fig.~\ref{fig_quantum_cooler}, this theoretical model accurately describes the cooling-power dependence on $\Delta$, as experimentally measured in~\citep{vogl2009laser,vogl2011collisional}.

\par
\par

\subsection{Summary}

The present cooling machine was shown to represent a very versatile scheme for manipulating the temperature of a bath such as a BG that causes pressure-broadening of laser-driven atoms, by simply changing the laser frequency. 

\par

The only requirements for cooling in this scheme are a red-detuned laser frequency and weak driving, without restriction on the bath spectra. This stands in contrast to the cooling models in Sec.~\ref{sec_periodically-modulated}, where not only the modulation frequency had to be above a critical frequency, but also non-overlapping hot and cold bath spectra were required. The present scheme has distinct advantages (in terms of the minimum cooling temperature and cooling efficiency) compared to standard cooling techniques such as sideband cooling or Doppler cooling~\citep{haensch1975cooling,wineland1975proposed,chu1998nobel,cohentannoudji1998nobel,phillips1998nobel,schliesser2008resolved,epstein1995observation,mungan1997laser,dehmelt1976entropy,wineland1979laser}.

\section{Cooling speed of quantum baths}\label{sec_cooling_speed}

\subsection{Motivation}

Here we are concerned with the third law of thermodynamics in its dynamical formulation, known as Nernst's unattainability theorem, which forbids physical systems to cool down to absolute zero ($T=0$) in finite time~\citep{nernst1906ueber,landsberg1956foundations,belgiorno2003notes}. In order to investigate this question~\citep{kolar2012quantum} we re-consider the quantum heat machine setup discussed in Sec.~\ref{sec_periodically-modulated}, consisting of a TLS whose transition frequency is periodically modulated by an external driving field, whilst being simultaneously coupled to cold and hot heat baths. The heat baths in Sec.~\ref{sec_periodically-modulated} were assumed to have an infinite heat capacity, meaning that their temperatures are not modified in the course of cooling, regardless of the amount of heat exchanged between the working fluid and the baths. In this section, however, we assume the cold bath to possess a \emph{finite heat capacity}, such that its temperature is progressively reduced as heat is extracted from it, i.e., when the heat machine is operated as a (quantum) refrigerator.

\subsection{Cooling-rate scaling with temperature}

Whilst the temperature of the hot bath remains constant throughout the time evolution, the temperature of the cold bath, whose heat capacity is finite, decreases due to heat being extracted to it via the work invested by the external (classical) driving field. From the definition of the heat capacity it follows that~\citep{reichlbook}
\begin{equation}\label{eq_cs_tc_derivative}
  \cv\frac{\dd \Tc}{\dd t}=-\Jc,
\end{equation}
where a positive sign of $\Jc$ corresponds to heat flowing from the cold bath to the working fluid.

\par

The heat capacity of the cold bath scales as~\citep{kolar2012quantum,kittelbook,reichlbook}
\begin{equation}\label{eq_cs_cv}
    \lim_{\Tc\rightarrow 0}\cv\sim \Tc^d
  \end{equation}
at low temperatures, where $d$ is the bath dimensionality. Under spectral separation of the two baths, the TLS is only coupled to the hot bath at $\omega\simeq\omega_0+\Omega$, and at $\omega-\Omega$ to both baths. The heat current $\Jc$ is maximized if the modulation rate $\Omega$ satisfies the condition $\hbar\omega_0-\hbar\Omega \simeq \kB\Tc$. This requires the modulation rate to be increased as the temperature decreases. The heat current~\eqref{eq_periodically_currents} from the cold bath can then be shown to scale as~\citep{kolar2012quantum}
\begin{equation}\label{eq_cs_jc_scaling}
  \Jc(\Tc)\propto -\Tc^{\gamma+d},
\end{equation}
where $\gamma$ is a bath-dependent exponent determined by the spectral dependence (dispersion) of the system-bath coupling at low frequencies,
\begin{equation}
  \lim_{\omega\rightarrow 0}|g(\omega)|^2\propto\omega^\gamma.
\end{equation}
Consequently, the time derivative of the temperature [Eq.~\eqref{eq_cs_tc_derivative}] scales as
\begin{equation}\label{eq_cs_tc_derivative_scaling}
  \frac{\dd \Tc}{\dd t}=-C \Tc^\gamma,
\end{equation}
where the positive constant $C$ is inversely proportional to the bath volume.

\subsection{Cooling-rate dependence on bath dispersion}

Remarkably, the cooling speed~\eqref{eq_cs_tc_derivative_scaling} is determined by a single parameter $\gamma$~\citep{kolar2012quantum,levy2012quantumrefrigerators}. For $\gamma=1$ the temperature $\Tc$ decreases exponentially with time, in accordance with Nernst's theorem~\citep{nernst1906ueber,landsberg1956foundations,belgiorno2003notes}. For smaller exponents $0\leq\gamma<1$, however, the temperature reaches absolute zero in finite time, in contradiction to Nernst's theorem.

\par

\begin{figure}
  \centering
  \includegraphics[width=9cm]{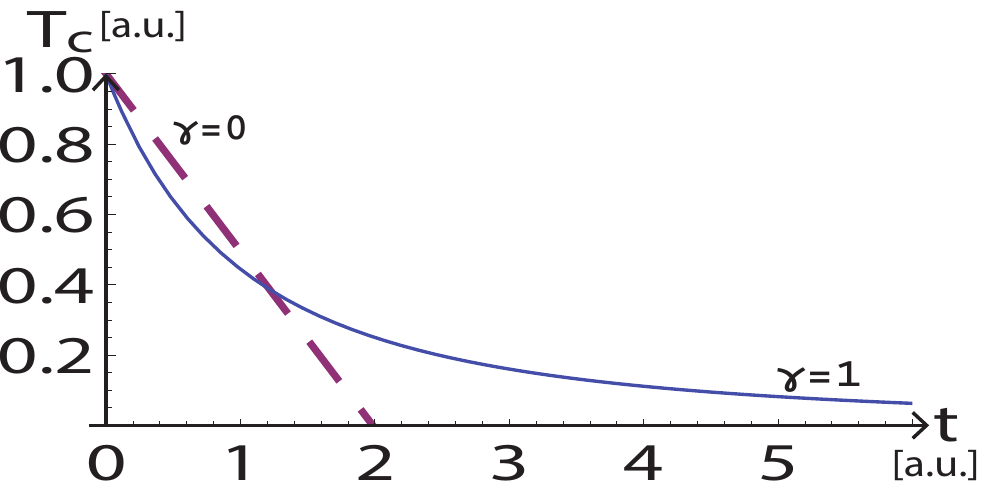}
  \caption{Cooling of a phonon bath ($\gamma=1$, solid line) and a magnon bath ($\gamma=0$, dashed line) in the scheme of Fig.~\ref{fig_universal}. The absolute zero is attained by a magnon bath at a finite time.}\label{fig_cooling_speed}
\end{figure}

As shown in~\citep{kolar2012quantum}, the value of $\gamma$ is related to the bath dispersion relation. Two typical cases can be discerned (see Fig.~\ref{fig_cooling_speed}):
\begin{itemize}
\item The dispersion relation of \emph{acoustic phonons} reads $\omega({\mathbf{k}}) \simeq v|{\mathbf{k}}|$, $v$ being the sound velocity. Their coupling to the system qubit scales linearly with the frequency, $|g(\omega)|^2~\sim |\mathbf{k}|^2/{\omega(\mathbf{k})}\sim \omega $, resulting in $\gamma = 1$. Consequently, a bath of acoustical phonons approaches $T=0$ exponentially slow [\cf Eq.~\eqref{eq_cs_tc_derivative_scaling}], in accordance with Nernst's principle~\citep{levy2012quantumrefrigerators}.
\item Spin-wave excitations (or \emph{magnons}) of a ferromagnetic spin lattice with nearest-neighbor interactions are bosons by virtue of the Holstein-Primakoff transformation~\citep{kittelbook}. The dipolar coupling of these bosons to the system is independent of the frequency, i.e., $\gamma=0$. Hence, the absolute zero is approached linearly in time [\cf Eq.~\eqref{eq_cs_tc_derivative_scaling}], thus violating Nernst's unattainability principle~\citep{kolar2012quantum}.
\end{itemize}

\par

\subsection{Summary}

The validity of the third law is challenged within the quantum refrigerator model presented in Sec.~\ref{sec_periodically-modulated}, when the cold bath is described by a bosonic bath with \emph{finite} heat capacity: Magnon baths may then be cooled down to absolute zero within a finite time. It is, however, important to bear in mind that the predicted violation of the third law has been derived within a simple model. It cannot be a-priori ruled out that in real systems modified interactions or couplings to additional baths arise at low temperatures, thereby effectively reducing the cooling rate. The fundamental nature of the third law and the possible implications of its invalidity calls for experimental verification of these predictions, preferably in ``clean'' (isolated) and controllable environments, e.g., within optical lattices that can emulate magnon baths~\citep{diehl2008quantum} in order to clarify these results.

\section{Control of non-Markovian thermodynamic processes}\label{sec_nonmarkovian}

Having analyzed various models of thermal machines based on Markovian dynamics, we present in the last parts of this review (Secs.~\ref{sec_nonmarkovian} and~\ref{sec_work_information}) the results of a non-Markovian model. To this end we consider a qubit (TLS) subject to frequent measurements or phase shifts on time scales shorter than the memory time of the bath to which it couples. This dynamical control affects the system-bath correlations and allows for work extraction.

\par

Quantum thermodynamics is usually based on long-time Markovian (Lindblad) master equations, which describe convergence to equilibrium at a constant rate. By contrast, the understanding of non-Markovian short-time effects in quantum thermodynamics is scanty at best. Here we explore one such non-Markovian process: An impulsive (brief) quantum non-demolition (QND) measurement of the qubit energy or its phase shift disturbs its equilibrium with the bath, thereby abruptly changing its temperature. The subsequent evolution of the qubit in the presence of the bath alternates between heating and cooling at times comparable to the qubit oscillation period~\citep{gordon2009cooling,erez2008thermodynamic}. Such effects are at odds with the standard Markovian notions of thermodynamics~\citep{spohn1978entropy,alicki1979quantum,kosloff2013quantum}, whereby temperature and entropy must monotonically converge to their equilibrium values.

\subsection{Model}

The basic model is the one already encountered within the present review, namely a qubit (TLS) with modulated transition frequency as in Eq.~\eqref{eq_periodically_Hs}. However, it is here permanently coupled to a \emph{single} heat bath $B$ via the spin-boson Hamiltonian and is subject to $\sigmaz$-control~\citep{kofman2004unified}
\begin{subequations}\label{eq_nm_H}
  \begin{equation}
    H_\mathrm{tot}=\Hs(t)+\Hb+\Hsb,
  \end{equation}
  with the constituents
  \begin{align}
    \Hs(t) &=\frac{1}{2}\hbar\omega(t)\sigmaz \label{eq_nm_HS}\\
    \Hb &=\hbar\sum_k\omega_k a_k^\dagger a_k \\
    \Hsb &=\sigmax B = \sigmax \hbar \sum_k\left(\eta_k a_k+\eta^*_k a_k^\dagger\right)\label{eq_nm_HSB}.
  \end{align}
\end{subequations}
Here $\omega(t)$ is the modulated (via phase shifts) or randomized (via impulsive measurements) qubit resonance frequency, $\sigmax$ is the $x$-Pauli operator, and $a_{k}$ ($a^\dagger_{k}$) are the annihilation (creation) operators of the bath. Finally, $\eta_{k}$ denotes the matrix element of the $\sigmax$-coupling of the system to the bath mode $k$. We stress that we \emph{do not} invoke the rotating-wave approximation (RWA)~\citep{cohenbook} in the interaction Hamiltonian $\Hsb$. Namely, we do not impose energy conservation on the excitation exchange between the system and the bath, on the time scales considered~\citep{kofman2004unified}. Nevertheless, the total energy is conserved by $H_\mathrm{tot}$ when $\Hs$ is time-constant.

\subsection{System-bath correlations at equilibrium}

At equilibrium, the qubit and the bath are inevitably in an entangled (correlated) state given by the density matrix for the total system and the bath, $\rhoeq = \exp(-\beta \Htot)/Z$, where $Z$ is the partition function and $1/\beta = \kB T$. Its off-diagonal elements express quantum correlations between the qubit and the bath. These correlations are concealed when observing only the qubit state, $\rho=\Tr_\mathrm{B}\rhoeq$. Yet these correlations increase the temperature (mixedness) of $\rho$ as the system-bath coupling grows~\citep{gordon2009cooling}. Even at \emph{zero temperature, the state-purity of a qubit is never complete, due to system-bath entanglement}. As the bath temperature rises, this purity exhibits a non-monotonic dependence on $\beta$.

\par

For a Lorentzian coupling spectrum of the bath of width $\Gamma$ (centered around $\nu_0$)
\begin{equation}
  \eta_{k}=\eta_\mathrm{max}\sqrt{\frac{\Gamma^2}{\Gamma^2+(\nu_0-\omega_{k})^2}},
\end{equation}
the mean interaction energy at $T=0$ is given by the bath-induced Lamb shift~\citep{cohenbook} in the form
\begin{equation}\label{eq_nm_Hsb_eq}
  \langle H_\mathrm{SB} \rangle_\mathrm{eq} \approx -\hbar \omega_0 \int^\infty_0 \dd\omega \frac{\Gamma^2}{\Gamma^2 + (\nu_0 - \omega)^2} \frac{1}{1+\omega/\omega_0},
\end{equation}
where $\omega_0$ denotes the unperturbed qubit resonance frequency. The \emph{negativity} of the mean system-bath interaction energy in equilibrium will play a crucial r\^ole in subsequent calculations.

\subsection{Impulsive perturbations of a qubit at equilibrium}\label{sec_nm_impulsive}

We consider the effect of quantum non-demolition (QND) operations performed on the qubit by a device through the Hamiltonian $H_\mathrm{SD}$, i.e., operations that commute with the operator $\sigmaz$~\citep{erez2008thermodynamic}. If performed at the right rate, these QND operations enable the gradual purification (cooling) of qubits coupled to a non-Markovian bath, although they are \emph{non-selective}, i.e., their results are neither read out nor acted upon. It is necessary that $H_\mathrm{SD}\propto\sigmaz$, i.e., that it does not commute with $\Hsb\propto\sigmax$, otherwise a classical-like force acts on $\rho$, changing it in a non-QND fashion. Examples of such QND operations are projective measurements of the qubit energy or its phase flips.

\par

In a projective measurement of $\sigmaz$, the state is projected onto the energy eigenstates $\ket{e}$ and $\ket{g}$, such that the joint system-bath ($SB$) state after the measurement ($M$) is changed from $\rhoeq$ to~\citep{gordon2009cooling}
\begin{equation}
  \rhosb=\rhoeq \mapsto \rhosb^\mathrm{M}=\frac{1}{2}(\rhoeq+\sigmaz\rhoeq\sigmaz).
\end{equation}
By contrast, a phase shift results in a rotation of the state about the $z$ axis by a phase $\phi$,
\begin{equation}
  \rhoeq\mapsto\rhosb^\phi=\exp(-i\phi\sigmaz/2)\rhoeq\exp(i\phi\sigmaz/2).
\end{equation}
Such a rotation may be realized by an ac Stark shift of the qubit's resonance frequency by an external field.

\par

By the definition of a QND operation, the qubit excitation would not be altered---if the qubit where isolated. However, due to its interaction with the bath (which does not commute with $H_\mathrm{SD}$), $\sigmaz$ operations result in alternate heating and cooling of the qubit, depending on the time interval between consecutive QND operations. As shown in~\citep{gordon2009cooling}, the system-bath correlation energy $\ew{\Hsb}$ is \emph{increased} by an impulsive QND projective measurement by $|\ew{\Hsb}_\mathrm{eq}|$, where the latter quantity [\cf Eq.~\eqref{eq_nm_Hsb_eq}] corresponds to the energy cost of \emph{erasing} the system-bath correlations (entanglement). The joint system-bath density matrix is thus transformed into an approximately factorized form~\citep{erez2008thermodynamic}. By contrast, phase shifts retain the system-bath correlations but modify the off-diagonal elements of the density matrix~\citep{gordon2009cooling}. However, as a result of either projective measurements or phase flips, the joint system-bath state is transformed into a non-equilibrium state that starts evolving.

\subsection{Post-measurement state and its free evolution: Alternating heating and cooling}

The joint system-bath post-measurement state satisfies~\citep{erez2008thermodynamic}
\begin{equation}
  \rhosb^\mathrm{M}\approx Z^{-1}e^{-\beta \left( \Hs+H_\mathrm{B} + \mathcal{O}(H_\mathrm{SB}^2) \right)},
\end{equation}
which for sufficiently weak coupling (small $\eta_\mathrm{max}$) can be approximated by the product state
\begin{equation}
  \rhosb^\mathrm{M}\approx \rho \otimes \rho_\mathrm{B}.
\end{equation}

\par

The evolution of $\rho$ at longer times can be approximately described by a non-Markovian master equation~\citep{gordon2009cooling,gordon2010equilibration,alvarez2010zeno}. The latter equation is equivalent to the rate equations~\citep{kofman2004unified,gordon2007universal}
\begin{subequations}
 \begin{align}
  \dot\rho_{ee}(t) &= -\dot\rho_{gg}(t) = R_g(t)\rho_{gg}-R_e(t)\rho_{ee}\label{eq_nm_rate_equations}\\
  R_{e(g)}(t) &= 2 t\int_{-\infty}^\infty \dd\omega G_T(\omega)\sinc\left[(\omega\mp\omega_0)t\right]
\end{align}
\end{subequations}
for the qubit populations. Here $\sinc(x)\coloneq\frac{\sin(x)}{x}$, and $G_T(\omega)$ is the temperature-dependent bath response. We shall assume that $G_0(\omega)$, the zero-temperature coupling spectrum, is peaked at $\omega_0$ and has spectral width $\sim 1/\tc$, the inverse memory time. The qubit dynamics is thus given by the time-dependent absorption (relaxation) rates $R_{e(g)}(t)$. We can identify three distinct regimes according to their typical time scales:
\begin{itemize}
\item At short times $t\ll 1/\omega_0 \ll \tc$ the $\sinc$ function is much broader than $G_T(\omega)$, resulting in equal rates $R_{e(g)}$ at \emph{any} temperature. In this regime the RWA [See Eq.~\eqref{eq_nm_HSB}] breaks down, as both excitation and de-excitation of the qubit do not require absorption or emission from/to the bath, but rather a change in the system-bath interaction energy. The transition rates are then linear in time, 
  \begin{subequations}
    \begin{gather}
      R_{e(g)}(t\ll \tc)\approx 2 \dot R_0 t\\
      \dot{R}_0\equiv\int_{-\infty}^\infty \dd\omega G_T(\omega),
    \end{gather}
  \end{subequations}
consistently with the quantum Zeno effect (QZE) of relaxation slowdown~\citep{kofman2000acceleration,kofman2004unified}. The heating rate in this QZE regime is~\citep{gordon2009cooling,gordon2010equilibration,erez2008thermodynamic}
 \begin{equation}
  \frac{\dd}{\dd t}\left(\rho_{ee}-\rho_{gg}\right) \approx 4 \dot{R}_0 t(\rho_{gg}-\rho_{ee}).
 \end{equation}
\item At intermediate (but still non-Markovian) times, $t \sim 1/\omega_0$, the width of the sinc function and the coupling spectrum become comparable, causing damped oscillations of the transition rates near the frequencies $\omega_0\pm\nu_0$, respectively. This oscillatory behavior neither conforms to the QZE nor to the anti-Zeno effect (AZE) that signifies relaxation speedup. It may be dubbed the \emph{oscillatory Zeno effect} (OZE). In this non-RWA regime the transition rates can adopt negative values (since the sinc function can become negative). Since $R_g(t)$ is more likely to become negative than its counterpart $R_e(t)$, there can be a larger growth of the ground-state population than allowed by detailed balance, i.e., \emph{transient cooling} can be observed at these intermediate times~\citep{gelbwaser2013can}.
\item At long times $t\gg \tc$, the relaxation rates attain their Golden-Rule (Markovian) values~\citep{kofman2004unified} $R_{e(g)}(t\gg \tc) \simeq 2\pi G_T(\pm\omega_0)$. The populations then approach those of an equilibrium Gibbs state whose temperature is equal to that of the thermal bath (see also Sec.~\ref{sec_cycles}). 
\end{itemize}

\par

How can one interpret the non-Markovian transient cooling due to the impulsive disturbance? As shown in~\citep{gordon2009cooling}, the energy change of the system can be related to the respective changes of the bath's internal energy and of the system-bath correlation energy according to
\begin{equation}\label{eq_nm_delta_HS}
  \delta\ew{\Hs(t)}=-[\delta\ew{\Hb(t)}+\delta\ew{\Hsb(t)}].
\end{equation}
Oscillations induced by co-rotating terms (confirming to the RWA) result in energy exchange between the system and the bath, $\delta\ew{\Hs}=-\delta\ew{\Hb}$, whilst counter-rotating terms cause a rapid variation of $-[\delta{\Hb(t)}+\delta\ew{\Hsb(t)}]$, resulting in the qubit heating, $\delta\ew{\Hs}>0$.

\par

\begin{figure}
  \centering
  \includegraphics[width=9cm]{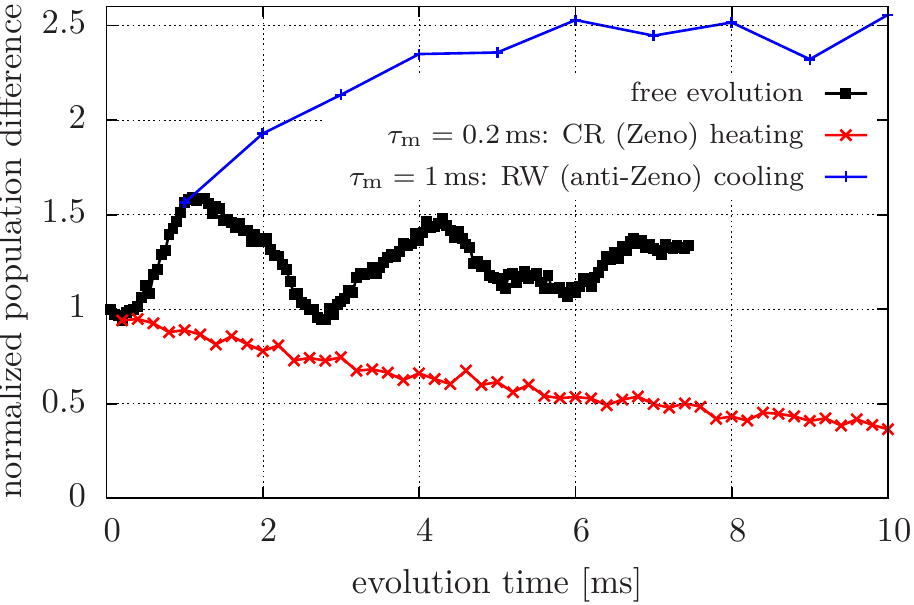}
  \caption{Demonstration of Zeno heating at short measurement intervals [due to counter-rotating terms in the Hamiltonian~\eqref{eq_nm_HSB}] and anti-Zeno cooling at longer measurement times (due to rotating-wave terms therein) in an NMR experiment~\citep{alvarez2010zeno}. The normalized population difference is $\frac{(\rho_{gg}-\rho_{ee})(t)}{(\rho_{gg}-\rho_{ee})(0)}$.}\label{fig_zeno}
\end{figure}

Upon repetition of this QND measurement procedure, the qubit is heated up or cooled down, depending on the time intervals between the measurements as verified experimentally (see Fig.~\ref{fig_zeno}). The bath and the system are affected differently and may thus acquire different temperatures. Remarkably, the system may heat up solely due to the QZE, although the \emph{bath is colder}, or cool down solely due to the OZE or AZE, although the \emph{bath is hotter}~\citep{gordon2008optimal,gordon2009cooling,gordon2010equilibration}. The bath, by contrast, may undergo changes in temperature and entropy only if its heat capacity is finite.

\par
\par

\subsection{Non-monotonic entropy evolution: Spohn's theorem violation}

One may introduce the ``entropy distance'', i.e., the entropy of a system-state $\rho$ \emph{relative} to its equilibrium state $\rho_0$ according to~\citep{alicki1979quantum,lindblad1974expectations}
\begin{equation}
  \mathcal{S}(\rho(t)||\rho_0)\coloneq\Tr\left[\rho(t) \ln \rho(t)\right]-\Tr\left[\rho(t) \ln \rho_0\right].
\end{equation}
The entropy-production rate then reads $\sigma(t) \equiv-\frac{\dd}{\dd t}\mathcal{S}(\rho(t)||\rho_0)$. In the Markovian regime, it follows from the second law that $\sigma(t) \geq 0$~\citep{alicki1979quantum,spohn1978entropy,lindblad1974expectations}. For diagonal $\rho$, $\sigma(t)$ is positive iff $\frac{\dd}{\dd t}\left|\rho_{ee}(t)-(\rho_0)_{ee}\right|\leq 0$. Yet, in the non-Markovian domain, $\rho_{ee}(t)$ may drift away from its equilibrium value, so that $\sigma(t)$ becomes \emph{negative}, thus violating Spohn's theorem [\cf Eq.~\eqref{eq_entropy_production_positive}]~\citep{gordon2009cooling,erez2008thermodynamic}: Equilibration need not be monotonic at non-Markovian times.

\subsection{Summary}

Periodic perturbations of a system that is initially at equilibrium with a thermal bath disrupt the equilibrium conditions, driving the system away from its original state, although these perturbations are quantum non-demolition (QND) operations (measurements or phase shifts) that would not change the system energy, if the system were isolated from the bath. The perturbed evolution strongly depends on the perturbation rate: Heating is obtained under extremely frequent perturbations that correspond to the quantum-Zeno domain, and cooling under less frequent ones associated with the anti-Zeno effect. These predictions were experimentally confirmed by an NMR experiment~\citep{alvarez2010zeno} in which a carbon nuclear spin coupled to a bath of three protons was either heated up or cooled down by noise pulses (that mimicked QND measurements), depending on their rate (see Fig.~\ref{fig_zeno}).

\section{Work-information relation under non-Markovian evolution: Violation of the Szilard-Landauer bound}\label{sec_work_information}

\subsection{The Szilard-Landauer principle revisited}

The Szilard-Landauer (SL) principle states that work obtainable from a measurement of a quantum system cannot exceed the energy cost of erasing its record from the observer's memory (\cf Sec.~\ref{sec_introduction})~\citep{szilard1929ueber,landauer1961irreversibility,bennett1973logical}. This means that an observer can extract work from a system upon \emph{modifying} it according to the results of its measurement. Such an omniscient observer is commonly referred to as ``Maxwell's demon''. But what happens if the information obtained from a measurement is \emph{not} read out, i.e., for non-selective measurements (NSM)? Is it still possible to extract work in such a scenario?

\par

Here we present a possibility to extract work from NSM and obtain a larger amount of work than that predicted by the SL principle, by exploiting the energy obtained from the change of the system-bath correlation energy through an impulsive measurement (\cf Eq.~\eqref{eq_nm_Hsb_eq} and Sec.~\ref{sec_nm_impulsive}). 

\par

\begin{figure}
  \centering
  \includegraphics[width=9cm]{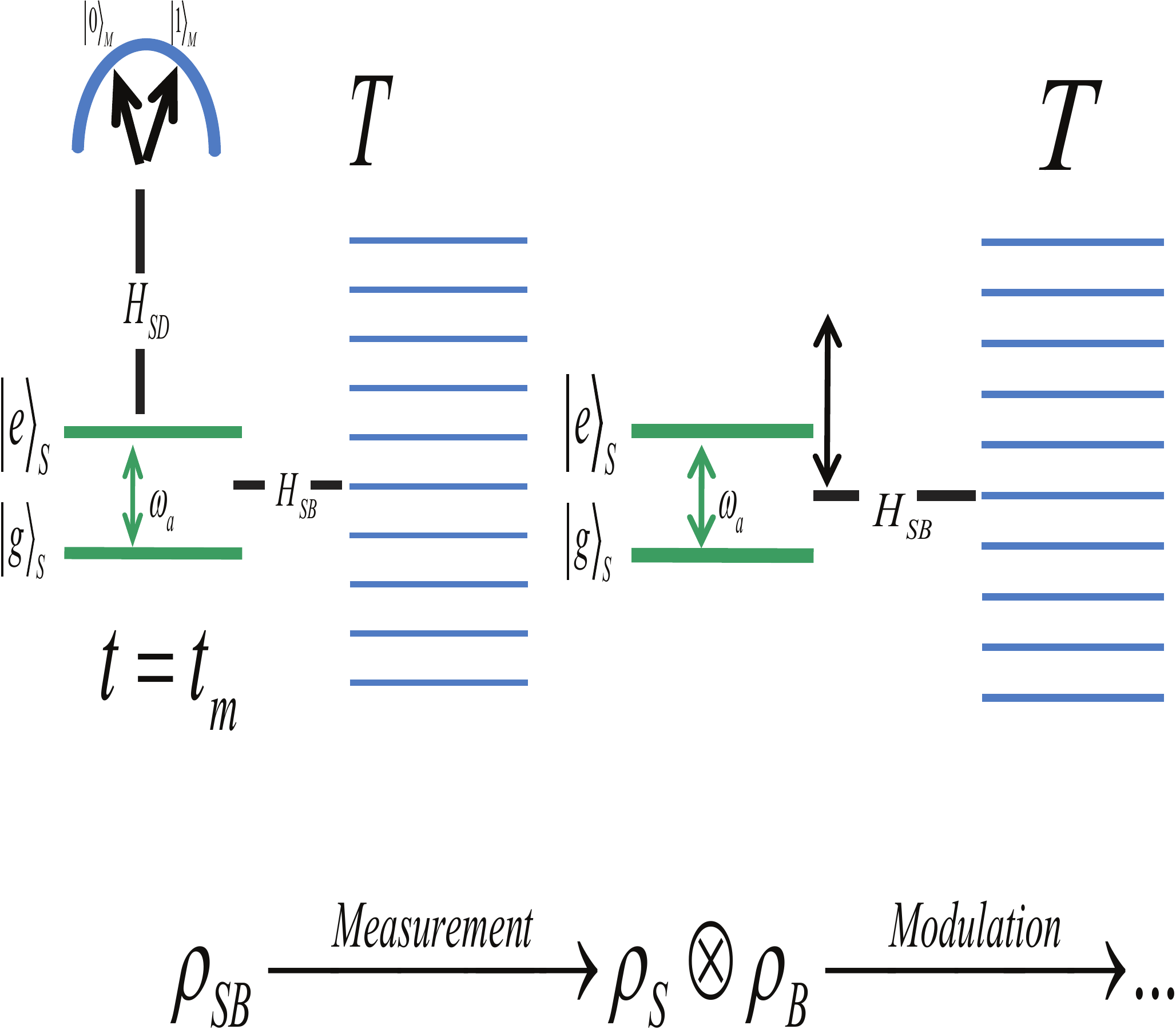}
  \caption{Sketch of a work-extraction cycle induced by a QND non-selective measurement (NSM) of the qubit energy, followed by its cyclic energy modulation effected by a classical piston. Throughout the cycle, the qubit is in contact with a bath at temperature $T$.}\label{fig_szilard_cycle}
\end{figure}

The protocol is as follows (see Fig.~\ref{fig_szilard_cycle})~\citep{gelbwaser2013work}: A brief QND measurement decorrelates the system from the bath, thus altering their correlation energy $\ew{\Hsb}$. Subsequently, a periodic modulation of the qubit's transition frequency allows for work extraction even if the measurement result remains \emph{unread}, i.e., for a NSM, provided that the cycle is completed within the bath memory time, so as to allow for temporal changes in the correlation energy that only take place on non-Markovian time scales.

\par
\par

\subsection{No work can be extracted from a single Markovian-bath in a closed cycle}

Let us revisit the rate equations~\eqref{eq_nm_rate_equations} under Markovian conditions, i.e., positive transition rates, for a thermal (Gibbs) state $\rho^\mathrm{eq}_{jj}(t)= Z^{-1}(t)e^{-\beta E_j(t)}$,
\begin{equation}
  R_{e}(t)\rho^\mathrm{eq}_{ee}(t) = R_{g}(t)\rho^\mathrm{eq}_{gg}(t).
\end{equation}
One can then prove~\citep{gelbwaser2013work} the following inequality for the von~Neumann entropy~\eqref{eq_entropy_general_definition},
\begin{equation}
    \dot{\mathcal{S}} \equiv -k_B\sum_{j} {\dot \rho}_{jj} \ln \rho_{jj} \geq -k_B\sum_{j} {\dot \rho}_{jj} \ln \rho^\mathrm{eq}_{jj} = {\frac{1}{T}} {\dot Q},
  \end{equation}
with $\dot Q= \sum_j {\dot \rho}_{jj} E_j$. For a closed cycle the initial and final state's entropy and energy are the same, and hence
\begin{equation}\label{eq_sl_no_work_single_bath}
  W_\mathrm{cycle}^\mathrm{ext}=Q\leq 0, 
\end{equation}
which corresponds to \emph{supplying} work to the system. Here the work extracted in a cycle is negative, i.e., it must be invested from the outside.

\subsection{Work extraction in a non-Markovian cycle}

The \emph{extractable} work in the cycle described above (Fig.~\ref{fig_szilard_cycle}), which is the negative of the work defined in Eq.~\eqref{eq_work_general_definition}, satisfies, on general grounds~\citep{alicki1979quantum},
\begin{equation}\label{eq_nm_oint}
  W^\mathrm{ext}_\mathrm{cycle}=-\oint \Tr\{\rho\dot{H}_\mathrm{S}\}\dd t=-\oint s(t)\dot{\omega}(t)\dd t.
\end{equation}
Here, as in Eq.~\eqref{eq_nm_HS}, $\omega(t)$ is the time-dependent qubit transition frequency and $s(t)$ is the qubit's level population difference. As shown by Eq.~\eqref{eq_sl_no_work_single_bath}, no work can be extracted in the Markovian regime. However, work can be extracted within such a cycle ($W^\mathrm{ext}_\mathrm{cycle}>0$) if the modulation happens on a non-Markovian time scale, i.e., for a modulation rate $\Omega$ that satisfies $\Omega\gg1/\tc$~\citep{gelbwaser2013work}. For this to happen, $\dot\omega(t)$ has to oscillate out of phase with $s(t)$.

\par

As discussed above in Sec.~\ref{sec_nm_impulsive}, an impulsive projective measurement decorrelates the system and the bath, thereby increasing their correlation energy by [\cf Eqs.~\eqref{eq_nm_Hsb_eq} and~\eqref{eq_nm_delta_HS}]
\begin{equation}\label{eq_nm_emeas}
  \Delta E_\mathrm{meas}=-\langle H_\mathrm{SB} \rangle _\mathrm{eq}>0.
\end{equation}
Here we encounter an important difference to the Szilard-Landauer (SL) scenario, where system-bath correlations are not accounted for. Assuming a cycle duration well below the bath memory time, $t_\mathrm{cycle}\ll\tc$, but longer than the time needed to perform the measurement that triggers the cycle, the maximal amount of extractable work is~\citep{erez2012thermodynamics}
\begin{equation}\label{eq_nm_wnsm_max}
  (W^\mathrm{ext}_\mathrm{NSM})_\mathrm{max}=\Delta E_\mathrm{meas}-T\Delta \mathcal{S}_\mathrm{meas},
\end{equation}
where $\Delta\mathcal{S}_\mathrm{meas}$ is the entropy increase due to the measurement. 

\par

\begin{figure}
  \centering
  \includegraphics[width=9cm]{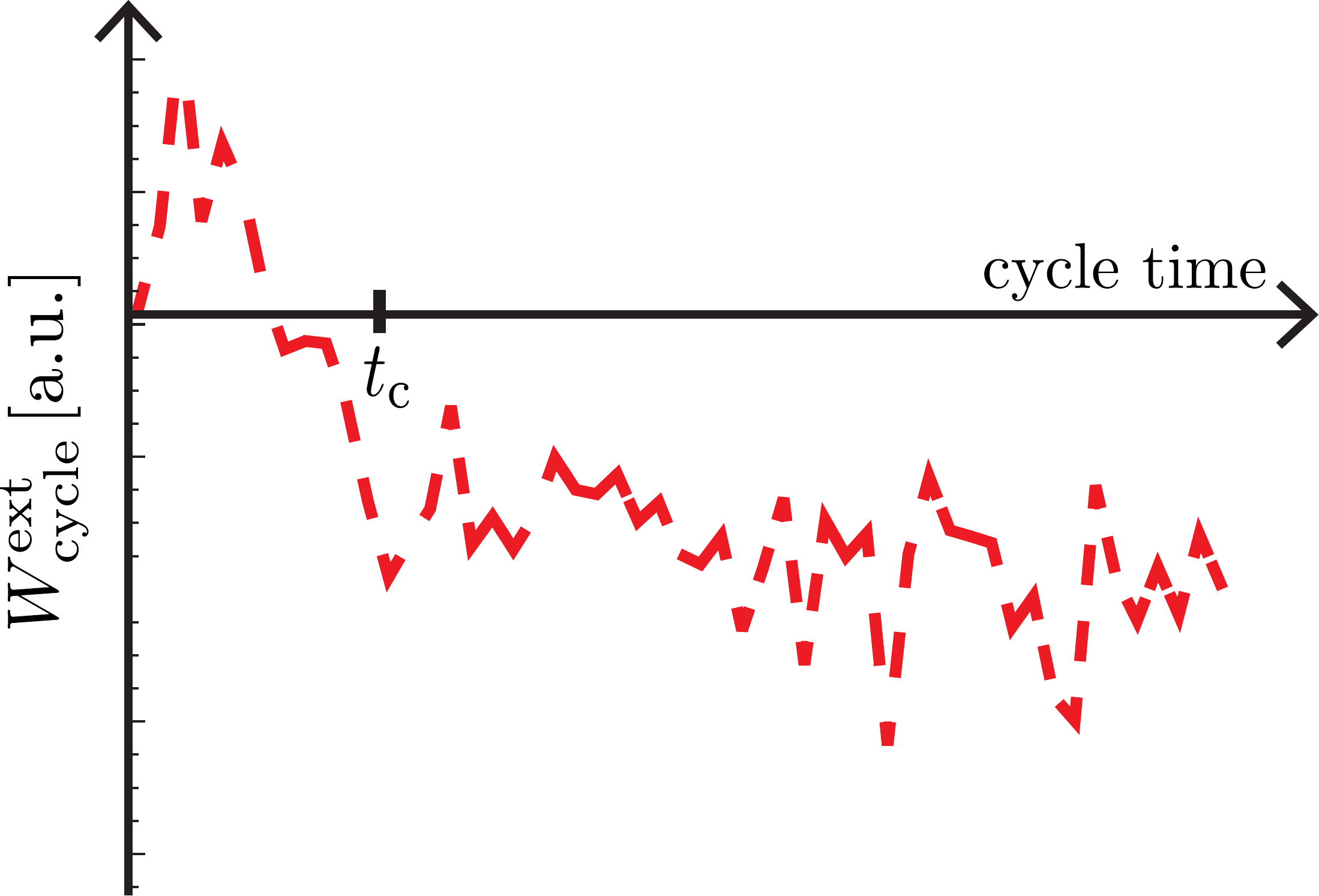}
  \caption{Simulation of the extractable work per cycle of the NSM-based scheme in Fig.~\ref{fig_szilard_cycle} as a function of the cycle duration (in units of the bath correlation/memory time $\tc$). It is seen that for cycles longer than $\tc$ positive work may not be extracted.}\label{fig_sl_work_extraction}
\end{figure}

A post-measurement cycle allowing to extract work ($W_\mathrm{cycle}^\mathrm{ext}>0$) can be designed. As in Secs.~\ref{sec_cycles},~\ref{sec_periodically-modulated}, and~\ref{sec_nlevels}, the ``piston'' is realized by a periodic modulation of the qubit's transition frequency of the form $\omega(t)=\omega_0+\kappa\sin\Omega t$. In the weak-modulation regime $\kappa\ll\Omega$, the work extracted within a cycle evaluates to~\citep{gelbwaser2013work}
\begin{equation}\label{eq_nm_w_ext_cycle}
  W^\mathrm{ext}_\mathrm{cycle}\approx- \kappa \int_{0}^{\frac{2\pi}{\Omega}}J_{g}(t)\Omega \cos(\Omega t) \dd t,
\end{equation}
with $J_{g}(t)\equiv\int_0^tR_g(t^\prime)\dd t^\prime$. Upon modulating the system on non-Markovian time scales (typically for $\Omega\gg\kappa\sim\omega_0$), Eq.~\eqref{eq_nm_w_ext_cycle} allows for either positive or negative work extraction, owing to its sign oscillation with $\Omega$. This means that the energy invested by performing the NSM [Eq.~\eqref{eq_nm_emeas}] can be partly extracted as work, but only in such a non-Markovian cycle for an appropriate choice of $\Omega$ and $\kappa$ (Fig.~\ref{fig_sl_work_extraction}). This invested energy can be in the form of \emph{noisy pulses}~\citep{alvarez2010zeno}.

\par
\par

\subsection{ Work-information relation for a non-Markovian cycle} 

Having shown that work extraction through a non-Markovian cycle following a NSM is possible, we now compare this result to the SL result. The relation between $(W^\mathrm{ext}_\mathrm{sel})_\mathrm{max}$ (for a selective measurement) and $(W^\mathrm{ext}_\mathrm{NSM})_\mathrm{max}$ (for a non-selective measurement) shows that the maximum work extractable from a selective measurement is \emph{higher} than the Szilard-Landauer (SL) bound $W_\mathrm{SL}$~\citep{gelbwaser2013work},
\begin{equation}\label{eq_nm_wsel_max}
  (W^\mathrm{ext}_\mathrm{sel})_\mathrm{max}=(W^\mathrm{ext}_\mathrm{NSM})_\mathrm{max}+W_\mathrm{SL}.
\end{equation}
As detailed above, the extra work $(W^\mathrm{ext}_\mathrm{NSM})_\mathrm{max}$ stems from system-bath correlations, which are not accounted for in the SL treatment. Equation~\eqref{eq_nm_wsel_max} demonstrates the difference between the Markovian and the non-Markovian case: System-bath entanglement is not included in the Markovian treatment and consequently $(W^\mathrm{ext}_\mathrm{NSM})_\mathrm{max}=0$ there. Hence, in the Markovian limit the maximum extractable work by a selective measurement adheres to the SL result.

\par

\begin{figure}
  \centering
  \includegraphics[width=9cm]{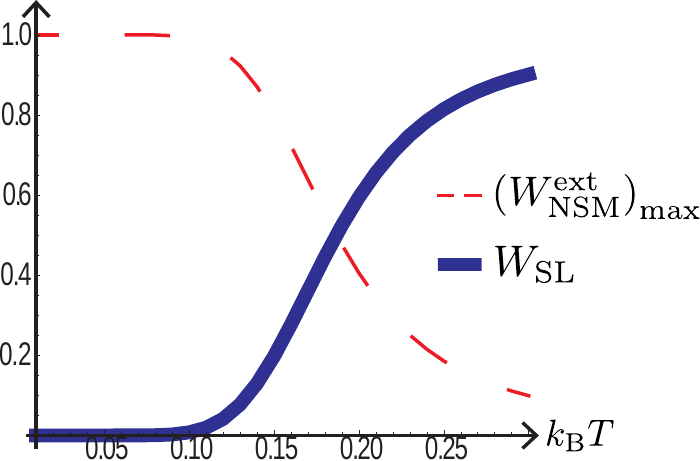}
  \caption{Comparison of the Szilard-Landauer (SL) bound on extractable work with that extractable in the NSM-based scheme of Fig.~\ref{fig_szilard_cycle}, as a function of the bath temperature $\kB T$. It is seen that at low temperatures the NSM-based work can greatly surpass the SL bound.}\label{fig_sl_work_comparison}
\end{figure}

Remarkably, the NSM allows for extracting work from a reservoir at $T=0$ without information gain~\citep{gelbwaser2013work}: The system-bath correlations~\eqref{eq_nm_Hsb_eq} are always negative, even for $T=0$. Decorrelation of the system and the bath through a measurement will increase this energy, such that the cycle is triggered, yielding $(W_\mathrm{sel}^\mathrm{ext})_\mathrm{max}=(W_\mathrm{NSM}^\mathrm{ext})_\mathrm{max}>0$. In this scenario all extractable work stems from this correlation energy change (see Fig.~\ref{fig_sl_work_comparison}).

\par
\par

\subsection{ Consistency with the second law}

The joint system-bath state at equilibrium $\rho_\mathrm{eq}$, which is entangled, is changed to a product state $\rho \otimes \rho_\mathrm{B}$ by the measurement (performed by a detector $D$). The total Hamiltonian describing this setup is
\begin{equation}
  H_\mathrm{tot}=H_\mathrm{S+B}+H_\mathrm{BD}+H_\mathrm{SD},
\end{equation}
with the system-detector interaction $H_\mathrm{SD}$ and the bath-detector interaction, $H_\mathrm{BD}$, respectively. This total Hamiltonian is $\tau$-periodic [$H_\mathrm{tot}(\tau)=H_\mathrm{tot}(0)$], such that the work extracted within a cycle amounts to~\citep{gelbwaser2013work}
\begin{equation}\label{eq_sl_wextmaxtot}
  W^\mathrm{ext}_\mathrm{tot}(\tau)=-\Tr \left[U(\tau)\rho_\mathrm{tot}(0)U^{\dagger}(\tau)H_\mathrm{tot}(0)-\rho_\mathrm{tot}(0)H_\mathrm{tot}(0)\right],
\end{equation}
which is the difference between the final and the initial mean energy. Owing to the unitary nature of the total time evolution, the entropy of the joint (system-bath-detector) density operator $\rho_\mathrm{tot}$ is fixed. Equation~\eqref{eq_sl_wextmaxtot} then yields
\begin{equation}\label{eq_nm_wtot}
  W^\mathrm{ext}_\mathrm{tot}= -\oint \Tr\left(\rho_\mathrm{tot}\dot{H}_\mathrm{tot}\right)\dd t=-\Delta E_\mathrm{meas}+W^\mathrm{ext}_\mathrm{cycle}<0.
\end{equation}
Here $\Delta E_\mathrm{meas}$ is as in Eq.~\eqref{eq_nm_emeas} and $W^\mathrm{ext}_\mathrm{cycle}$ is as in Eq.~\eqref{eq_nm_w_ext_cycle} and we have used the fact that the mean energy at fixed entropy is minimized for a thermal equilibrium state. Hence, no work can be extracted from a single bath by the entangled evolution of the \emph{joint} (system-bath-detector) ``supersystem'', consistently with the second law.

\subsection{Realizations and practical consequences}

Consider TLS (atoms or molecules) with resonance frequency $\omega_0$ in the microwave domain in a high-$Q$ cavity with controllable finite-temperature. QND measurements of the atomic-level population can be performed by injected optical probe pulses. Such QND probing may be performed at time intervals shorter than $\omega_0^{-1}$~\citep{gordon2009cooling}. 

\par

\begin{figure}
  \centering
  \includegraphics[width=9cm]{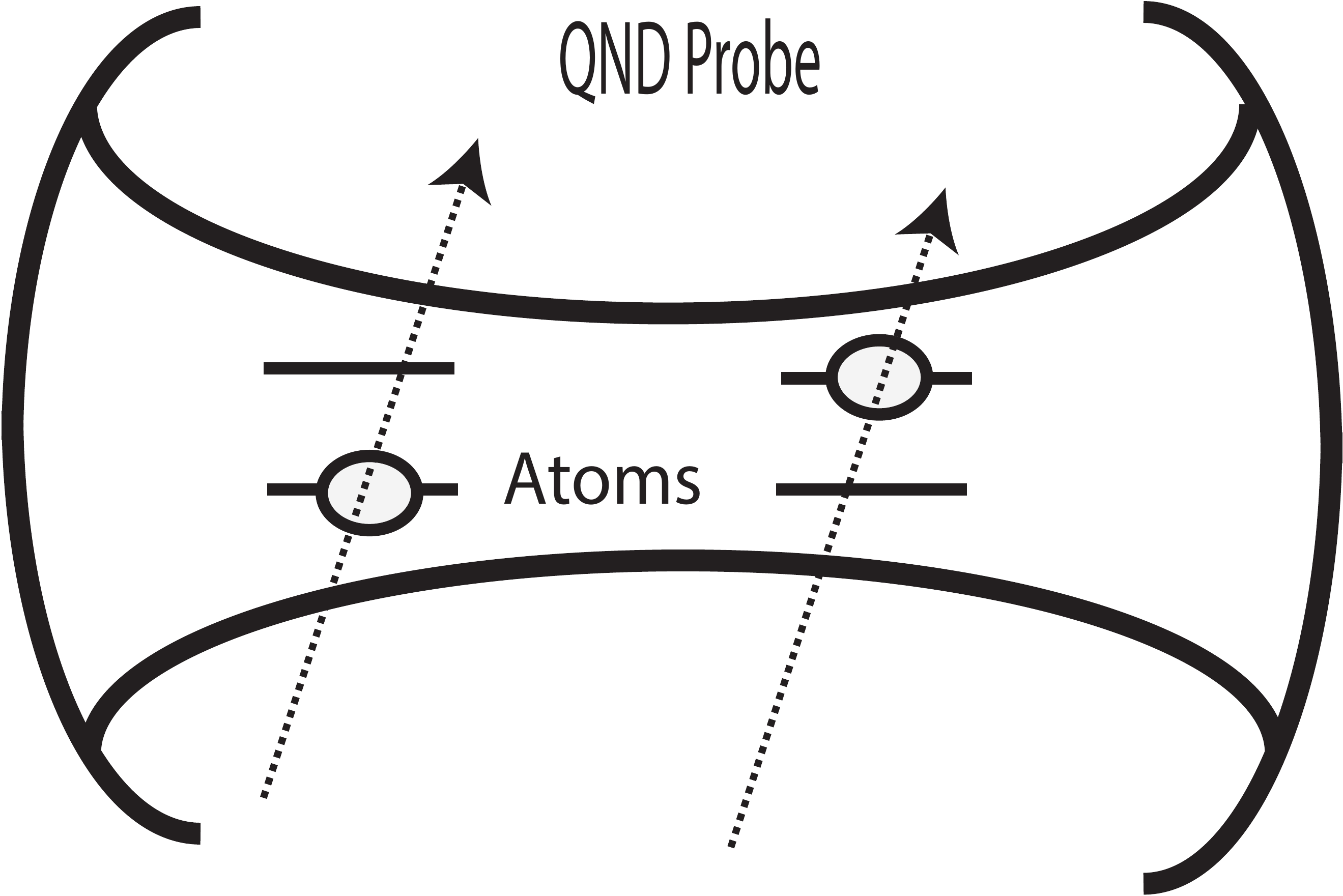}
  \caption{Schematic view of a possible setup for the realization of the NSM-based scheme in Fig.~\ref{fig_szilard_cycle}, using two-level atoms in a cavity (that serves as a bath) subject to periodic QND measurements and off-resonant energy modulation by optical (laser) probes.}\label{fig_qnd_cavity}
\end{figure}

The cavity modes constitute a bath with memory (correlation) times $\tc\sim10^{-4}\text{\,s}$~\citep{petrosyan2009reversible,verdu2009strong}. The system-bath correlation energy $\ew{\Hsb}$ (and thus the measurement-induced work) may then attain the GHz range. In this setup the piston is realized by an off-resonant coherent signal that modulates the atomic transition frequency at a rate $\Omega\gg1/\tc$. The QND measurement of the atomic population is performed by injected pulses (much shorter than the bath memory time). An amplification (lasing) of the off-resonant coherent piston mode will then signify work extraction (see Fig.~\ref{fig_qnd_cavity}).

\par
\par

\subsection{Summary}

The predicted work extraction above the Szilard-Landauer bound is at the expense of the change in the system-bath correlation energy (consistently with the first law). The second law is also adhered to when including the measuring device in the calculation of the energy and entropy changes.

\par

An application of this scenario might be the transformation of (possibly very noisy) energy input into work, which is manifest by a coherent amplification of the piston field. It might also be useful in situations where the coupling of the system to two different baths (as required by standard heat machines) is very hard to implement experimentally.

\section{Discussion and outlook}\label{sec_discussion}

\subsection{Motivation and main issues}

The study of models of thermal quantum machines is a necessary step in the strive to clarify the interplay between quantum mechanics and thermodynamics, and the limits set by thermodynamics to the performance of quantum devices. As discussed in this review, thermodynamics does impose limits on quantum processes and these limits are in general similar to the classical ones. Yet, \emph{extra resources} associated with ``quantumness'' may cause quantum machines to outperform their classical counterparts, while adhering to the second law: Multilevel degeneracy (quantum-correlated multiple thermalization pathways) can cause a power boost compared to a two-level machine (Sec.~\ref{sec_nlevels}); quantum-state preparation of the ``piston'' may endow a quantum heat machine with efficiency higher than the standard Carnot bound, in contrast to its semiclassical limit that conforms to the Carnot bound (Secs.~\ref{sec_quantum_piston} and~\ref{sec_refrigerator}); system-bath entanglement allows an engine to transgress the Szilard-Landauer bound (Sec.~\ref{sec_work_information}); Nernst's unattainability principle is challenged by the finite cooling speed of a quantum (magnon) bath as the absolute zero is approached (Sec.~\ref{sec_cooling_speed}). In what follows we shall dwell on the key aspects of these quantum resources and their potential applications.

\subsection{Continuous-cycle modulation-driven heat machines}

The Floquet analysis of periodically-driven open quantum systems presented here (Secs.~\ref{sec_cycles},~\ref{sec_periodically-modulated}, and~\ref{sec_nlevels}) shows that continuous-cycle quantum thermal machines may have an important edge over reciprocating (strokes-cycle) machines, of which the Carnot or the Otto cycles are known examples (Sec.~\ref{sec_introduction_basics}). Reciprocating-cycle operation incurs the entropic and energetic costs of abrupt consecutive strokes, especially when these strokes become rapid and non-adiabatic. This non-adiabaticity is not only difficult to keep track of, it also causes a decrease in the reciprocating-cycle efficiency as their non-adiabaticity increases~\citep{geva1992quantum,geva1992classical,curzon1975efficiency,kosloff2013quantum}. By contrast, the continuous-cycle $\sigmaz$-driven quantum heat machines (QHM) may attain the Carnot efficiency limit precisely because of its cycle non-adiabaticity (high modulation rate). Namely, it can be as efficient as the ideal, infinitely-slow, reciprocating heat engines.

\par

Yet, the periodically-driven continuous QHM, while being conceptually simple and performing efficiently under extreme non-adiabaticity, has a basic limitation: It reaches the Carnot efficiency bound provided that the hot and cold baths are spectrally separated. This ``price'' to be paid for having both baths permanently coupled to the system (working fluid) can be understood as follows. The ``natural'' energy-flow direction is from the hot bath to the cold one, without producing any work. In quantum oscillator baths, however, the heat will flow from the mode with a larger thermal occupancy to the one with a lower occupancy. Then, heat will flow from bath $1$ to bath $2$ if their respective products of inverse temperature and frequency satisfy $\beta_1\hbar\omega_1<\beta_2\hbar\omega_2$. Thus, spectral separation of the baths is required to select the modes and temperatures that produce work or cooling. Without modulation, the heat flows (currents) from the hot ($\indexh$) and the cold ($\indexc$) baths have the same absolute value with opposite signs, $\Jh=-\Jc>0$. As the modulation frequency increases, $\Jh$ decreases while $\Jc$ increases and may allow for work extraction, until they switch signs, thereby triggering refrigeration (\cf Fig.~\ref{fig_universal} in Sec.~\ref{sec_periodically-modulated}). These two operation regimes are separated by a critical modulation frequency, at which the machine reaches the maximum allowed efficiency, the Carnot bound, corresponding to zero power or cooling power (Sec.~\ref{sec_periodically-modulated}). 

\par

On the practical side, these continuous-cycle models may be useful for understanding and optimizing future nanomachines. At nanoscales, systems are completely embedded by their surrounding, making on-off switching of the coupling to thermal baths impossible. Hence, ``strokes'' cycle models may not be useful and systems permanently coupled to the thermal baths may become indispensable. 

\par

Are there alternative continuous-cycle models? In Sec.~\ref{sec_quantum_cooler} a scheme that does not require spectrally separated baths for cooling was presented. It aims at cooling a bath that dephases two-level systems continuously driven by a field, e.g., it achieves buffer-gas (BG) cooling via its collisions with laser-driven two-level atoms~\citep{vogl2009laser,vogl2011collisional}. The thermodynamic analysis of this process (Sec.~\ref{sec_quantum_cooler}) reveals that it may be viewed as a laser-driven heat distributor rather than a refrigerator---it uses the laser power to cool down the BG without any restriction on the bath spectra, as long as the laser is red-detuned from the atomic resonance. This cooling may attain high efficiency at ultralow temperatures.

\subsection{Multilevel degeneracy as a resource in heat machines}

The generalization of the minimal universal QHM based on a two-level system (Sec.~\ref{sec_periodically-modulated}) to the degenerate multilevel case reveals the possibility of a significant enhancement of the engine's output power (Sec.~\ref{sec_nlevels}). An enhancement (boost) is achievable for any transition-dipole orientation, but its temperature dependence is determined by the dipole alignment. The maximally possible enhancement over the entire temperature range is achieved if all dipole transition vectors are aligned, provided that coherently-superposed multilevel dark states are initially unpopulated. The initial population that yields maximal enhancement can then be an arbitrary mixture of the ground and the bright (coherently-superposed) state populations. In particular, an initial full population of the ground state ensures maximal enhancement. Hence, initial (or steady-state) coherence is neither required nor is it always beneficial for the performance boost. In fact, the power boost depends on the \emph{ground-state population} of the working fluid in steady state. Notwithstanding this boost, the efficiency or COP of the heat machine is not altered compared to its TLS counterpart.

\subsection{Fully quantized versus semiclassical heat machines}\label{sec_discussion_fully_quantized_vs_semiclassical}

In analogy to atom-light interaction theory, the machines surveyed above are termed semiclassical, because they are driven by an external classical field. A step towards the complete quantization of heat engines was taken here: A ``quantum piston'' was used as the drive instead of a classical external field (Secs.~\ref{sec_quantum_piston} and~\ref{sec_refrigerator}). Yet, fully-quantized heat machines pose conceptual challenges. Whilst it is clear how to calculate the energy exchange between interacting quantum systems, it is unclear how to divide it into work and heat (Sec.~\ref{sec_introduction_carnot}). Some of the different proposals for defining the work exchange at the quantum level claim an efficiency higher than the standard Carnot bound~\citep{boukobza2006thermodynamics,schroeder2010work,horodecki2013fundamental,goswani2013thermodynamics,allahverdyan2005minimal}. Yet, the origin of the extra efficiency in these proposals is not clear, casting doubts on the definitions they use.

\par

The definition of work based on \emph{passivity}~\citep{lenard1978thermodynamical,pusz1978passive} was shown here to comply with thermodynamics and retrieve in the semiclassical limit all the results of classical external driving (by a semiclassical piston). By contrast, work extraction by a quantum piston was shown not only to depend on the heat currents and the spectral separation of the baths, but also on the piston state's non-passivity: High work efficiency requires sustainable non-passive, stable piston states (resilient to thermalization) that yield a low entropy production, for example an initial coherent state. Conversely, no work is extracted for a state that rapidly loses its non-passivity, e.g., an initial Fock state (Sec.~\ref{sec_quantum_piston}). By contrast, in the refrigeration regime, the cooling COP is increased when the piston is initially in a Fock state (that evolves with a high entropy production) (Sec.~\ref{sec_refrigerator}). Remarkably, the work or the cooling efficiency in a fully quantized heat machine may surpass the standard Carnot limit. The origin of this extra efficiency is the possibility of the piston to absorb heat, adding an extra thermodynamic resource to the heat machine~\citep{gelbwaser2014heat}. A new classification of quantum states based on their work extraction and cooling capacities need be pursued. In the case of work extraction, there are indications that pointer states~\citep{zurek1981pointer} are the best candidates in terms of their sustainable non-passivity. 

\par

Fully quantum heat machines may play a special r\^ole in nanotechnological applications. Their quantized piston can allow maximal compactness (miniaturization) and completely autonomous (self-contained) operation.

\subsection{Applications to quantum refrigeration and work extraction}

An important potential application of the present results is towards overcoming the main bottleneck of electronics miniaturization, which is the heat produced in the microprocessor. As microprocessors become smaller and denser, heat generation may render them unusable. Hence, to continue the miniaturization trends, new and more advanced cooling methods are required~\citep{petrosyan2009reversible,pop2006heat,taylor2011laser,oconnell2010quantum,zhao2012atomic,gordon2009cooling} than standard cooling of gases and solids~\citep{haensch1975cooling,chu1998nobel,cohentannoudji1998nobel,phillips1998nobel,djeu1981laser,zander1995cooling,epstein1995observation,hoyt2003advances,thiede2005cooling,sheik2007optical}. The sought methods must be thermodynamically efficient at quantum scales.

\par
 
This question has largely motivated the recent interest in models of quantum refrigerators (QR)~\citep{kosloff2013quantum}. Yet, most of these models have been inspired by the Carnot or Otto cycle~\citep{palao2001quantum,linden2010how,levy2012quantum,geva1996quantum,correa2013performance,venturelli2013minimal,brunner2014entanglement,geusic1967quantum,gordonbook,velasco1997new,allahverdyan2010optimal}. 

\par

By contrast, impurities and double-well qubits embedded in appropriate environments may act, depending on their energy modulation, in either the engine (QHE) or the refrigerator (QR) mode of the proposed continuous cycle QHM~\citep{gelbwaser2013minimal}, with potentially significant technological advantages. In particular, an impurity or quantum-dot fast-modulated qubit ``sandwiched'' between two nanosize solid layers (Fig.~\ref{fig_qdot_heat_machine}) may act as a nanoscopic refrigerator (heat-pump) of a transistor (chip), that is much more miniaturized and less power-consuming than currently available microelectronic refrigerators~\citep{pop2006heat}. 

\par

Under slower modulation rate, the same setup may act as electron-current generator without external voltage bias, and as a substitute for phase-coherent electron-current control~\citep{kurizki1989phase}. The realization of these novel machines in NV-center spin ensembles or nanomechanical platforms will constitute a major technological leap, as their present counterparts are based on traditional thermodynamical principles.

\par

\begin{figure}
  \centering
  \includegraphics[width=5cm]{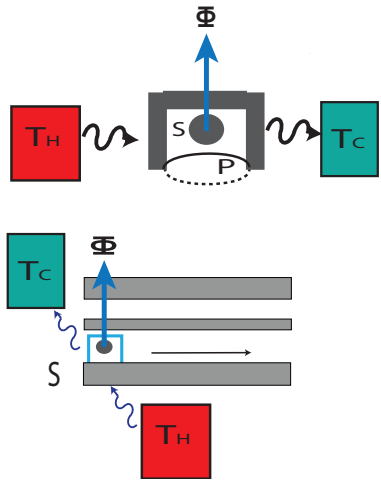}
  \caption{Possible realizations of a fully quantum heat machine. Above: Coupling of a qubit ($S$) in a cavity to a piston ($P$) realized by mirror strain, and to two heat baths. Below: A superconducting qubit coupled to magnetic flux in a superconducting (coplanar waveguide) cavity whose mode serves as piston, and to two heat baths.}\label{fig_realizations}
\end{figure}

\begin{figure}
  \centering
  \includegraphics[width=9cm]{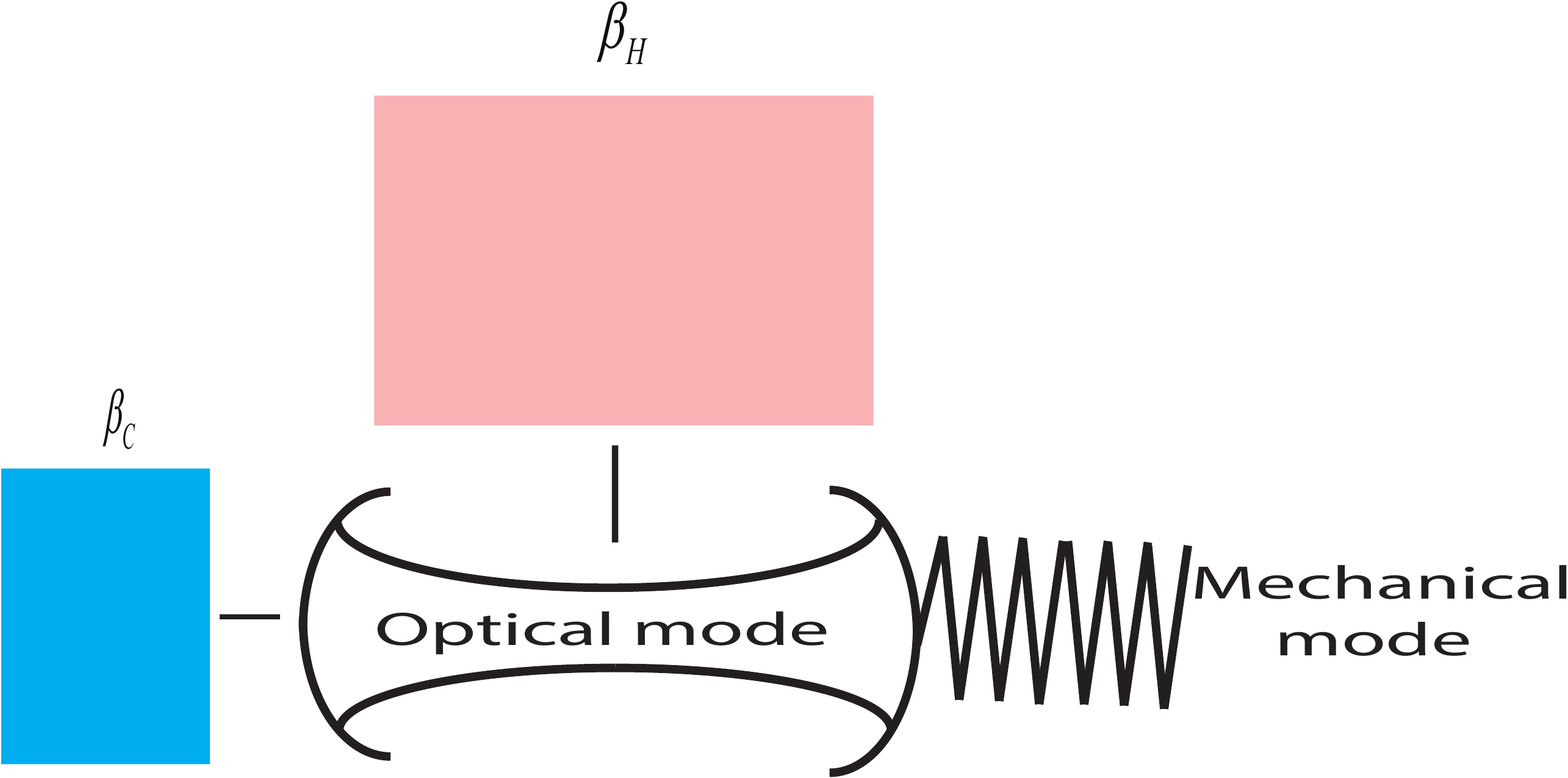}
  \caption{Optomechanical realization of a quantum heat machine in a cavity setup.}\label{fig_optomechanical}
\end{figure}

Fully-quantized system-piston couplings analyzed in Secs.~\ref{sec_quantum_piston} and~\ref{sec_refrigerator} are realizable in well-investigated experimental setups: 
\begin{itemize}
\item A superconducting flux qubit $S$ dispersively coupled to the piston $P$ may be realized by a high-$Q$ (phonon) mode of a nanomechanical cavity (or cantilever), as shown in Fig.~\ref{fig_realizations}. The quantized position of $P$ (cavity-mirror strain) affects the magnetic flux through the qubit and thereby modulates its frequency $\omega_0$.
\item Alternatively, $P$ can be a field mode of a coplanar resonator~\citep{blais2004cavity} whose quantized electromagnetic field amplitude affects the magnetic flux through a superconducting qubit (see Fig.~\ref{fig_realizations}). In such scenarios, it should be possible to amplify the output signal within the $P$-mode coherence time. Similar considerations may apply to cold atoms~\citep{petrosyan2009reversible} or spin ensembles~\citep{kubo2010strong} in high-$Q$ cavities.
\item In an optomechanical cavity-setup the mechanical degree of freedom may play the r\^ole of the piston, the optical field mode be the working fluid, and the electromagnetic vacuum may act as the cold bath. Optical couplers may be used to inject radiation into the cavity from a hot bath, e.g., solar radiation (see Fig.~\ref{fig_optomechanical}). The ability to prepare the quantum state of the mechanical piston~\citep{aspelmeyer2014cavity,vanner2011pulsed,rimberg2014cavity} identify optomechanical setups as a leading platform for exploiting the thermodynamic properties of quantum states surveyed here~\citep{gelbwaser2015work}.
\end{itemize}

\par
\par
\par
\par

\subsection{Third law and unattainability challenged}

Quantum machines must comply with the second law of thermodynamics, but may differ from their classical counterparts concerning compliance with the third law formulated as Nernst's unattainability principle (Sec.~\ref{sec_cooling_speed}). A quantum refrigerator with the appropriate coupling may approach the absolute zero at finite time, as predicted for certain realistic quantized baths, e.g., quantized spin-waves in ferromagnets (magnons)~\citep{kolar2012quantum}. However, any other baths unaccounted for by the model, as well as deviations from the weak-coupling dynamics assumed here may restore the third law. Experimental verification is needed to show whether the unattainability principle may indeed be broken or at least whether the predicted non-vanishing of the cooling speed near the absolute zero holds in reality.

\subsection{The Szilard-Landauer bound challenged}

Another principle that may be challenged for quantum machines is the Szilard-Landauer work-information tradeoff (Sec.~\ref{sec_work_information}). In the non-Markovian regime (Sec.~\ref{sec_nonmarkovian}), where the system-bath separability fails, the proposed engine may provide more work than what was originally expected by Szilard-Landauer if the cycle is completed within the memory time of the bath. This result demonstrates the need of further investigating thermodynamic effects not only at the quantum level, but also in the non-Markovian regime. Besides this fundamental aspect, the present engine model, in which the system is always coupled to a single bath and yet may perform useful work, is potentially important for nanoscale systems totally embedded in a single bath, where conventional thermodynamic cycles may be impossible to implement. The model reveals the possibility of extracting work from changes in the system-bath quantum-correlation energy, which constitute a hitherto unexploited thermodynamic (work) resource.

\subsection{Open issues}

Despite the important advances in quantum thermodynamics in the recent years, certain fundamental questions remain still open. Below we present a short and non-exhaustive, list of possible future investigation paths.
\begin{itemize}
\item Although many quantum heat machine models have been proposed and studied~\citep{geusic1967quantum,ford2006quantum,ford1985quantum,geva2002irreversible,quan2007quantum,vandenbroeck2005thermodynamic,lin2003performance,esposito2009universality,kieu2004second,linden2010how,scully2011quantum,blickle2011realization,alicki1979quantum,allahverdyan2005minimal,talkner2007fluctuation,campisi2009fluctuation,campisi2011colloquium,jarzynski2007comparison,vandenbroeck2012efficiency}, important systems with inherent quantum effects, such as solar cells and photosynthesis~\citep{wuerfelbook,nelsonbook,engel2007evidence,collini2010coherently,panitchayangkoon2010long,lee2007coherence} have thus far eluded full analysis from a quantum-thermodynamic perspective; see recent discussion by~\citep{alicki2015solar,einax2011heterojunction}.
\item Most of the quantum heat machine models assume that the working fluid interacts with heat reservoirs that are at thermal equilibrium. Yet, pumping by incident radiation or collisions, which are often treated as thermal baths, must be described as non-equilibrium baths. Kelvin's formulation of the second law restricts work extraction to engines which operate between two equilibrium (thermal) baths, but is it possible to extract work from a quantum system coupled to a \emph{single non-equilibrium bath}? 

\par

Among the proposed non-equilibrium baths, \emph{squeezed baths} have been claimed to yield higher than Carnot efficiencies in a QHM, because they supply more energy than non-squeezed thermal baths~\citep{rossnagel2014nanoscale,abah2014efficiency}. Yet, our non-passivity analysis suggests that such a bath may \emph{increase the non-passivity} of the working fluid, and thus endow it with extra work rather than heat. The resulting efficiency is then expected to comply with the Carnot bound, because the extra work produced by the working fluid is compensated by the extra work provided by the bath. However, a more comprehensive analysis of such scenarios is called for, resorting to the notion of non-passivity.
\item The controversy discussed above concerning the ability to surpass the Carnot bound by coupling the system (working fluid) to nonthermal baths expresses the present lack of consensus on the proper definition of work in quantized setups (\cf Sec.~\ref{sec_discussion_fully_quantized_vs_semiclassical}). The criteria for such a definition should be the possibility to convert it into mechanical work and its compliance with the first and second laws of thermodynamics. We contend that non-passivity satisfies these criteria.
\item The models described here and elsewhere assume a weak coupling between the system and the bath. This assumption not only simplifies the analysis, but also reflects the established notion that system-bath separability is essential for thermodynamics. On the other hand, the output of the machines is proportional to this coupling, hence negligible coupling renders them useless for practical applications. Furthermore, as shown in Sec.~\ref{sec_nonmarkovian}, finite coupling may have qualitative implications, such as the transgression of the Szilard-Landauer bound. Hence, the strong-coupling limit should be explored in order to find out if quantum heat machines are at all possible in this limit and, if so, what would be their performance bound.
\item The open issues raised above indicate that our results represent only a first step towards the clarification of the thermodynamic r\^ole of quantum resources, as there is currently no theory that addresses such issues in a general fashion. Consequently, the dependence of the ultimate thermodynamic bounds on quantum resources must be further explored.
\end{itemize}

\section*{Acknowledgments}
This work has been supported by the US-Israel BSF, ISF, AERI and CONACYT.

\end{document}